\documentclass[11pt]{article}
\usepackage{jcappub}
\usepackage{bm}
\usepackage{color}
\usepackage{array}
\usepackage{graphicx}
\usepackage{multirow}
\newcolumntype{P}[1]{>{\centering\arraybackslash}p{#1}}
\newcolumntype{M}[1]{>{\centering\arraybackslash}m{#1}}
\def\nn{\nonumber}
\newcommand{\be}{\begin{equation}}
\newcommand{\ee}{\end{equation}}
\newcommand{\een}{\end{subequations}}
\newcommand{\ben}{\begin{subequations}}
\newcommand{\beq}{\begin{eqalignno}}
\newcommand{\eeq}{\end{eqalignno}}

\newcommand{\lsim}{\mathrel{\mathop{\kern 0pt \rlap
      {\raise.2ex\hbox{$<$}}}\lower.9ex\hbox{\kern-.190em $ \sim$}}}
\newcommand{\gsim}{\mathrel{\mathop{\kern 0pt
      \rlap{\raise.2ex\hbox{$>$}}}\lower.9ex\hbox{\kern-.190em $\sim$}}}

\newcommand{\CO}{\mathcal{O}}

\newcommand{\erf}{\mbox{erf}}

\title{DAMA/LIBRA-phase2 in WIMP effective models}

\author{Sunghyun Kang,}
\author{Stefano Scopel,}
\author{Gaurav Tomar,}
\author{Jong--Hyun Yoon}
\affiliation{Department of Physics, Sogang University, 
Seoul, Korea, 121-742}

\emailAdd{scopel@sogang.ac.kr}
\emailAdd{tomar@sogang.ac.kr}
\emailAdd{francis735@naver.com}
\emailAdd{jyoon@sogang.ac.kr}

\abstract{The DAMA/LIBRA collaboration has recently released updated
  results from their search for the annual modulation signal expected
  from Dark Matter (DM) scattering in their NaI detectors. We have
  fitted the updated DAMA result for the modulation amplitudes in
  terms of a Weakly Interacting Massive Particle (WIMP) signal,
  parameterizing the interaction with nuclei in terms of the most
  general effective Lagrangian for a WIMP particle spin up to 1/2,
  systematically assuming dominance of one of the 14 possible
  interaction terms, and assuming for the WIMP velocity distribution a
  standard Maxwellian.  We find that most of the couplings of the
  non--relativistic effective Hamiltonian can provide a better fit
  compared to the standard Spin Independent interaction case, and with
  a reduced fine--tuning of the three parameters (WIMP mass,
  WIMP--nucleon effective cross-section and ratio between the
  WIMP--neutron and the WIMP--proton couplings).  Moreover, effective
  models for which the cross section depends explicitly on the WIMP
  incoming velocity can provide a better fit of the DAMA data at large
  values of $m_{\chi}$ compared to the standard velocity--independent
  cross--section due to a different phase of the modulation
  amplitudes.  All the best fit solutions are in tension with
  exclusion plots of both XENON1T and PICO60.}

\begin{document}

\maketitle

\section{Introduction}
\label{sec:introduction}
Weakly Interacting Massive Particles (WIMPs) are the most popular
candidates to provide the Dark Matter (DM) which is believed to make
up 27\% of the total mass density of the Universe~\cite{planck} and
more than 90\% of the halo of our Galaxy, and a worldwide experimental
effort is under way to detect them. In particular the DAMA
experiment~\cite{dama_2008,dama_2010,dama_2013} has been measuring for
more than 15 years a yearly modulation effect with a sodium iodide
target consistent with that expected due to the Earth rotation around
the Sun from the elastic scattering of WIMPs, claiming a statistical
significance of more than 9 $\sigma$.  Many experimental
collaborations using nuclear targets different from $NaI$ and various
background--subtraction techniques to look for WIMP--elastic
scattering (XENON1T~\cite{xenon_1t}, LUX~\cite{lux},
XENON100~\cite{xenon100}, XENON10~\cite{xenon10},
KIMS~\cite{kims,kims_modulation}, CDMS-$Ge$~\cite{cdms_ge},
CDMSlite~\cite{cdms_lite}, SuperCDMS~\cite{super_cdms}, CDMS
II~\cite{cdms_2015}, SIMPLE~\cite{simple}, COUPP~\cite{coupp},
PICASSO~\cite{picasso}, PICO-2L~\cite{pico2l}, PICO-60~\cite{pico60})
have failed to observe any anomaly so far, implying severe constraints
on the most popular WIMP scenarios used to explain the DAMA excess.

Recently the DAMA collaboration has released first result from the
upgraded DAMA/ LIBRA-phase2 experiment \cite{dama_2018}. The two most
important improvements compared to the previous data is that now the
exposure has almost doubled and the energy threshold has been lowered
from 2 keV electron--equivalent (keVee) to 1 keVee. In particular,
this latter feature has improved the chances to exploit the DAMA
annual modulation amplitudes spectral energy shape to test specific
WIMP models. The most popular of them, predicted in ultraviolet
completions of the Standard Model such as Supersymmety or Large Extra
Dimensions, implies a Spin--Independent (SI) WIMP--nucleus scattering
cross section $\sigma_{\chi N}$ that scales with the square of the
number of nucleon targets in the nucleus:

\begin{equation}
  \sigma_{\chi N}\propto \left [  c^p Z + (A-Z) c^n\right ]^2,
\label{eq:si}
  \end{equation}

\noindent with $A$ the nuclear mass number, $Z$ the nuclear charge and
$c^{p,n}$ the WIMP couplings to protons and neutrons, with $c^n$=$c^p$
(i.e. an isoscalar interaction) in the most natural realizations. On
the other hand, the expected WIMP--induced scattering spectrum depends
on a convolution on the velocity distribution $f(\vec{v})$ of the
incoming WIMPs, usually described by a thermalized non--relativistic
gas described by a Maxwellian distribution whose r.m.s. velocity
$v_{rms}\simeq$ 270 km/s is determined from the galactic rotational
velocity by assuming equilibrium between gravitational attraction and
WIMP pressure.  Indeed, such model, usually referred to as Isothermal
Sphere, is confirmed by numerical simulations\cite{kelso}, although
the detailed merger history of the Milky Way is not known, allowing
for the possibility of the presence of sizable non--thermal components
for which the density, direction and speed of WIMPs are hard to
predict\cite{green}.

The combination of a SI isoscalar cross section with the Isothermal
Sphere model has provided for a long time a good fit to the DAMA
results in Refs.\cite{dama_2008,dama_2010,dama_2013}, either for a
light WIMP mass, $m_{\chi}\simeq$ 10 GeV
\cite{bottino_low_mass1,bottino_low_mass2}, or for a heavy WIMP mass
$m_{\chi}\lsim$ 100 GeV. However, as pointed out in
Refs.\cite{freese_2018} and \cite{Kahlhoefer:2018}, with the new DAMA
data the goodness of fit of such scenario has considerably worsened,
and is now disfavored (at 5.1 $\sigma$ for the low-mass solution,
$m_{\chi}\simeq$ 8 GeV and at 3.2 $\sigma$ for the high mass solution,
$m_{\chi}\simeq$ 53 GeV~\cite{freese_2018}). On the other hand an
acceptable fit can be obtained by allowing for a substantial isovector
component in the WIMP--nucleon interaction, although at the price of
tuning the coupling ratio in order to suppress the WIMP-iodine
interaction.

Although theoretically motivated, a SI WIMP--nucleus cross section is
not the only possible WIMP--nucleus interaction.  Actually, the
non-observation so far of new physics at the Large Hadron Collider has
strongly prompted for the necessity to go beyond this ``top-down"
approach in order to extend the search of Dark Matter candidates to a
wider range of properties through an alternative ``bottom-up" strategy
not biased by theoretical prejudice.  In particular the WIMP--nucleus
cross section can be parameterized in terms of the most general
non--relativistic effective theory complying with Galilean symmetry,
including a possible explicit dependence of $\sigma_{\chi N}$ on the
transferred momentum and on the WIMP incoming
velocity~\cite{dobrescu_eft,reece_eft,haxton1,haxton2}.

Also within such effective framework it has been shown that a strong
tension persists between an interpretation of the DAMA modulation
effect in terms of a WIMP signal and the results of null experiments
\cite{catena_dama}, if a Maxwellian velocity distribution for the
WIMPs is assumed. Nevertheless no fit of the DAMA result is available
in the literature in terms of non--relativistic EFT models.  Moreover,
in addition to increasing the exposure, the phase2 result also
includes a lower energy threshold, and the new spectrum of modulation
amplitudes no longer shows a maximum, but is rather monotonically
decreasing with energy. In light of these differences it is
significant to extend an assessment of the goodness of fit of the new
DAMA result to such scenarios.  To this aim, and making the same
assumptions on the WIMP velocity distribution, in the present paper we
wish to discuss how effective WIMP--nucleus interactions can fit the
new DAMA data. On top of that, in the analysis of
Ref.\cite{catena_dama} it was shown that a combination of xenon and
fluorine targets (namely XENON1T and PICO60) was needed to exclude all
the effective theory parameter space. Such assessment made only use of
the size of the modulation amplitudes in the first three bins of the
experimental result in Refs.\cite{dama_2008,dama_2010, dama_2013}, but
did not exploit the goodness of fit information. So in the present
paper we will also compare the best--fit parameter space of a WIMP
interpretation of the DAMA result to the present constraints from
XENON1T and PICO60.

In our approach we will consider the most general WIMP--nucleus
effective Lagrangian for a WIMP particle of spin 0 or spin 1/2
scattering elastically off nuclei, systematically assuming the
dominance of one of the 14 possible interaction terms of the most
general non--relativistic Hamiltonian invariant by Galilean
transformations~\cite{haxton1,haxton2}, fitting the new DAMA data to
the three parameters $m_{\chi}$ (WIMP mass), $\sigma_p$ (WIMP--nucleon
effective cross-section) and $c^n/c^p$. Specifically, the goals of our
analysis are i) to check if a fit to the DAMA data better than in the
standard SI case can be obtained by using any of the non--standard
interactions of the non--relativistic effective theory; ii) if this is
possible with less fine--tuning of the parameters (in particular of
the ratio $c^n/c^p$) compared to the SI case; iii) to check in each
case the level of tension between a DAMA interpretation in terms of a
WIMP annual modulation signal and the null results from other
experiments (in the following we will consider the two representative
bounds from XENON1T and PICO60).

The paper is organized as follows. In Section \ref{sec:eft} we
summarize the non--relativistic Effective Field Theory (EFT) approach
of Ref.\cite{haxton1,haxton2} and provide the formulas to calculate
expected rates for WIMP--nucleus scattering; in Section
\ref{sec:analysis} we analyze the DAMA result including the latest
upgrade by comparing the measured modulation amplitudes to the
calculated ones in a chi square analysis where in a systematic way
each of the couplings of the effective non--relativistic Hamiltonian
is assumed to be the dominant one. We will provide our conclusions in
Section \ref{sec:conclusions}. In Appendix \ref{app:wimp_eft} we
provide for completeness the WIMP response functions for the
non--relativistic effective theory while in Appendix \ref{app:exp} we
provide the details of the calculations of the constraints from
XENON1T and PICO60.

\section{WIMP rates in non--relativistic effective models}
\label{sec:eft}

Making use of the non--relativistic EFT approach of
Ref.\cite{haxton1,haxton2} the most general Hamiltonian density
describing the WIMP--nucleus interaction can be written as:

\begin{eqnarray}
{\bf\mathcal{H}}({\bf{r}})&=& \sum_{\tau=0,1} \sum_{j=1}^{15} c_j^{\tau} \mathcal{O}_{j}({\bf{r}}) \, t^{\tau} ,
\label{eq:H}
\end{eqnarray}

\noindent where:

\begin{eqnarray}
  \CO_1 &=& 1_\chi 1_N; \;\;\;\; \CO_2 = (v^\perp)^2; \;\;\;\;  \CO_3 = i \vec{S}_N \cdot ({\vec{q} \over m_N} \times \vec{v}^\perp) \nn\\
  \CO_4 &=& \vec{S}_\chi \cdot \vec{S}_N;\;\;\;\; \CO_5 = i \vec{S}_\chi \cdot ({\vec{q} \over m_N} \times \vec{v}^\perp);\;\;\;\; \CO_6=
  (\vec{S}_\chi \cdot {\vec{q} \over m_N}) (\vec{S}_N \cdot {\vec{q} \over m_N}) \nn \\
  \CO_7 &=& \vec{S}_N \cdot \vec{v}^\perp;\;\;\;\;\CO_8 = \vec{S}_\chi \cdot \vec{v}^\perp;\;\;\;\;\CO_9 = i \vec{S}_\chi \cdot (\vec{S}_N \times {\vec{q} \over m_N}) \nn\\
  \CO_{10} &=& i \vec{S}_N \cdot {\vec{q} \over m_N};\;\;\;\;\CO_{11} = i \vec{S}_\chi \cdot {\vec{q} \over m_N};\;\;\;\;\CO_{12} = \vec{S}_\chi \cdot (\vec{S}_N \times \vec{v}^\perp) \nn\\
  \CO_{13} &=&i (\vec{S}_\chi \cdot \vec{v}^\perp  ) (  \vec{S}_N \cdot {\vec{q} \over m_N});\;\;\;\;\CO_{14} = i ( \vec{S}_\chi \cdot {\vec{q} \over m_N})(  \vec{S}_N \cdot \vec{v}^\perp )  \nn\\
  \CO_{15} &=& - ( \vec{S}_\chi \cdot {\vec{q} \over m_N}) ((\vec{S}_N \times \vec{v}^\perp) \cdot {\vec{q} \over m_N}),
\label{eq:ops}
\end{eqnarray}

\noindent In the above equation $1_{\chi N}$ is the identity operator,
$\vec{q}$ is the transferred momentum, $\vec{S}_{\chi}$ and
$\vec{S}_{N}$ are the WIMP and nucleon spins, respectively, while
$\vec{v}^\perp = \vec{v} + \frac{\vec{q}}{2\mu_{\chi N}}$ (with
$\mu_{\chi N}$ the WIMP--nucleon reduced mass) is the relative
transverse velocity operator satisfying $\vec{v}^{\perp}\cdot
\vec{q}=0$. For a nuclear target $T$ it can also be written as:

\begin{equation}
(v^{\perp}_T)^2=v^2_T-v_{min}^2.
\label{eq:v_perp}
\end{equation}

\noindent where, for WIMP--nucleus elastic scattering:

\begin{equation}
v_{min}^2=\frac{q^2}{4 \mu_{T}^2}=\frac{m_T E_R}{2 \mu_{T}^2},
\label{eq:vmin}
\end{equation}

\noindent represents the minimal incoming WIMP speed required to
impart the nuclear recoil energy $E_R$, while $v_T\equiv|\vec{v}_T|$
is the WIMP speed in the reference frame of the nuclear center of
mass, $m_T$ the nuclear mass and $\mu_{T}$ the WIMP--nucleus reduced
mass. Moreover $t^0=1$, $t^1=\tau_3$
denote the the $2\times2$ identity and third Pauli matrix in isospin
space, respectively, and the isoscalar and isovector (dimension -2)
coupling constants $c^0_j$ and $c^{1}_j$, are related to those to
protons and neutrons $c^{p}_j$ and $c^{n}_j$ by
$c^{p}_j=(c^{0}_j+c^{1}_j)/2$ and $c^{n}_j=(c^{0}_j-c^{1}_j)/2$.

Operator ${\cal O}_2$ is of higher order in $v$ compared
to all the others, implying a cross section suppression of order
${\cal O}(v/c)^4)\simeq 10^{-12}$ for the non--relativistic WIMPS in
the halo of our Galaxy. Moreover it cannot be obtained from the
leading-order non relativistic reduction of a manifestly relativistic
operator \cite{haxton1}.  So, following Ref.\cite{haxton1,haxton2},
we will not include it in our analysis.

Assuming that the nuclear interaction is the sum of the interactions
of the WIMPs with the individual nucleons in the nucleus the WIMP
scattering amplitude on the target nucleus $T$ can be written in the
compact form:

\begin{equation}
  \frac{1}{2 j_{\chi}+1} \frac{1}{2 j_{T}+1}|\mathcal{M}|^2=
  \frac{4\pi}{2 j_{T}+1} \sum_{\tau=0,1}\sum_{\tau^{\prime}=0,1}\sum_{k} R_k^{\tau\tau^{\prime}}\left [c^{\tau}_j,(v^{\perp}_T)^2,\frac{q^2}{m_N^2}\right ] W_{T k}^{\tau\tau^{\prime}}(y).
\label{eq:squared_amplitude}
\end{equation}

\noindent In the above expression $j_{\chi}$ and $j_{T}$ are the WIMP
and the target nucleus spins, respectively, $q=|\vec{q}|$ while the
$R_k^{\tau\tau^{\prime}}$'s are WIMP response functions (that we
report for completeness in Eq.(\ref{eq:wimp_response_functions}))
which depend on the couplings $c^{\tau}_j$ as well as the transferred
momentum $\vec{q}$ and 
$(v^{\perp}_T)^2$. In equation (\ref{eq:squared_amplitude}) the
$W^{\tau\tau^{\prime}}_{T k}(y)$'s are nuclear response functions
and the index $k$ represents different effective nuclear operators,
which, crucially, under the assumption that the nuclear ground state
is an approximate eigenstate of $P$ and $CP$, can be at most eight:
following the notation in \cite{haxton1,haxton2}, $k$=$M$,
$\Phi^{\prime\prime}$, $\Phi^{\prime\prime}M$,
$\tilde{\Phi}^{\prime}$, $\Sigma^{\prime\prime}$, $\Sigma^{\prime}$,
$\Delta$,$\Delta\Sigma^{\prime}$. The $W^{\tau\tau^{\prime}}_{T k}(y)$'s are function of $y\equiv (qb/2)^2$, where $b$ is the size of the nucleus. For the target nuclei $T$ used in
most direct detection experiments the functions
$W^{\tau\tau^{\prime}}_{T k}(y)$, calculated using nuclear shell
models, have been provided in Refs.\cite{haxton2,catena}.

For a given recoil energy imparted to the target the differential rate
for the WIMP--nucleus scattering process is given by:

\be
\frac{d R_{\chi T}}{d E_R}(t)=\sum_T N_T\frac{\rho_{\mbox{\tiny WIMP}}}{m_{\mbox{\tiny WIMP}}}\int_{v_{min}}d^3 v_T f(\vec{v}_T,t) v_T \frac{d\sigma_T}{d E_R},
\label{eq:dr_de}
\ee

\noindent where $\rho_{\mbox{\tiny WIMP}}$ is the local WIMP mass density in the
neighborhood of the Sun, $N_T$ the number of the nuclear targets of
species $T$ in the detector (the sum over $T$ applies in the case of
more than one target), while

\be
\frac{d\sigma_T}{d E_R}=\frac{2 m_T}{4\pi v_T^2}\left [ \frac{1}{2 j_{\chi}+1} \frac{1}{2 j_{T}+1}|\mathcal{M}_T|^2 \right ],
\label{eq:dsigma_de}
\ee

\noindent with the squared amplitude in parenthesis given explicitly
in Eq.(\ref{eq:squared_amplitude}). Finally, $f(\vec{v}_T)$ is the
WIMP velocity distribution, for which we assume a standard isotropic
Maxwellian at rest in the Galactic rest frame truncated at the escape
velocity $u_{esc}$, and boosted to the Lab frame by the
velocity of the Earth. So for the former we assume:

\begin{eqnarray}
  f(\vec{v}_T,t)&=&\frac{1}{N}\left(\frac{3}{ 2\pi v_{rms}^2}\right )^{3/2}
  e^{-\frac{3|\vec{v}_T+\vec{v}_E|^2}{2 v_{rms}^2}}\Theta(u_{esc}-|\vec{v}_T+\vec{v}_E(t)|)\\
  N&=& \left [ \erf(z)-\frac{2}{\sqrt{\pi}}z e^{-z^2}\right ]^{-1},  
  \label{eq:maxwellian}
  \end{eqnarray}

\noindent with $z=3 u_{esc}^2/(2 v_{rms}^2)$. In the isothermal sphere
model hydrothermal equilibrium between the WIMP gas pressure and
gravity is assumed, leading to $v_{rms}$=$\sqrt{3/2}v_0$ with $v_0$
the galactic rotational velocity. The yearly modulation effect is due
to the time dependence of the Earth's speed with respect to the
Galactic frame:

\begin{equation}
\vec{v}_E(t)=v_{Sun}+v_{orb}\cos\gamma \cos\left [\frac{2\pi}{T_0}(t-t_0)
  \right ],
\label{eq:modulation}
  \end{equation}

\noindent where $\cos\gamma\simeq$0.49 accounts for the inclination of
the ecliptic plane with respect to the Galactic plane, $T_0$=1 year,
$t_0$=2 June, $v_{orb}$=2$\pi r_{\oplus}/(T_0)\simeq$ 29 km/s
($r_{\oplus}$=1 AU, neglecting the small eccentricity of the Earth's
orbit around the Sun) while $v_{Sun}$=$v_0$+12, accounting for a
peculiar component of the solar system with respect to the galactic
rotation. For the two parameters $v_0$ and $u_{esc}$ we take $v_0$=220
km/s \cite{v0_koposov} and $u_{esc}$=550 km/s \cite{vesc_2014}.
In the isothermal model the time dependence of
Eq. (\ref{eq:modulation}) induces an expected rate with the functional
form $S(t)=S_0+S_m \cos(2\pi/T-t_0)$, with $S_m>0$ at large values of
$v_{min}$ and turning negative when $v_{min}\lsim$ 200 km/s.  In such
regime of $v_{min}$ and below the phase is modified by the focusing
effect of the Sun's gravitational potential \cite{gf}, while when
$S_m\ll S_0$ the time dependence differs from a simple cosine due the
contribution of higher harmonics\cite{freese_review}.

The expected rate in a given visible energy bin $E_1^{\prime}\le
E^{\prime}\le E_2^{\prime}$ of a direct detection experiment is given
by:

\begin{eqnarray}
R_{[E_1^{\prime},E_2^{\prime}]}(t)&=&MT_{exp}\int_{E_1^{\prime}}^{E_2^{\prime}}\frac{dR}{d
  E^{\prime}}(t)\, dE^{\prime} \label{eq:start}\\
 \frac{dR}{d E^{\prime}}(t)&=&\sum_T \int_0^{\infty} \frac{dR_{\chi T}(t)}{dE_{ee}}{\cal
   G}_T(E^{\prime},E_{ee})\epsilon(E^{\prime})\label{eq:start2}\,d E_{ee} \\
E_{ee}&=&q(E_R) E_R \label{eq:start3},
\end{eqnarray}

\noindent with $\epsilon(E^{\prime})\le 1$ the experimental
efficiency/acceptance. In the equations above $E_R$ is the recoil
energy deposited in the scattering process (indicated in keVnr), while
$E_{ee}$ (indicated in keVee) is the fraction of $E_R$ that goes into
the experimentally detected process (ionization, scintillation, heat)
and $q(E_R)$ is the quenching factor, ${\cal
  G_T}(E^{\prime},E_{ee}=q(E_R)E_R)$ is the probability that the
visible energy $E^{\prime}$ is detected when a WIMP has scattered off
an isotope $T$ in the detector target with recoil energy $E_R$, $M$ is
the fiducial mass of the detector and $T_{exp}$ the live--time
exposure of the data taking.

In particular, in each visible energy bin DAMA is sensitive to the
yearly modulation amplitude $S_m$, defined as the cosine transform of
$R_{[E_1^{\prime},E_2^{\prime}]}(t)$:

\begin{equation}
S_{m,[E_1^{\prime},E_2^{\prime}]}\equiv \frac{2}{T_0}\int_0^{T_0}
\cos\left[\frac{2\pi}{T_0}(t-t_0)\right]R_{[E_1^{\prime},E_2^{\prime}]}(t)dt,
\label{eq:sm}
\end{equation}  

\noindent while other experiments put upper bounds on the time average
$S_0$:

\begin{equation}
S_{0,[E_1^{\prime},E_2^{\prime}]}\equiv \frac{1}{T_0}\int_0^{T_0}
R_{[E_1^{\prime},E_2^{\prime}]}(t)dt.
\label{eq:s0}
\end{equation}  

In the present paper we will systematically consider the possibility
that one of the couplings $c_{j}$ dominates in the effective
Hamiltonian of Eq. (\ref{eq:H}). In this case it is possible to
factorize a term $|c_j^p|^2$ from the squared amplitude of
Eq.(\ref{eq:squared_amplitude}) and express it in terms of the {\it
  effective} WIMP--proton cross section:

\begin{equation}
\sigma_p=(c_j^p)^2\frac{\mu_{\chi{\cal N}}^2}{\pi},
  \label{eq:conventional_sigma}
\end{equation}

\noindent (with $\mu_{\chi{\cal N}}$ the WIMP--nucleon reduced mass)
and the ratio $r\equiv c_j^n/c_j^p$. It is worth pointing out here
that among the generalized nuclear response functions arising from the
effective Hamiltonian (\ref{eq:H}) only the ones corresponding to $M$
(SI interaction), $\Sigma^{\prime\prime}$ and $\Sigma^{\prime}$ (both
related to the standard spin--dependent interaction) do not vanish for
$q\rightarrow$0, and so allow to interpret $\sigma_p$ in terms of a
long--distance, point--like cross section. In the case of the other
interactions $\Phi^{\prime\prime}$, $\Phi^{\prime\prime}M$,
$\tilde{\Phi}^{\prime}$, $\Delta$ and $\Delta\Sigma^{\prime}$ the
quantity $\sigma_p$ is just a convenient alternative to directly
parameterizing the interaction in terms of the $c_j^p$ coupling.
\begin{table}[ht!]
\begin{center}
\begin{tabular}{|M{1cm}|M{3cm}|M{2cm}|M{2cm}|M{2cm}|} \hline
\rule[-12.5pt]{0pt}{30pt}  $\mathbf{c_j}$  & $\mathbf{m_{\chi,min}}$ \textbf{(GeV)} & $\mathbf{r_{\chi,min}}$ & $\mathbf{\sigma~(\mbox{\bf cm}^2)}$ & $\mathbf{\chi^2_{min}}$ \\\hline
  \multirow{ 2}{*}{$c_1$}     &  11.17 & -0.76 &  2.67e-38 & 11.38 \\ 
                 &  45.19 & -0.66 &  1.60e-39 & 13.22 \\\hline
   \multirow{ 2}{*}{$c_3$}      &  8.10 & -3.14 &  2.27e-31 & 11.1 \\ 
                 &  35.68 & -1.10 & 9.27e-35 & 14.23 \\\hline
 \multirow{ 2}{*}{$c_4$}      &  11.22 & 1.71 &  2.95e-36 &  11.38\\ 
                 &  44.71 & -8.34 & 5.96e-36 & 27.7 \\\hline
   \multirow{ 2}{*}{$c_5$}     &  8.34 & -0.61 &  1.62e-29 &  10.83\\ 
                 &  96.13 & -5.74 & 3.63e-34 &  11.11\\\hline  
 \multirow{ 2}{*}{$c_6$}      &  8.09 & -7.20 &  5.05e-28 &  11.11\\ 
                 &  32.9 & -6.48 & 5.18e-31  &  12.74\\\hline
  \multirow{ 2}{*}{$c_7$}      &  13.41 & -4.32 & 4.75e-30  &  13.94\\ 
                 &  49.24 & -0.65 & 1.35e-30  & 38.09\\\hline 
  \multirow{ 2}{*}{$c_8$}      &  9.27 & -0.84 & 8.67e-33  &  10.82\\ 
                 &  42.33 & -0.96 & 1.30e-34  &  11.6\\\hline
  \multirow{ 2}{*}{$c_9$}      &  9.3 & 4.36 & 8.29e-33  &  10.69\\ 
                 &  37.51 & -0.94 & 1.07e-33  & 15.23 \\\hline
  \multirow{ 2}{*}{$c_{10}$}   &  9.29 &  3.25 & 4.74e-33   &  10.69\\ 
                 &  36.81 & 0.09 &  2.25e-34 & 12.40 \\\hline
  \multirow{ 2}{*}{$c_{11}$}   &  9.27 &  -0.67 & 1.15e-34   &  10.69\\ 
                 &  38.51 & -0.66 &  9.17e-37 & 13.02 \\\hline
  \multirow{ 2}{*}{$c_{12}$}   &  9.26 &  -2.85 & 3.92e-34   &  10.69\\ 
                 &  35.22 & -1.93 &  2.40e-35 & 12.47 \\\hline
  \multirow{ 2}{*}{$c_{13}$}   &  8.65&  -0.26 & 1.21e-26   & 10.76\\ 
                 &  29.42 & 0.10 &  5.88e-29  & 14.28 \\\hline
  \multirow{ 2}{*}{$c_{14}$}   &  10.28 & -0.59  & 2.61e-26   & 11.21\\ 
                 &  38.88 &  -1.93 &  2.19e-27   & 14.48  \\\hline
  \multirow{ 2}{*}{$c_{15}$}   &  7.32 &  -3.58 &  2.04e-27  & 12.91\\ 
                 &  33.28 & 4.25  &  2.05e-33  &  16.26\\\hline
\end{tabular}
\caption{Absolute and local minima of the $\chi^2$ (see
  Eq.(\ref{eq:chi2})) for each of the couplings $c_j$ of the effective
  Hamiltonian (\ref{eq:H}).}
\label{tab:best_fit_values}
\end{center}
\end{table}

\section{Analysis}
\label{sec:analysis}

\begin{figure}
\begin{center}
 \includegraphics[width=0.8\columnwidth]{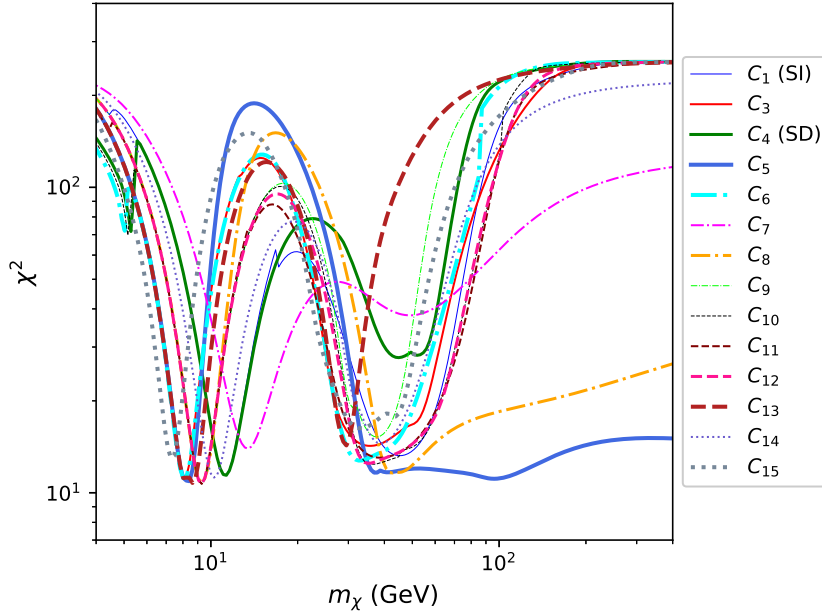}
\end{center}
\caption{Minimum of the $\chi^2$ of Eq.(\ref{eq:chi2}) at fixed WIMP
  mass $m_\chi$ as a function of $m_\chi$ for different WIMP-nucleus
  interactions.}
\label{fig:chi2_m}
\end{figure}

\begin{figure}
\begin{center}
  \includegraphics[width=0.32\textwidth]{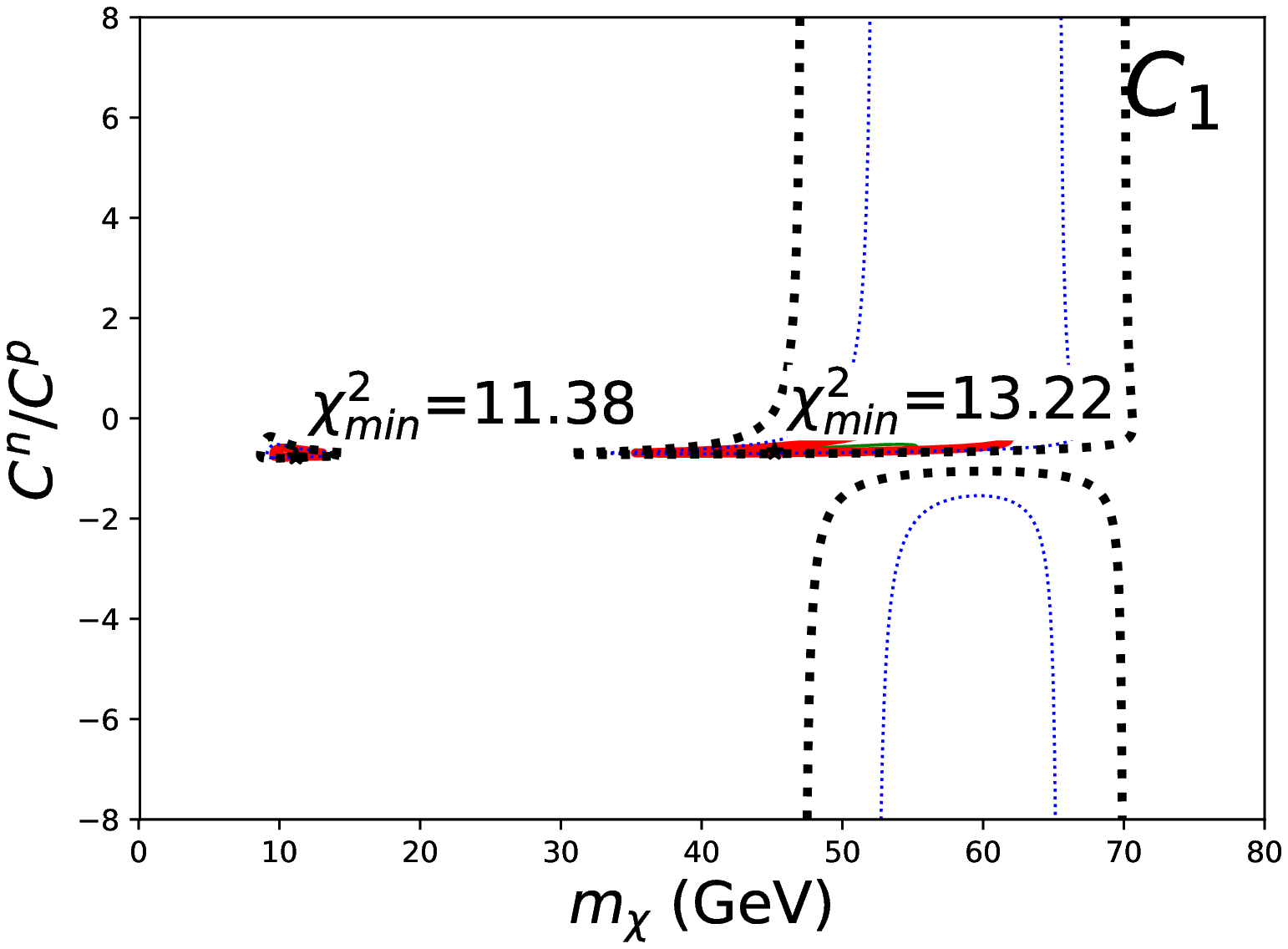}
   \includegraphics[width=0.32\textwidth]{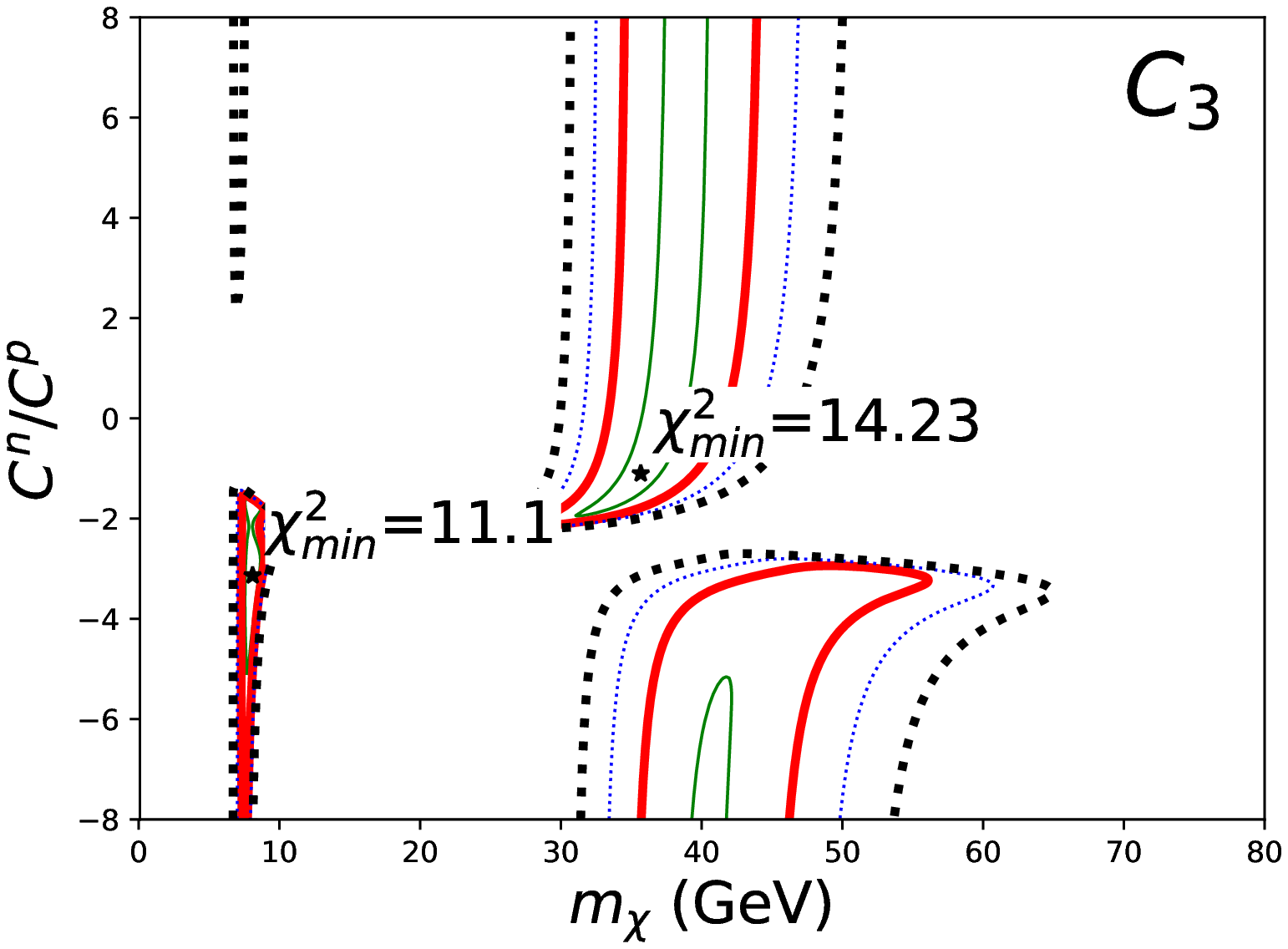}
  \includegraphics[width=0.32\textwidth]{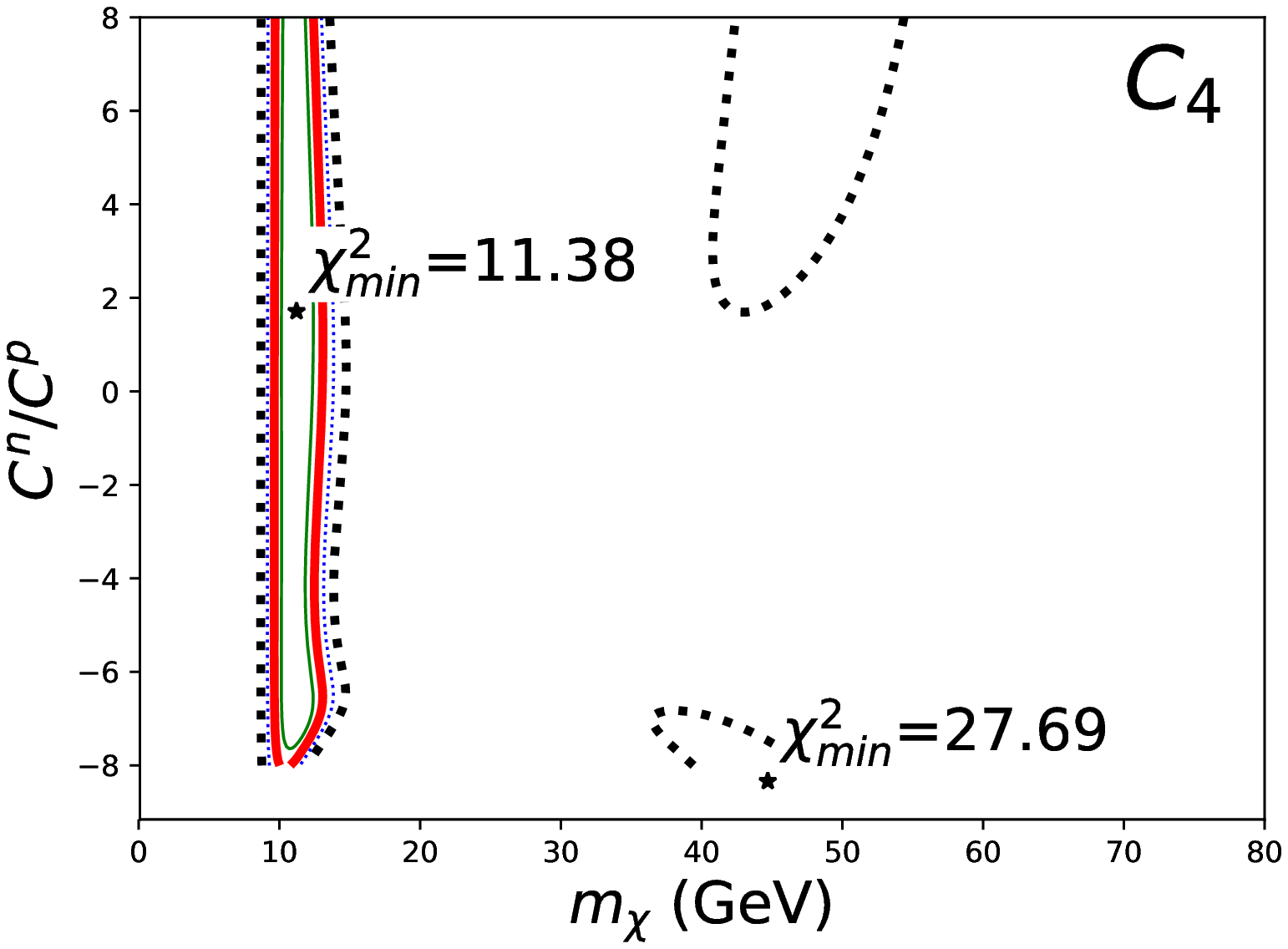}
   \includegraphics[width=0.32\textwidth]{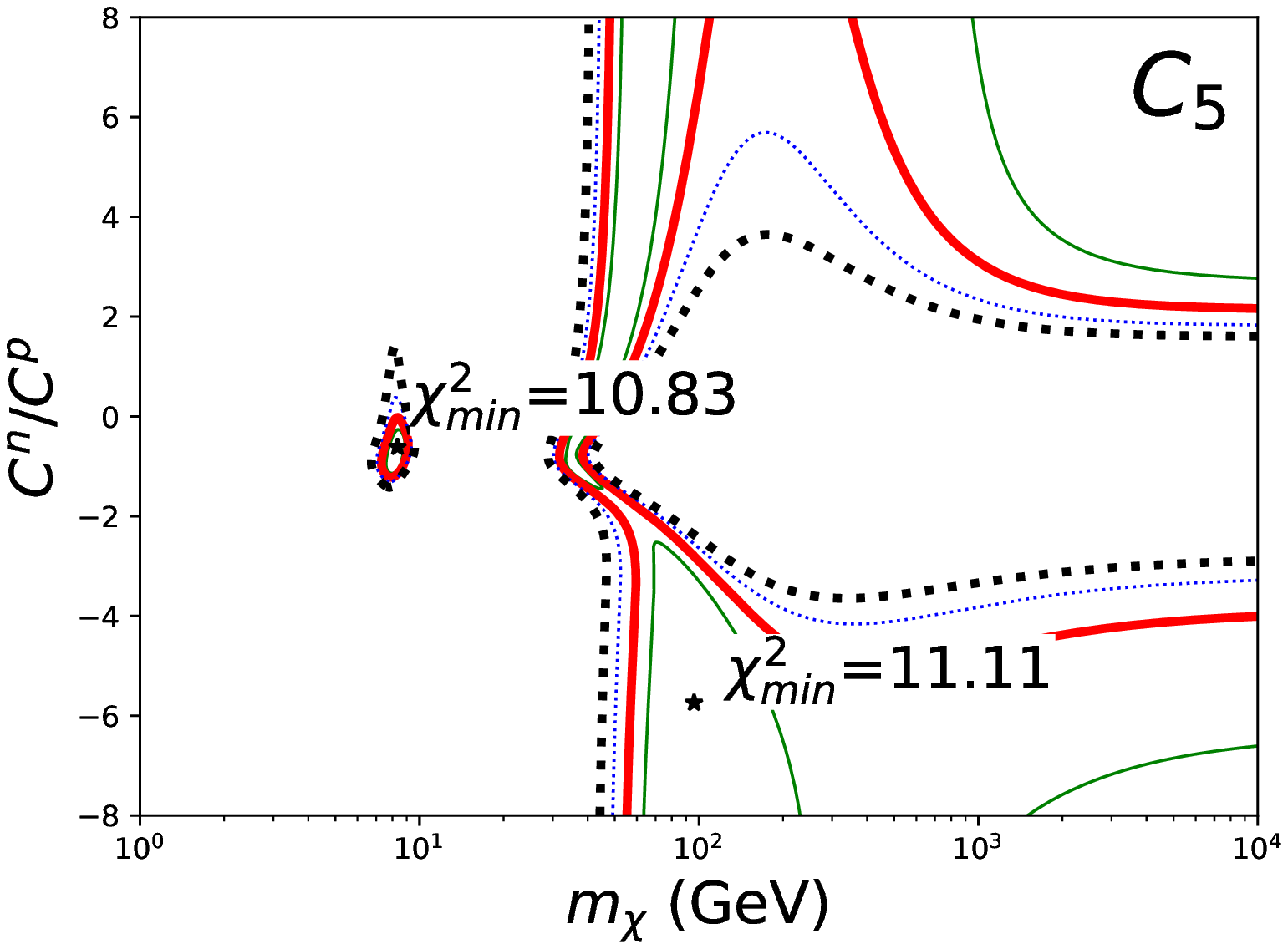}
  \includegraphics[width=0.32\textwidth]{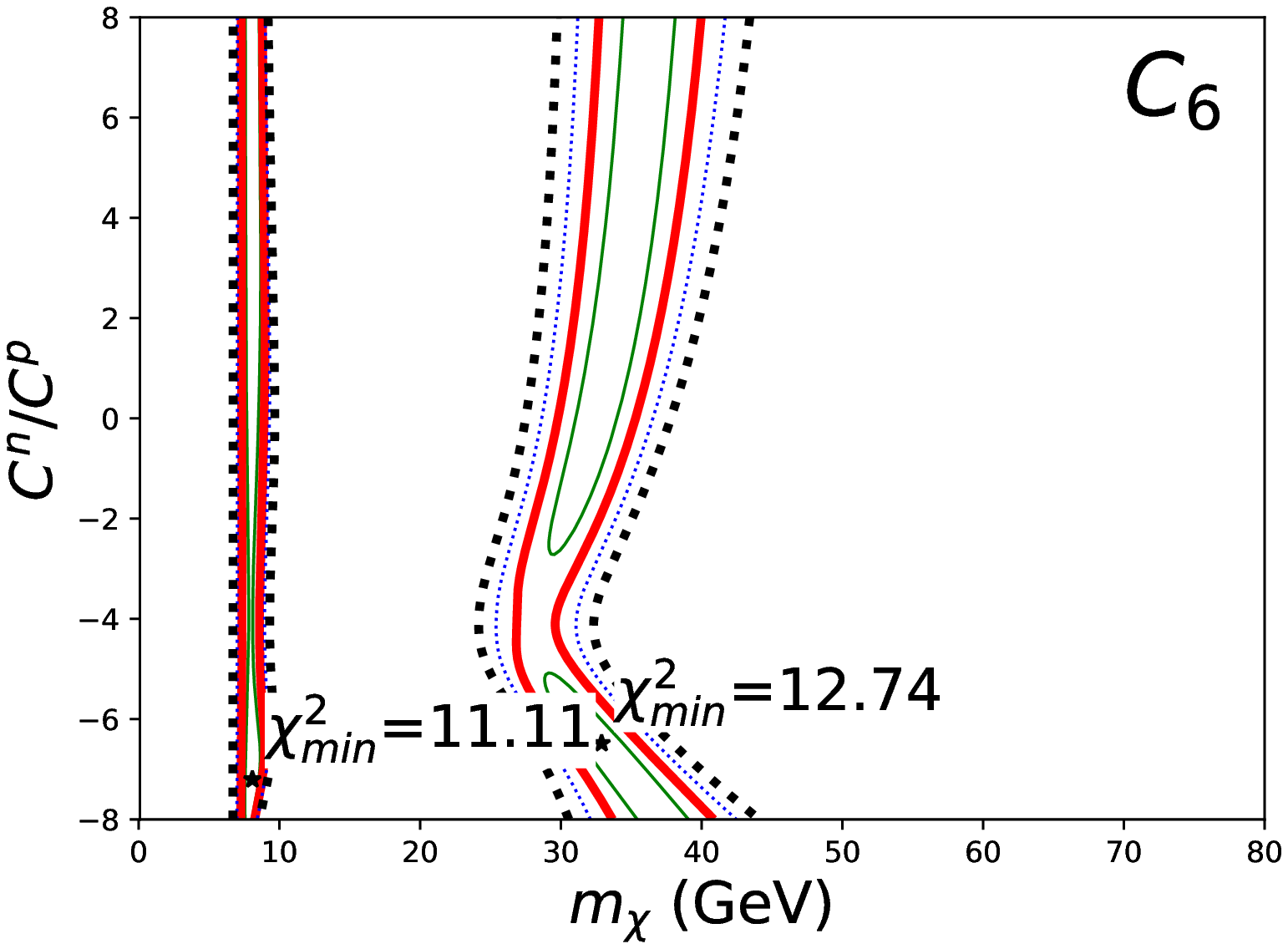}
   \includegraphics[width=0.32\textwidth]{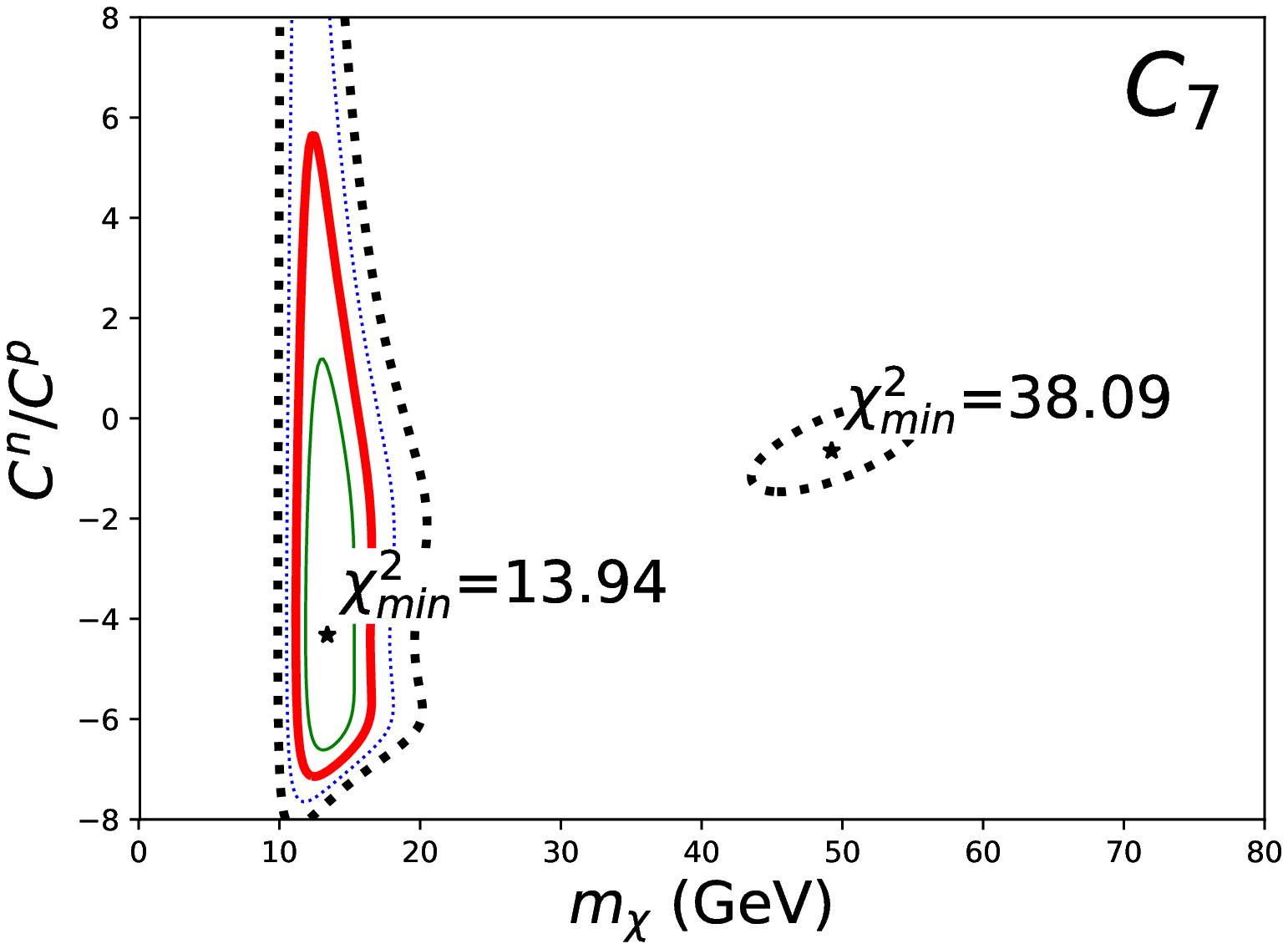}
  \includegraphics[width=0.32\textwidth]{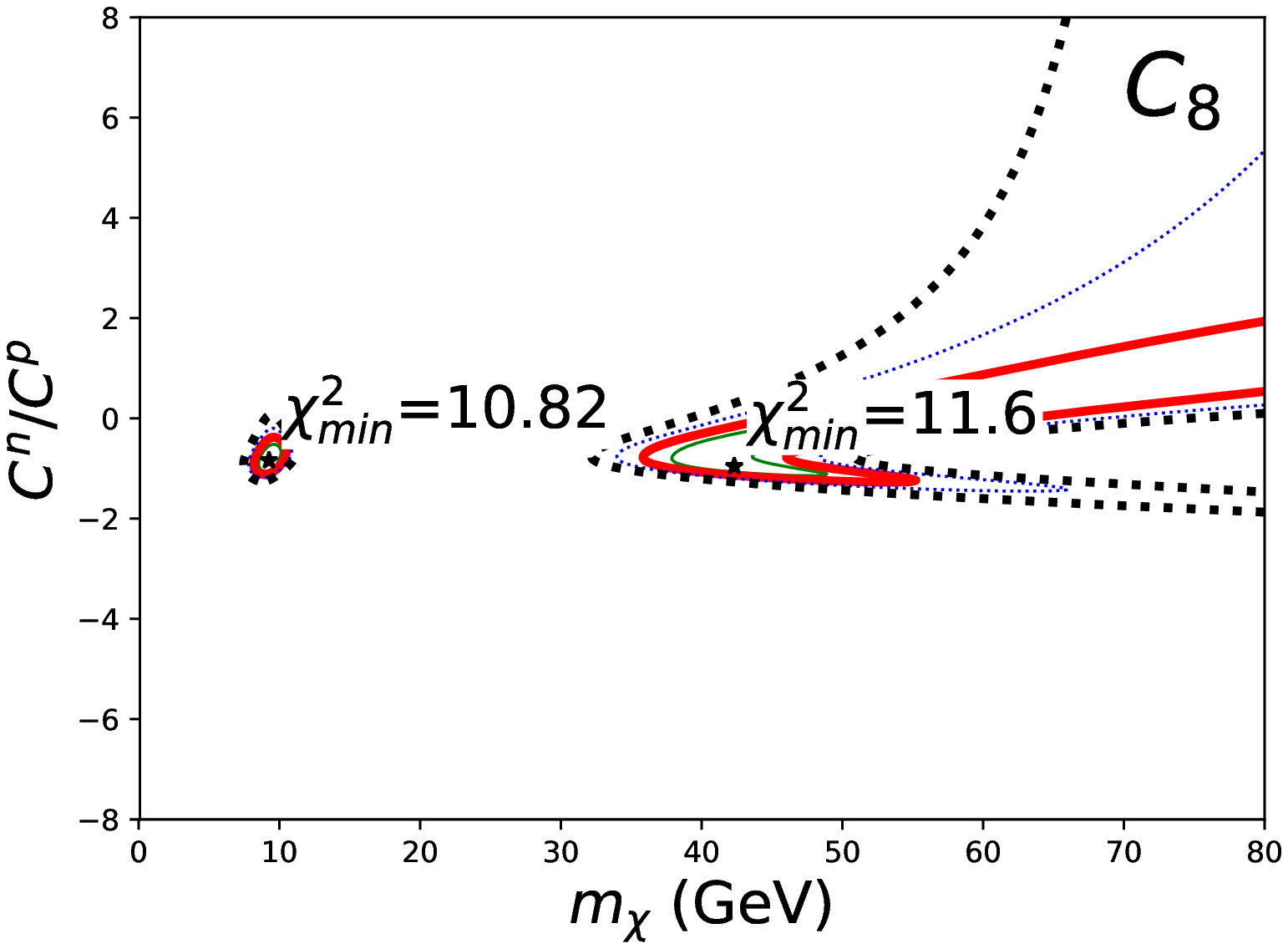}
   \includegraphics[width=0.32\textwidth]{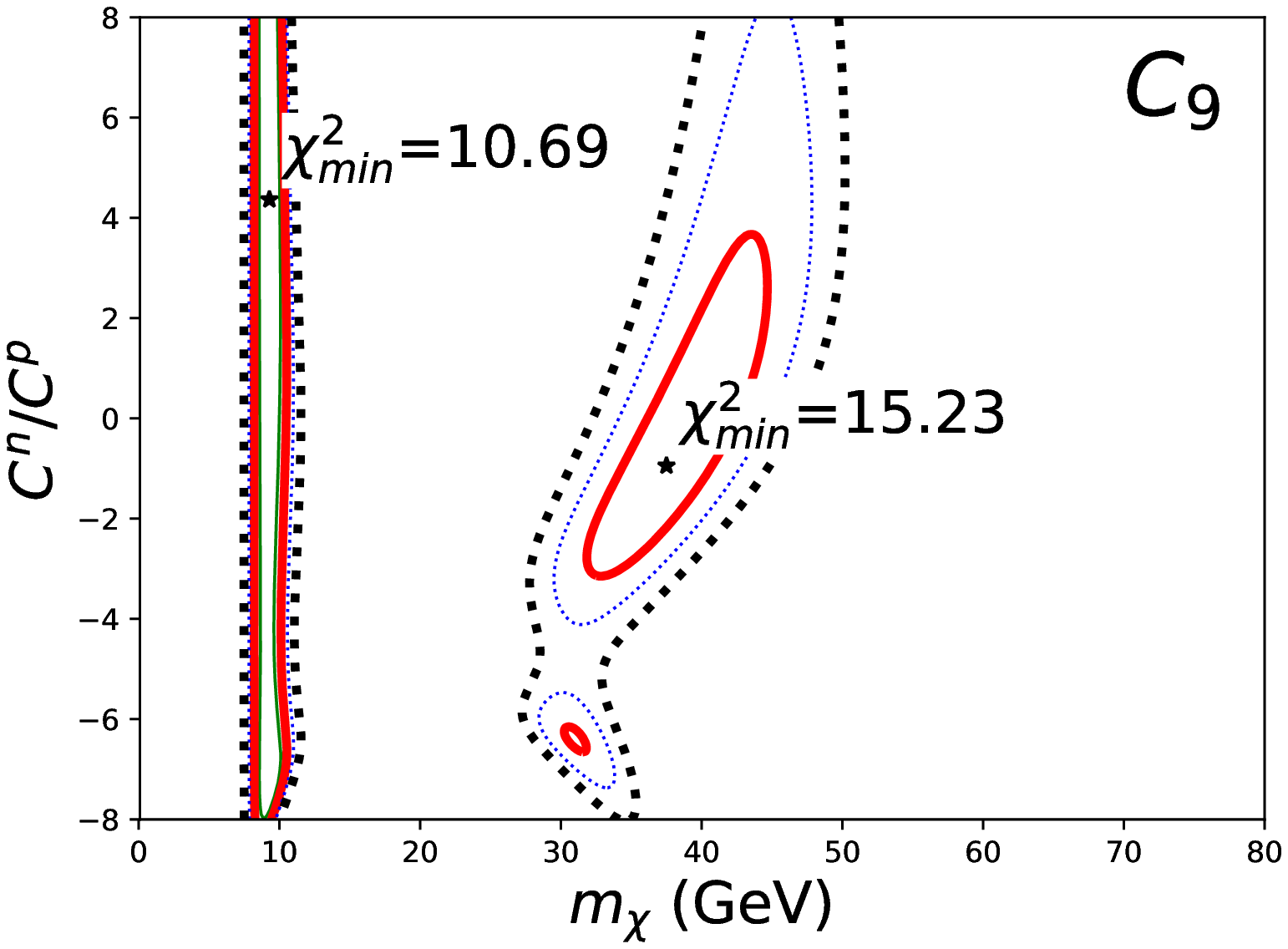}
  \includegraphics[width=0.32\textwidth]{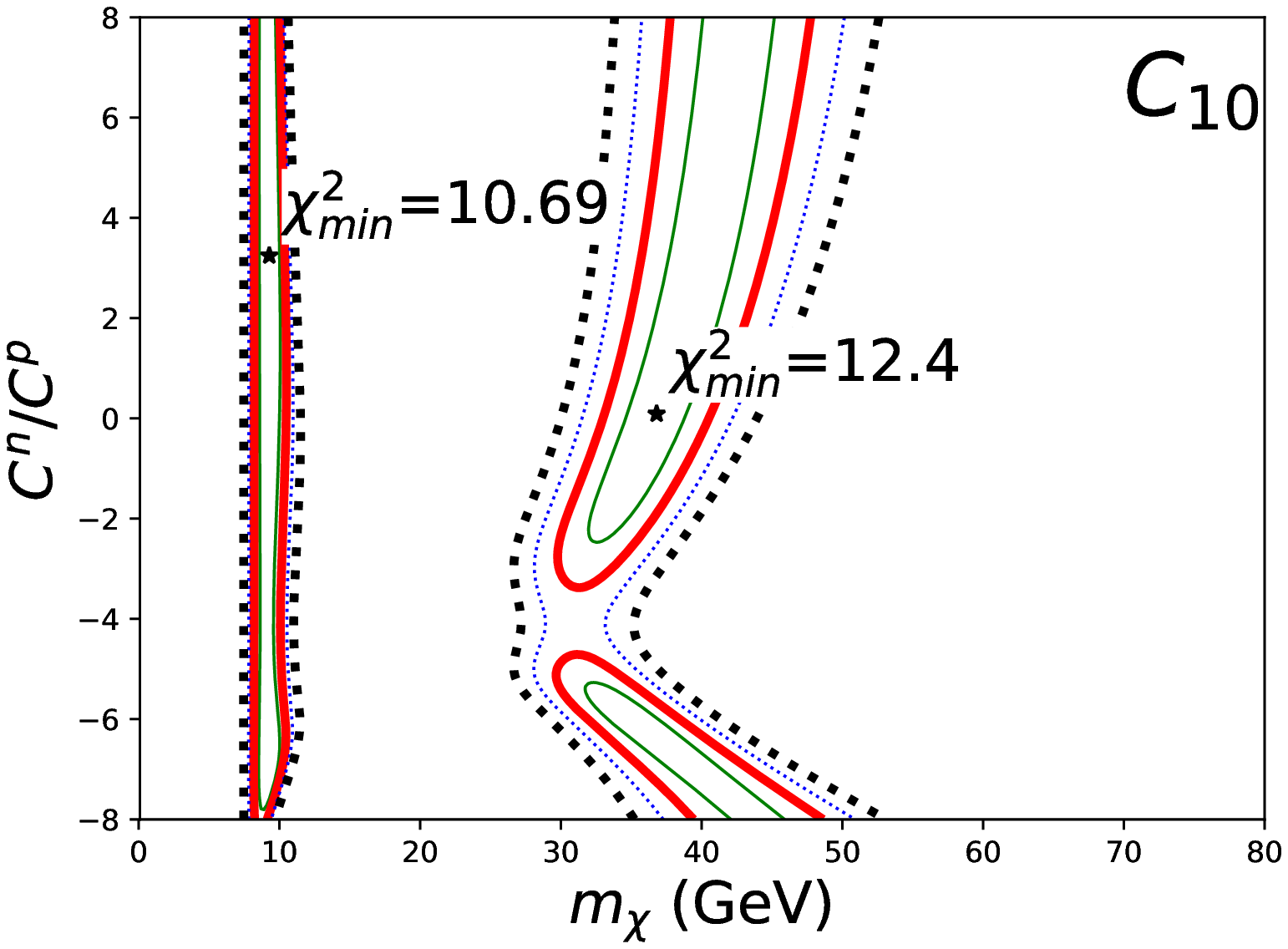}
   \includegraphics[width=0.32\textwidth]{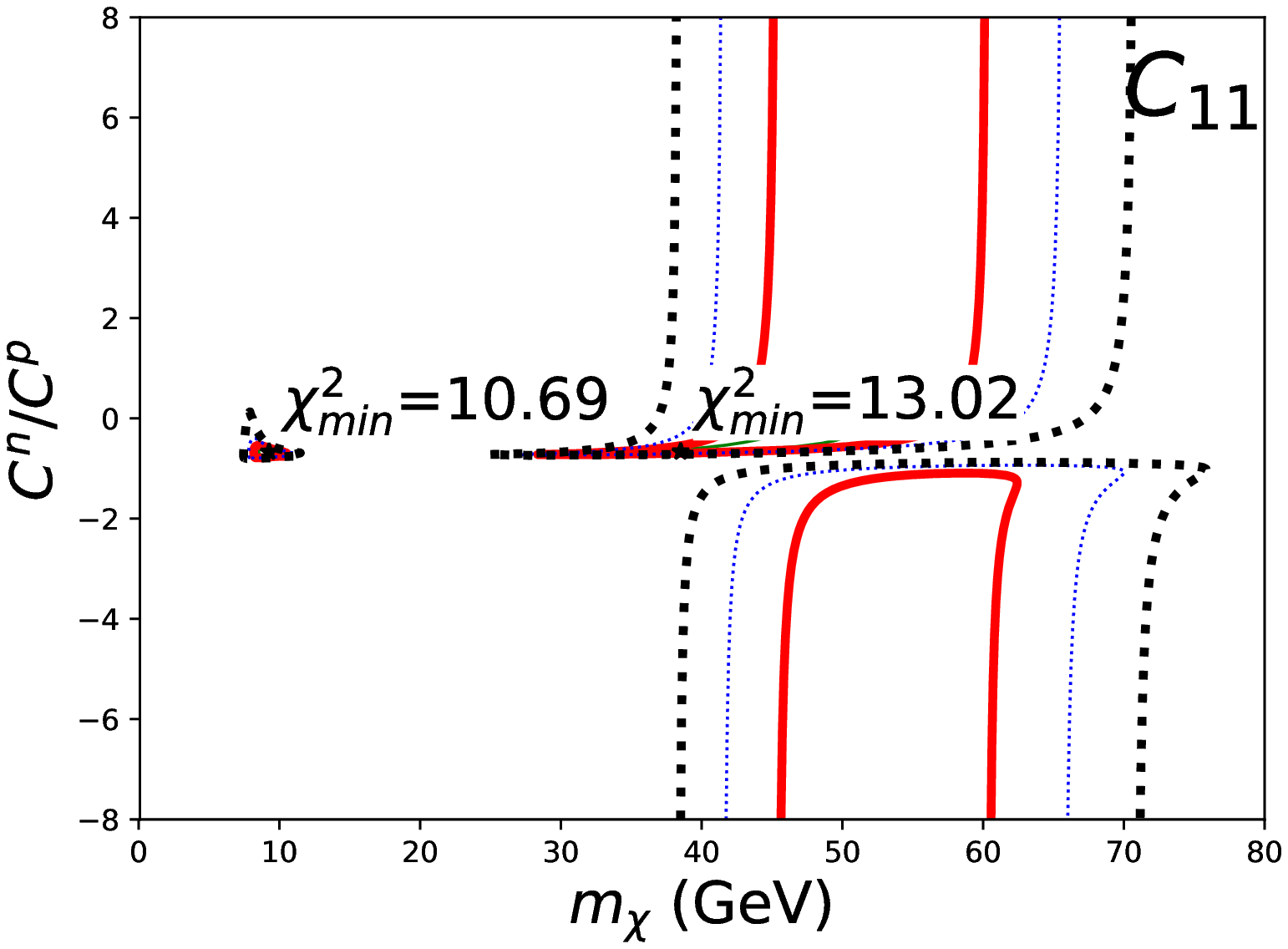}
  \includegraphics[width=0.32\textwidth]{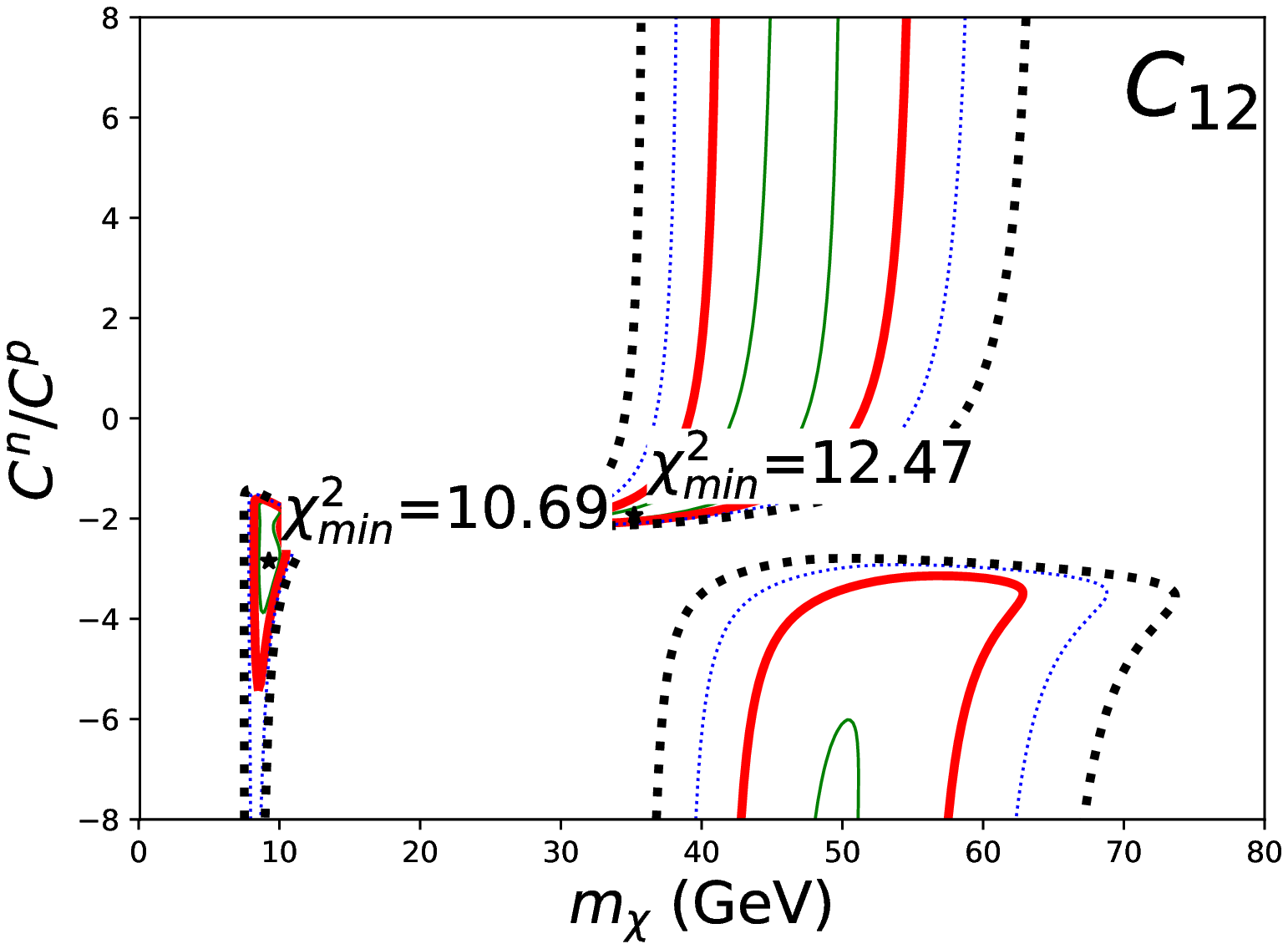}
  \includegraphics[width=0.32\textwidth]{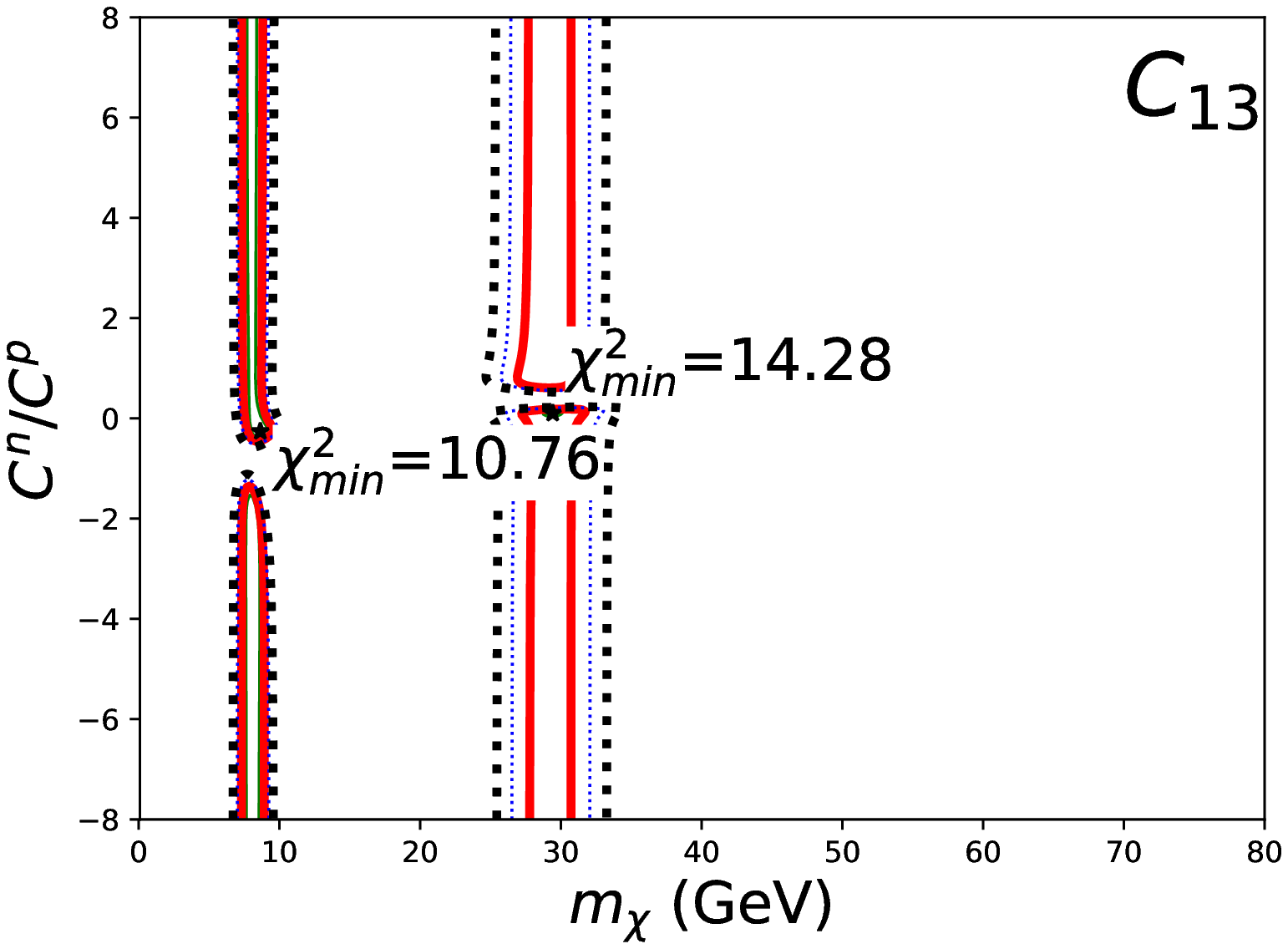}
   \includegraphics[width=0.32\textwidth]{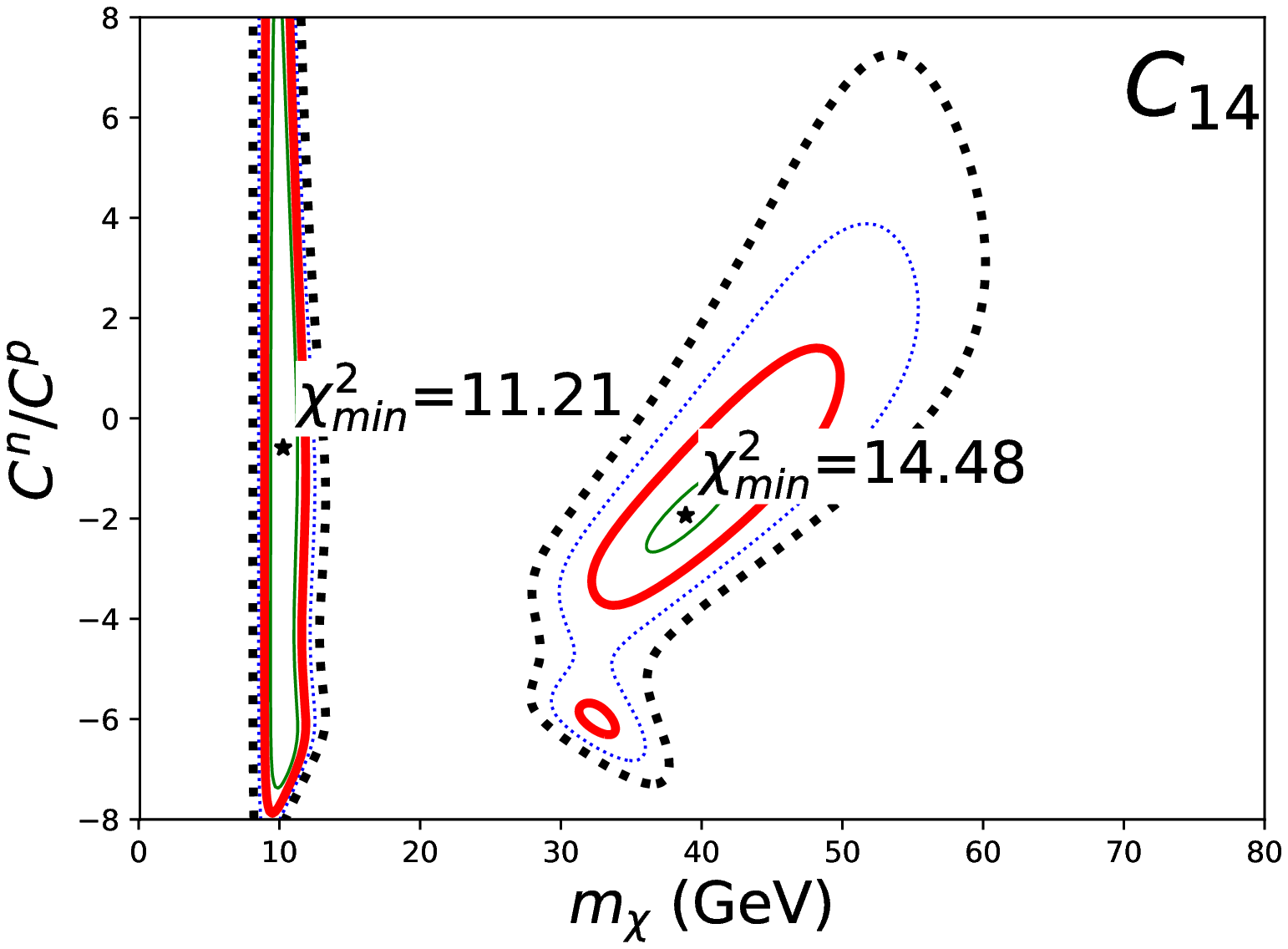}
   \includegraphics[width=0.32\textwidth]{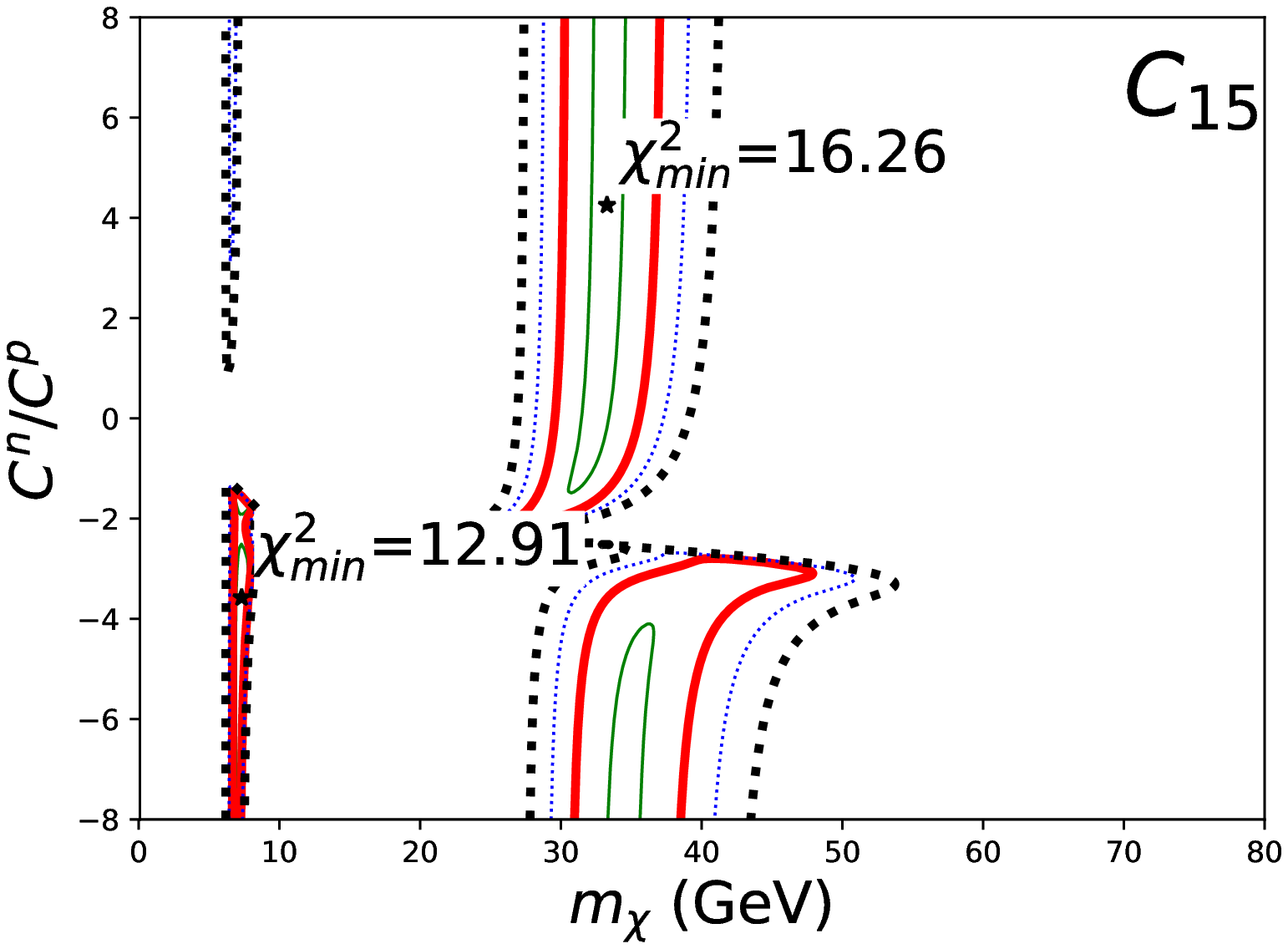}
\end{center}
\caption{Contour plots of the $\chi^2$ of Eq.(\ref{eq:chi2}) minimized
  with respect to $\sigma_p$ in the $m_\chi$-$r$ plane for each of the
  interaction terms of Eq.(\ref{eq:H}). The thin (green) solid lines,
  the thick (red) solid lines, the thin (blue) dotted lines, and the
  thick (black) dotted lines correspond to 2, 3, 4, and 5 $\sigma$
  regions respectively. The best-fit points in the low and high mass
  regions are shown by the star and the corresponding values of the
  best--fit parameters are quoted in Table~\ref{tab:best_fit_values}.}
\label{fig:chi2_m_r_planes}
\end{figure}

\begin{figure}
\begin{center}
\includegraphics[width=0.32\textwidth]{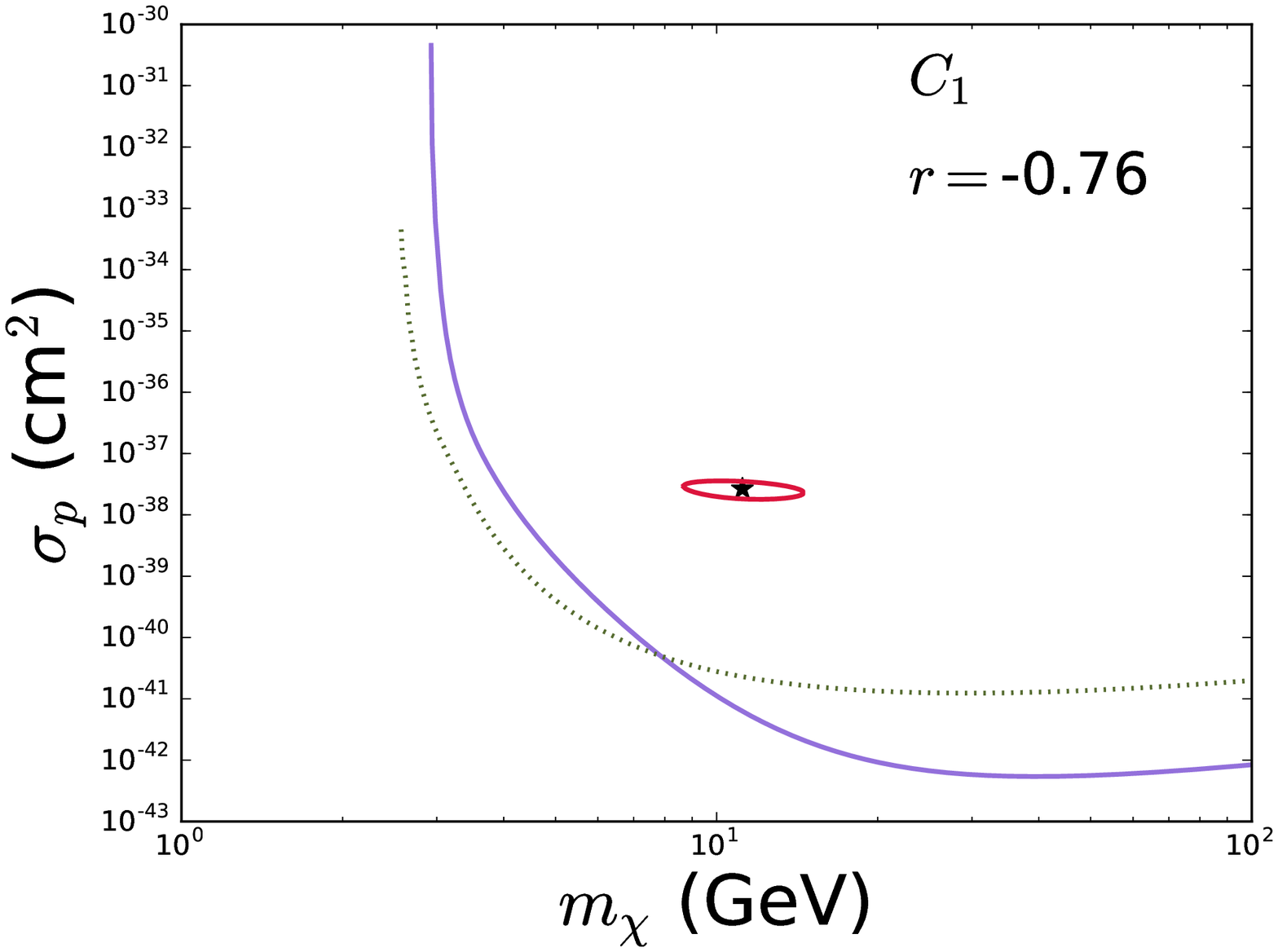}
\includegraphics[width=0.32\textwidth]{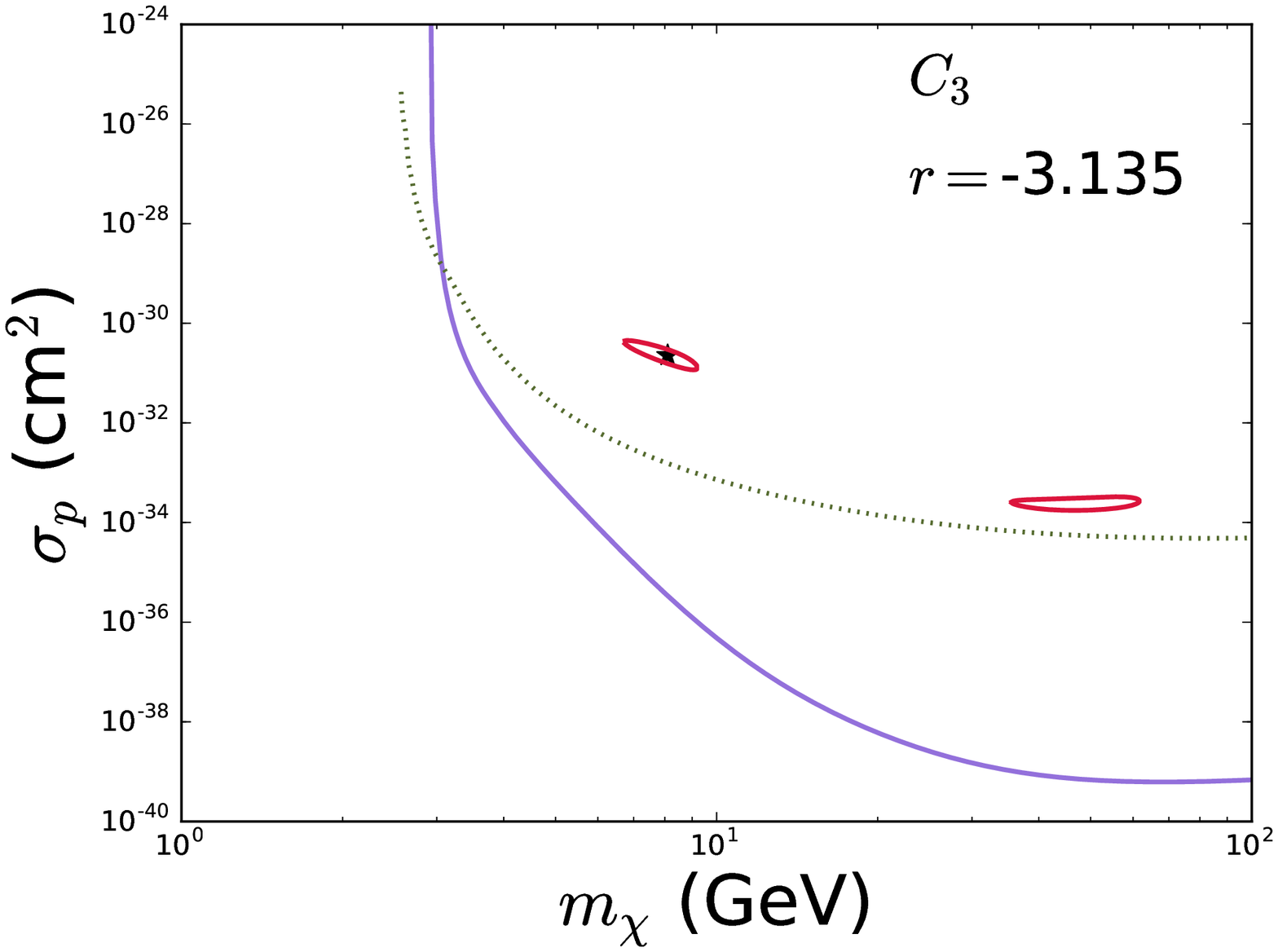}      
\includegraphics[width=0.32\textwidth]{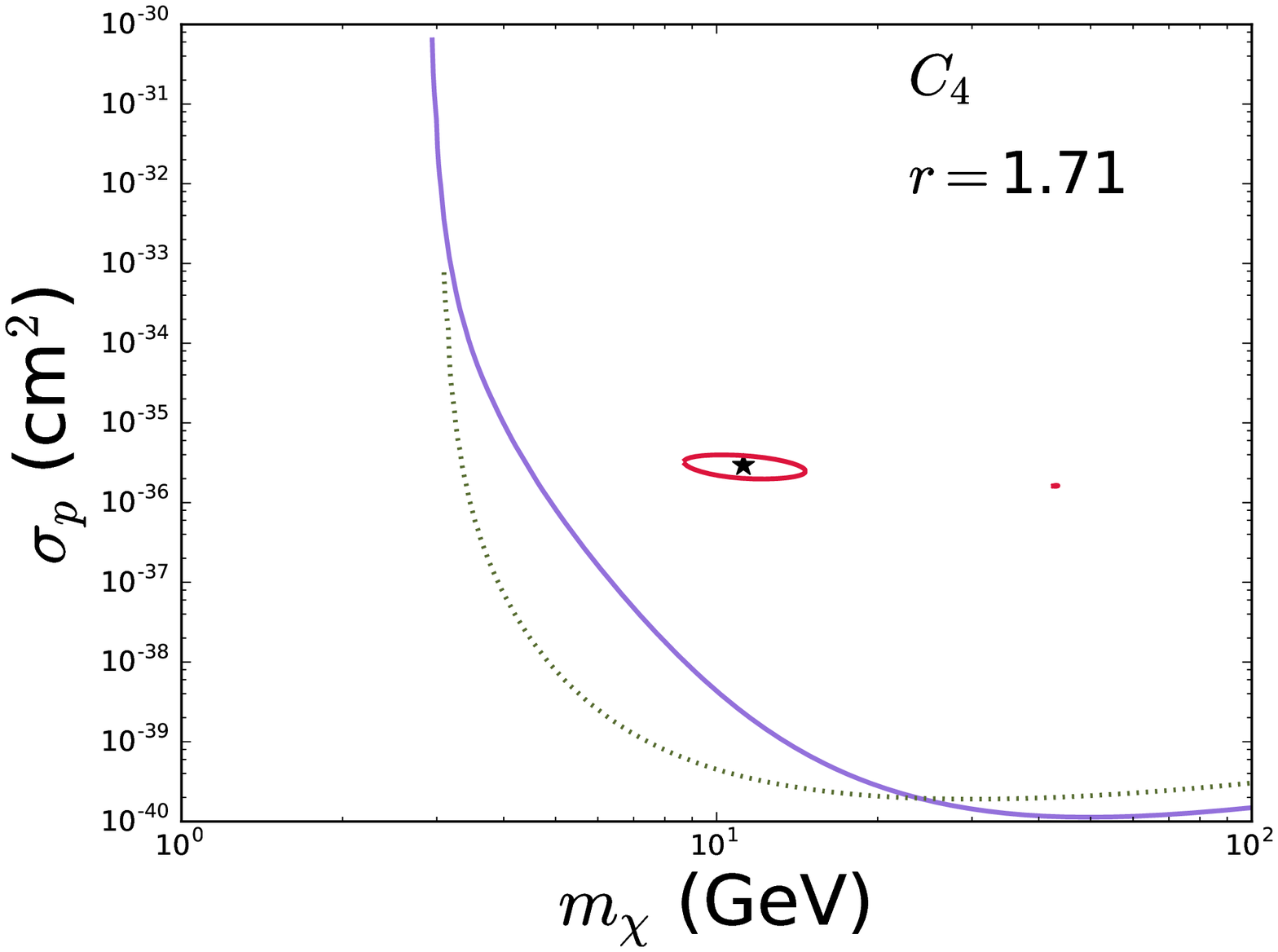}      
\includegraphics[width=0.32\textwidth]{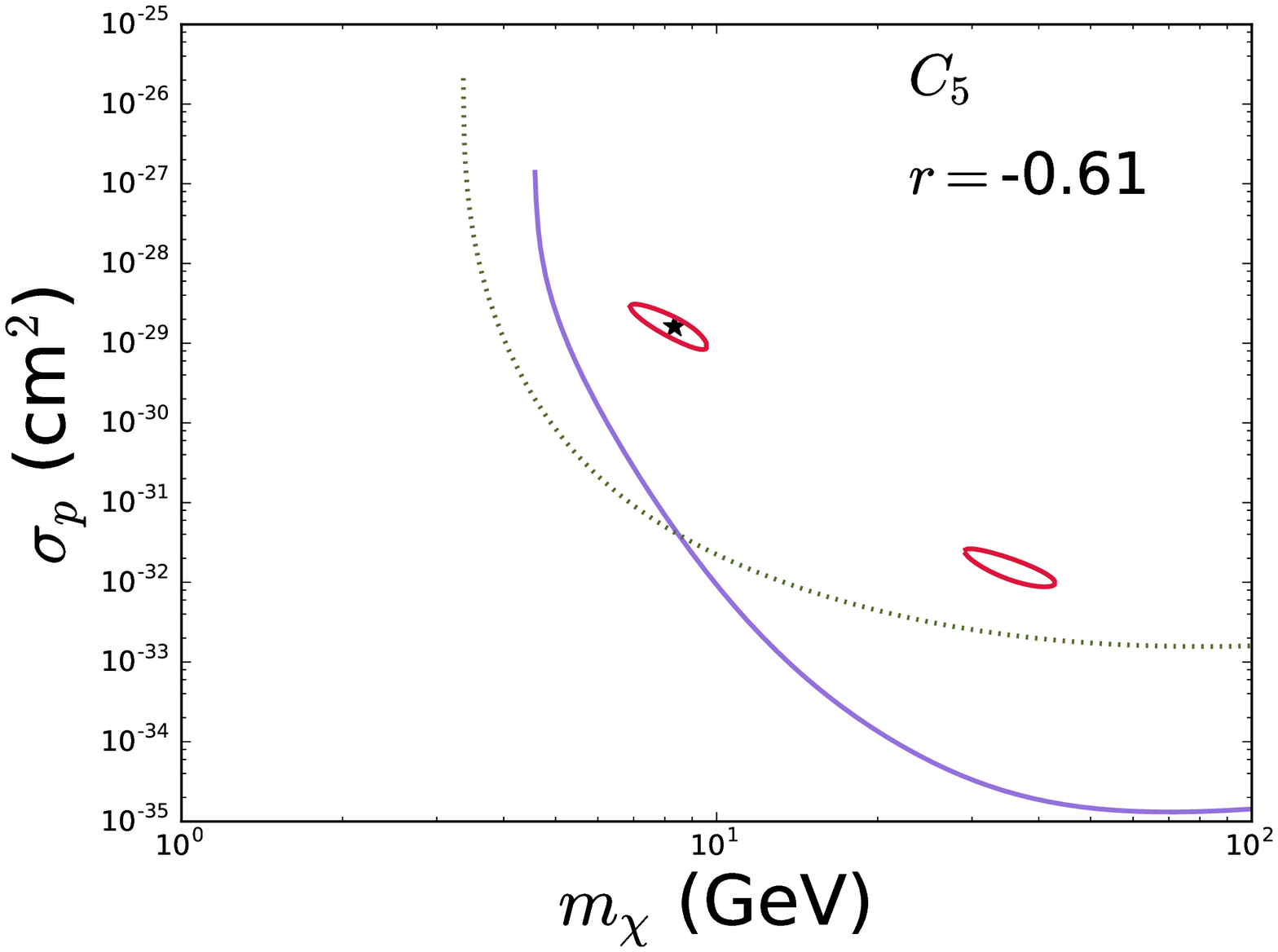}      
\includegraphics[width=0.32\textwidth]{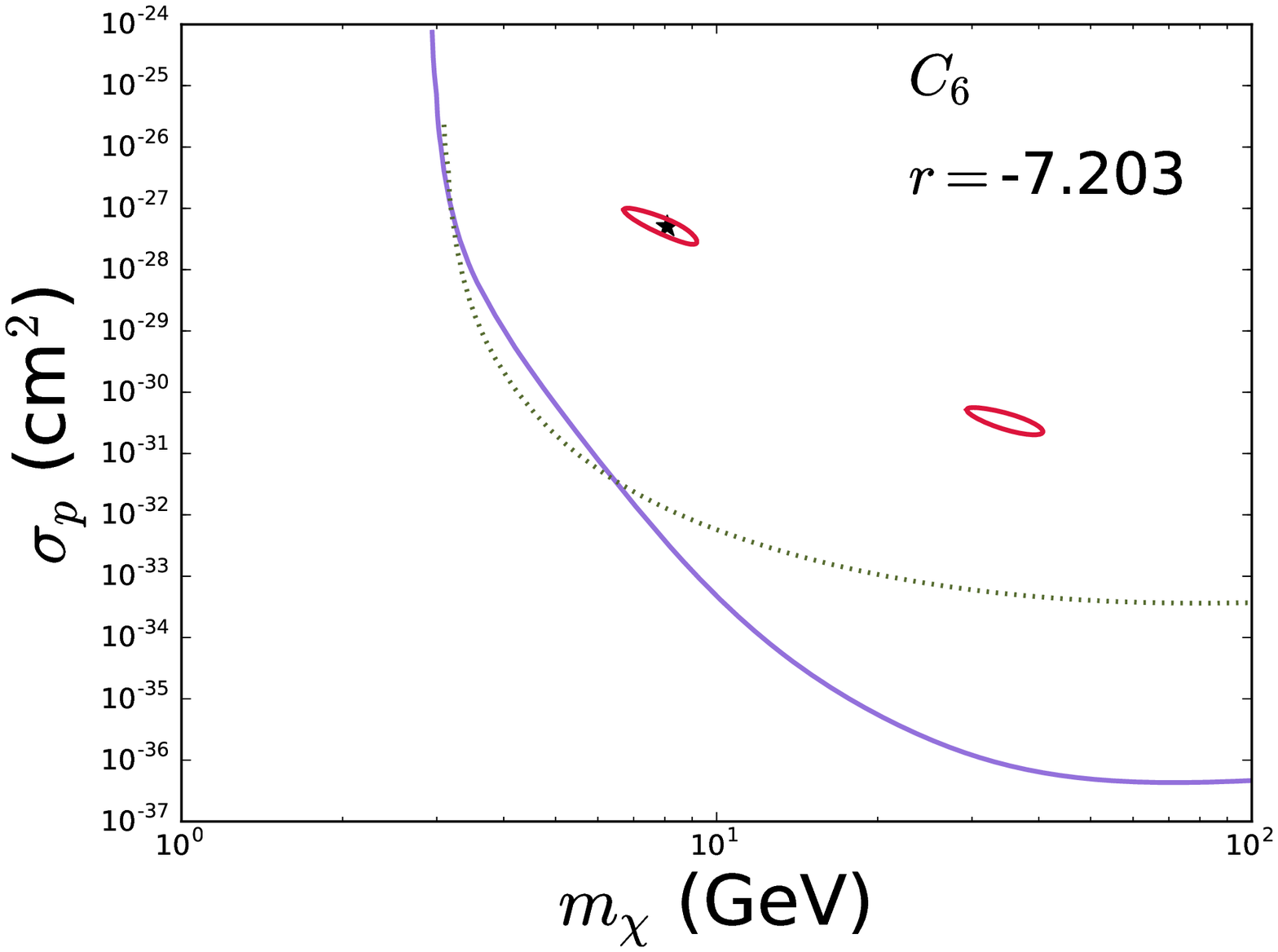}      
\includegraphics[width=0.32\textwidth]{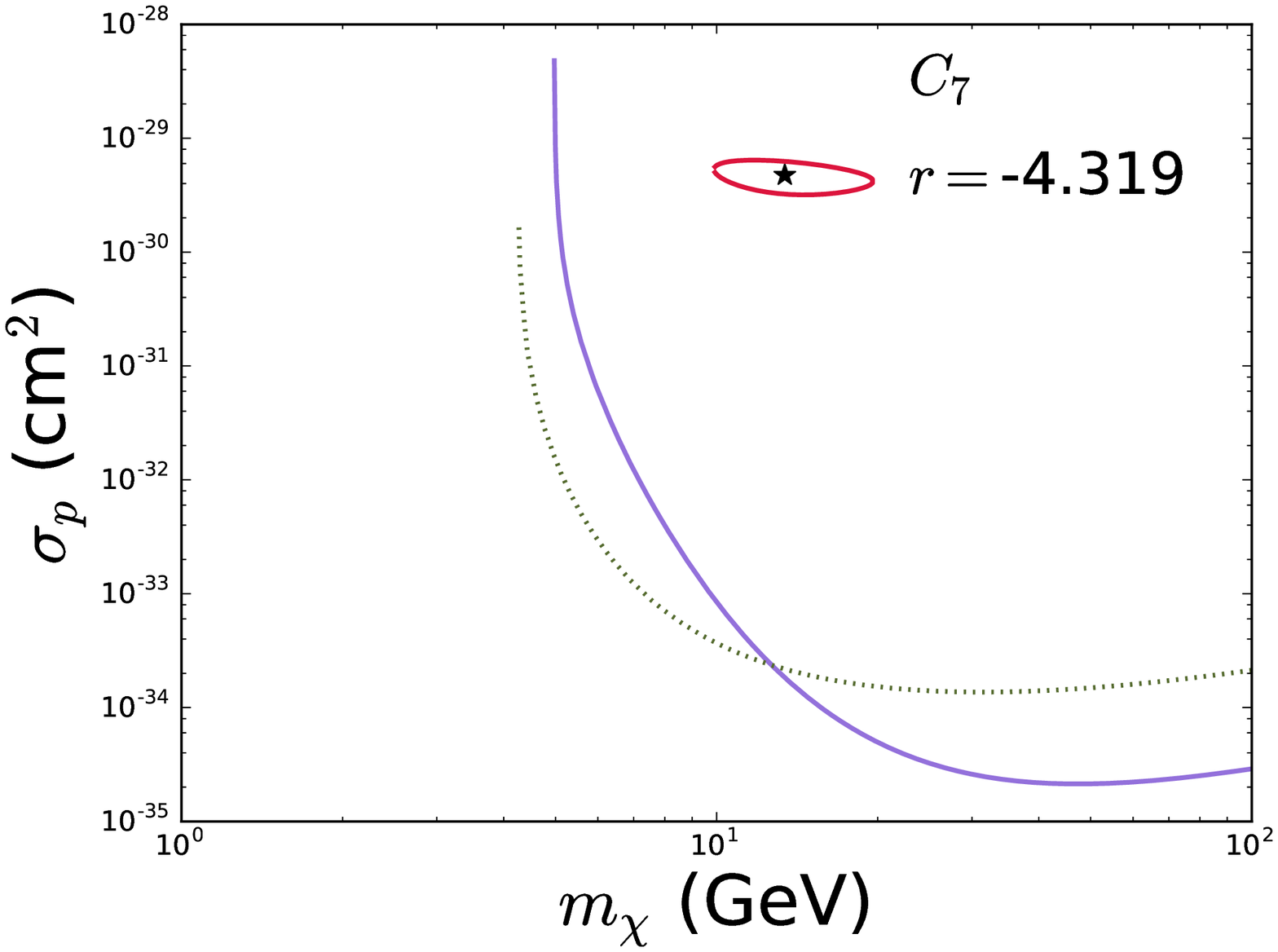}      
\includegraphics[width=0.32\textwidth]{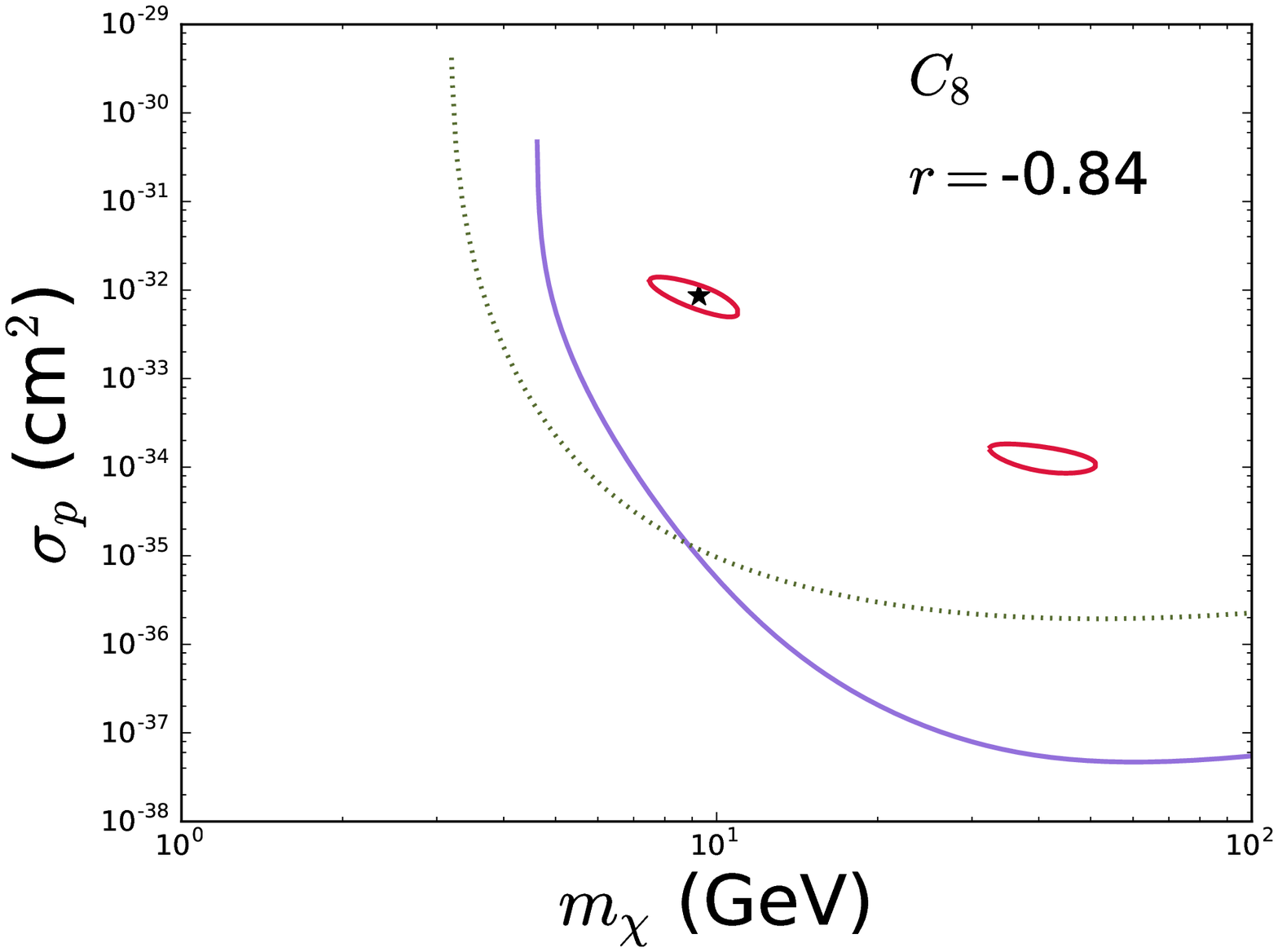}      
\includegraphics[width=0.32\textwidth]{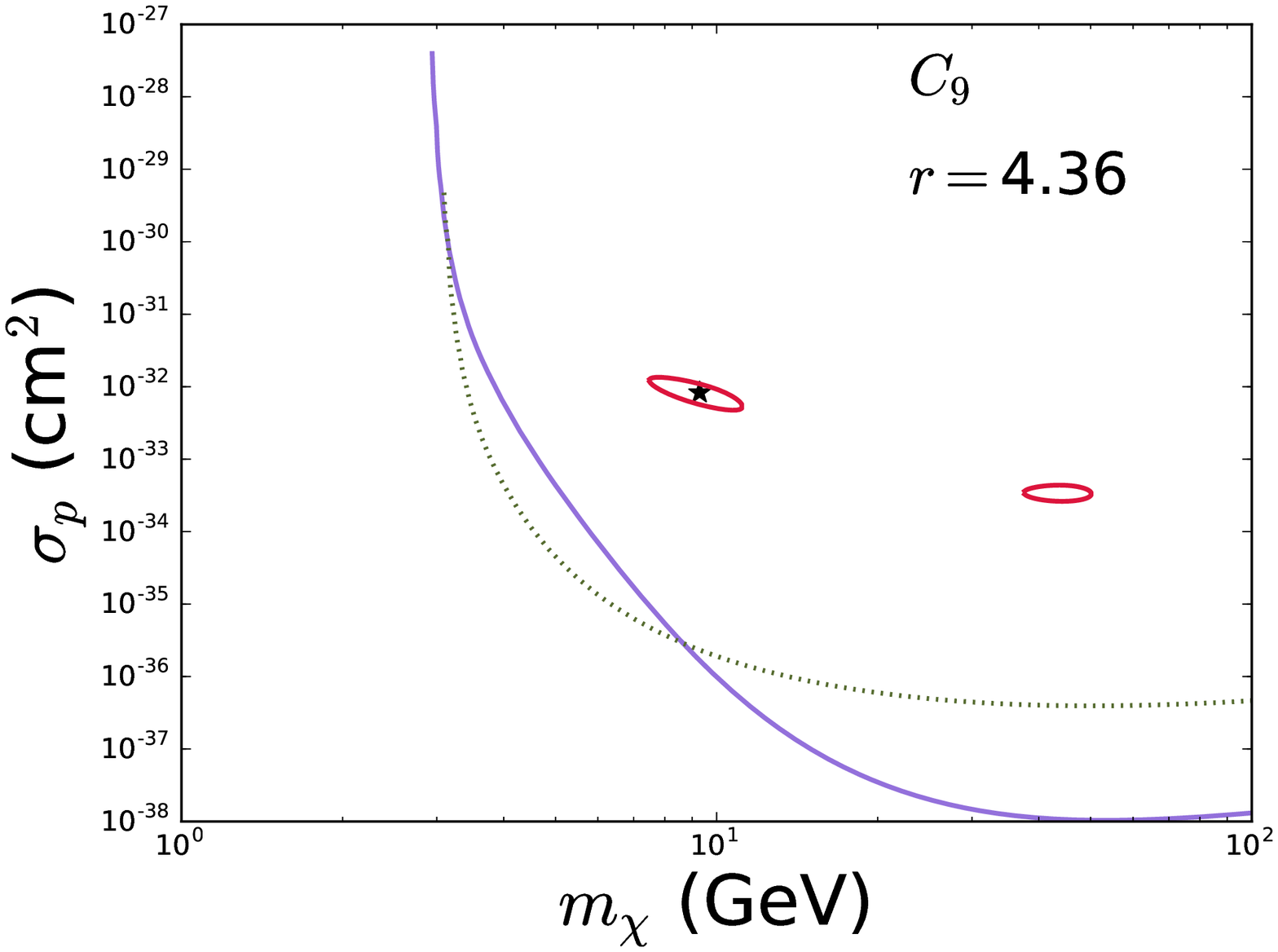}      
\includegraphics[width=0.32\textwidth]{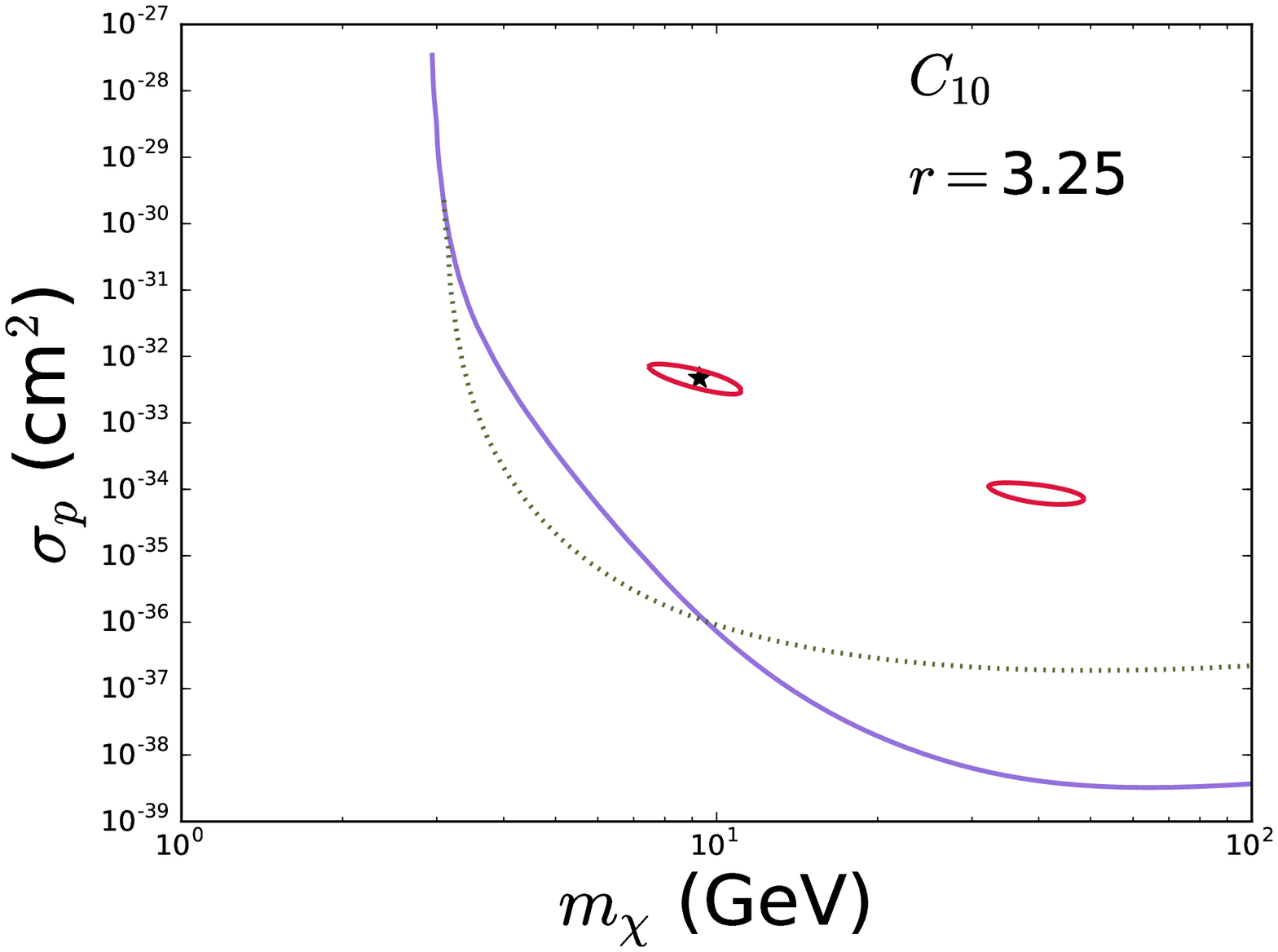}      
\includegraphics[width=0.32\textwidth]{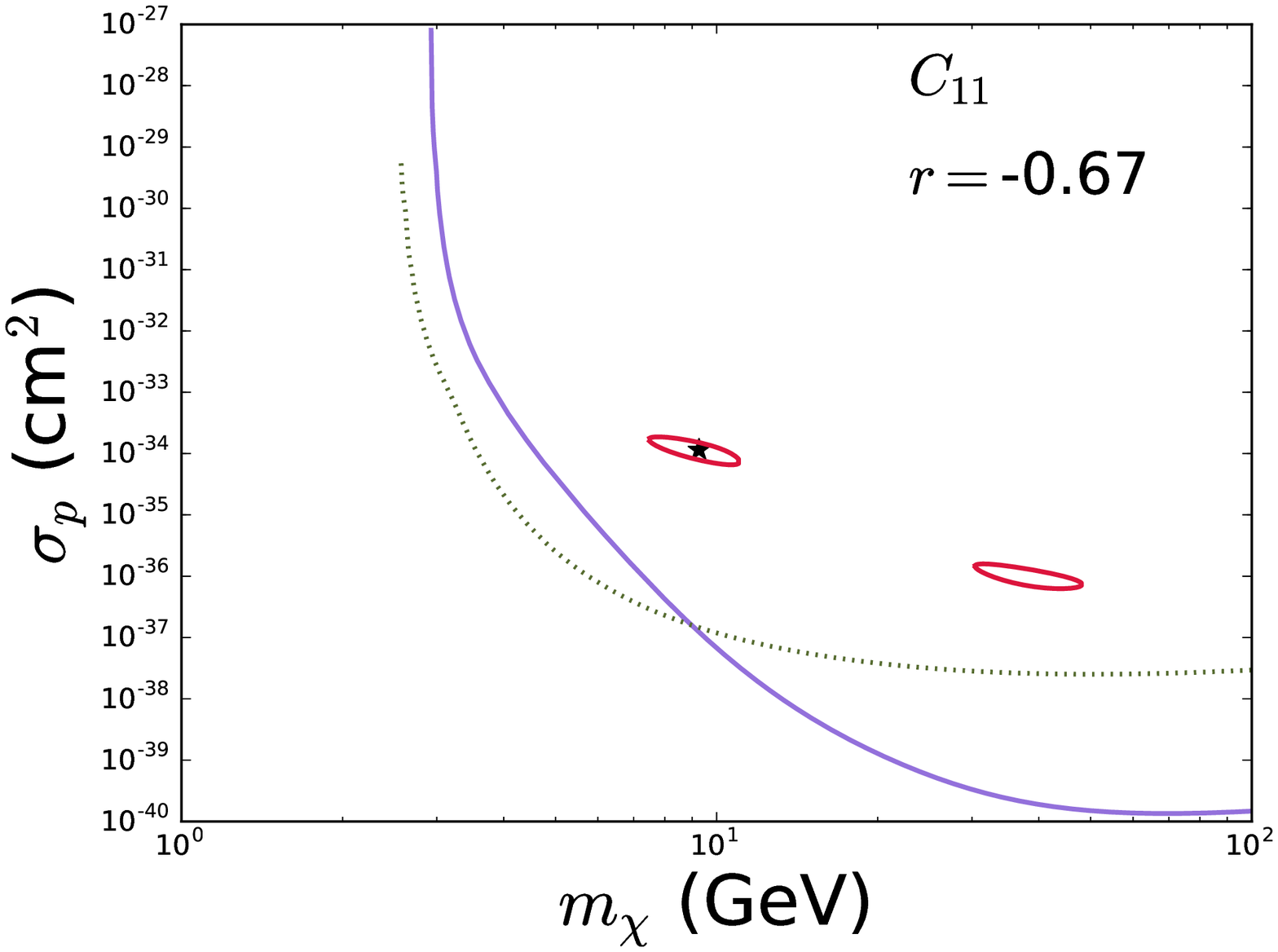}     
\includegraphics[width=0.32\textwidth]{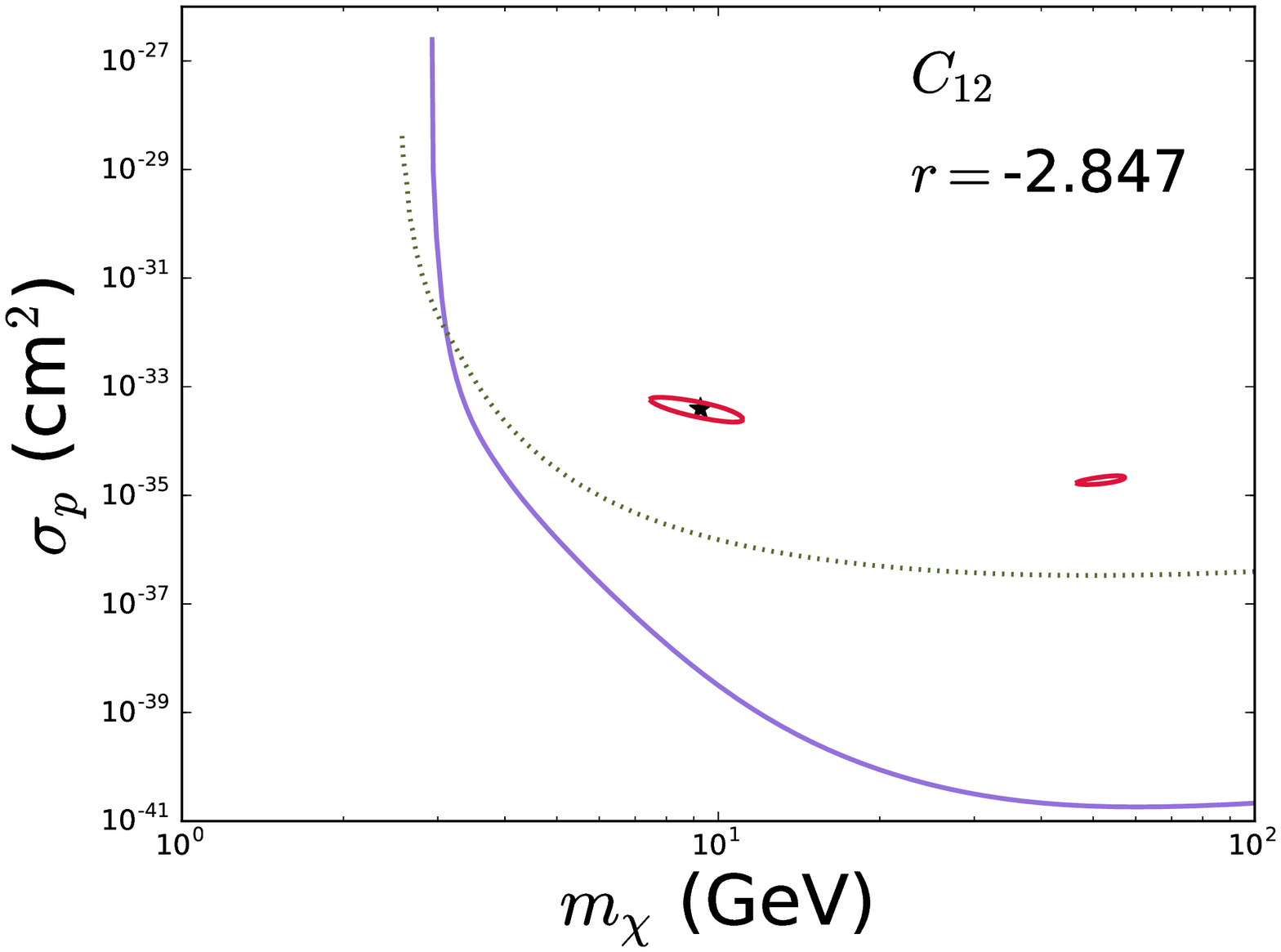}     
\includegraphics[width=0.32\textwidth]{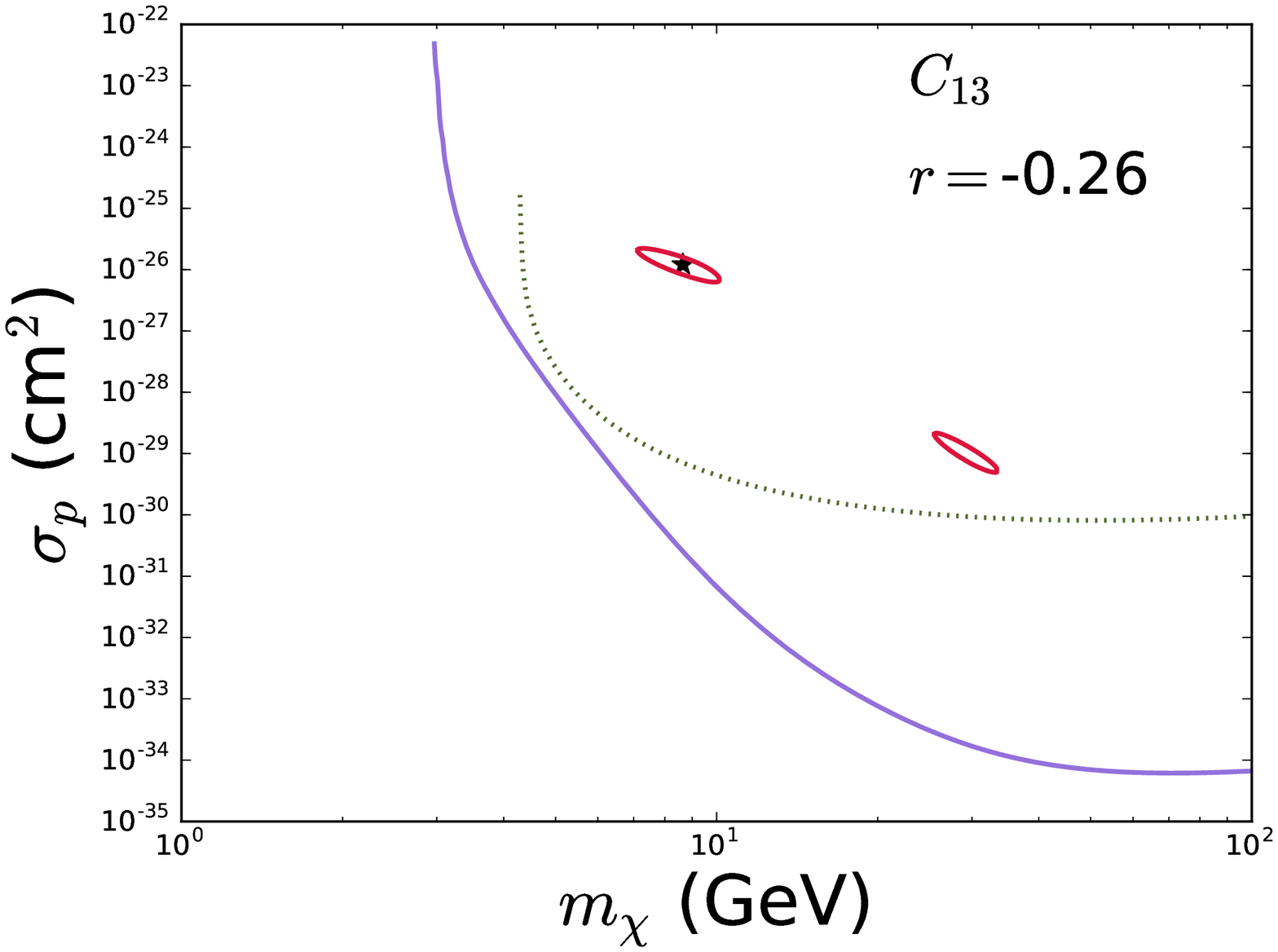}       
\includegraphics[width=0.32\textwidth]{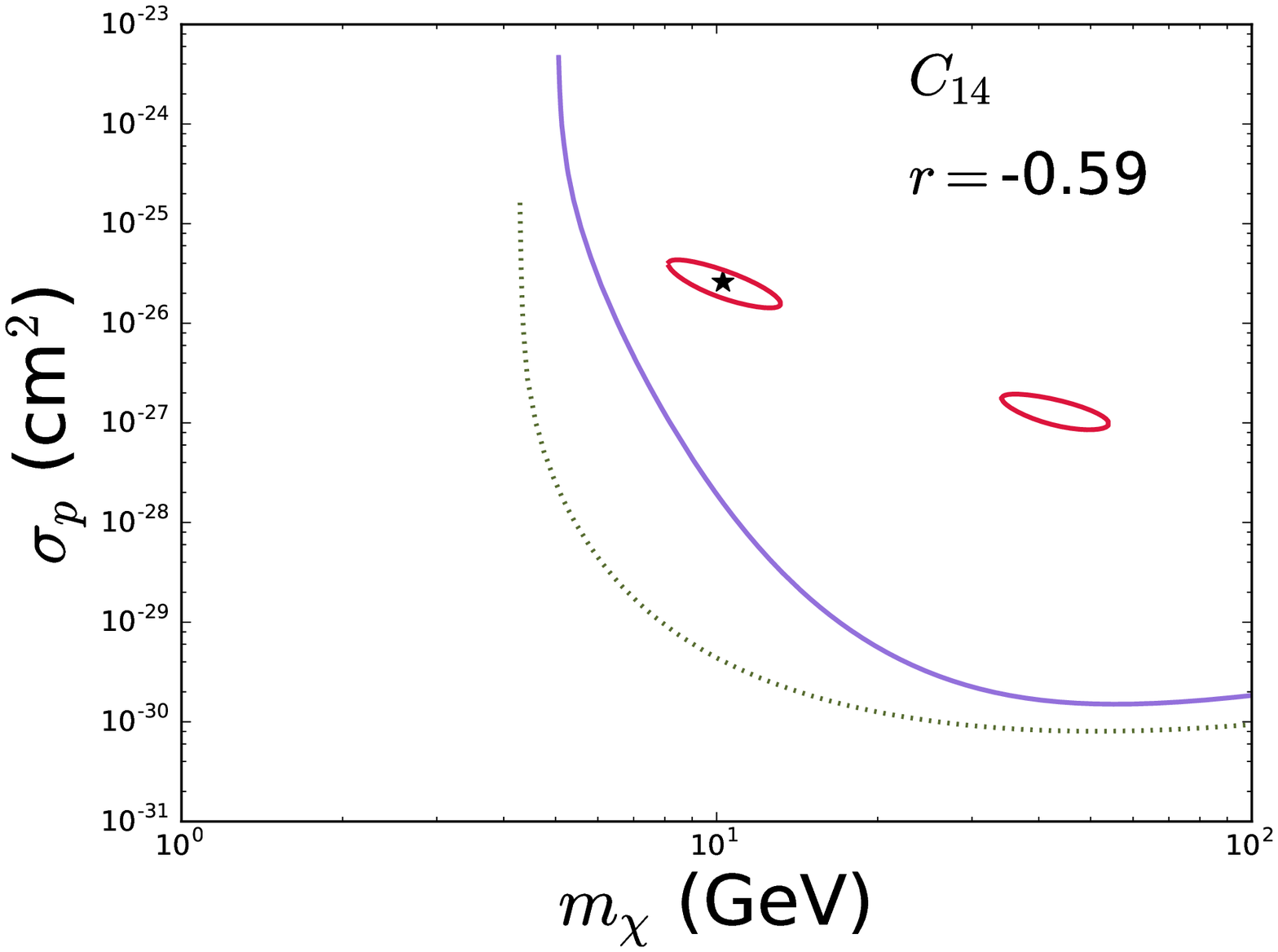}     
\includegraphics[width=0.32\textwidth]{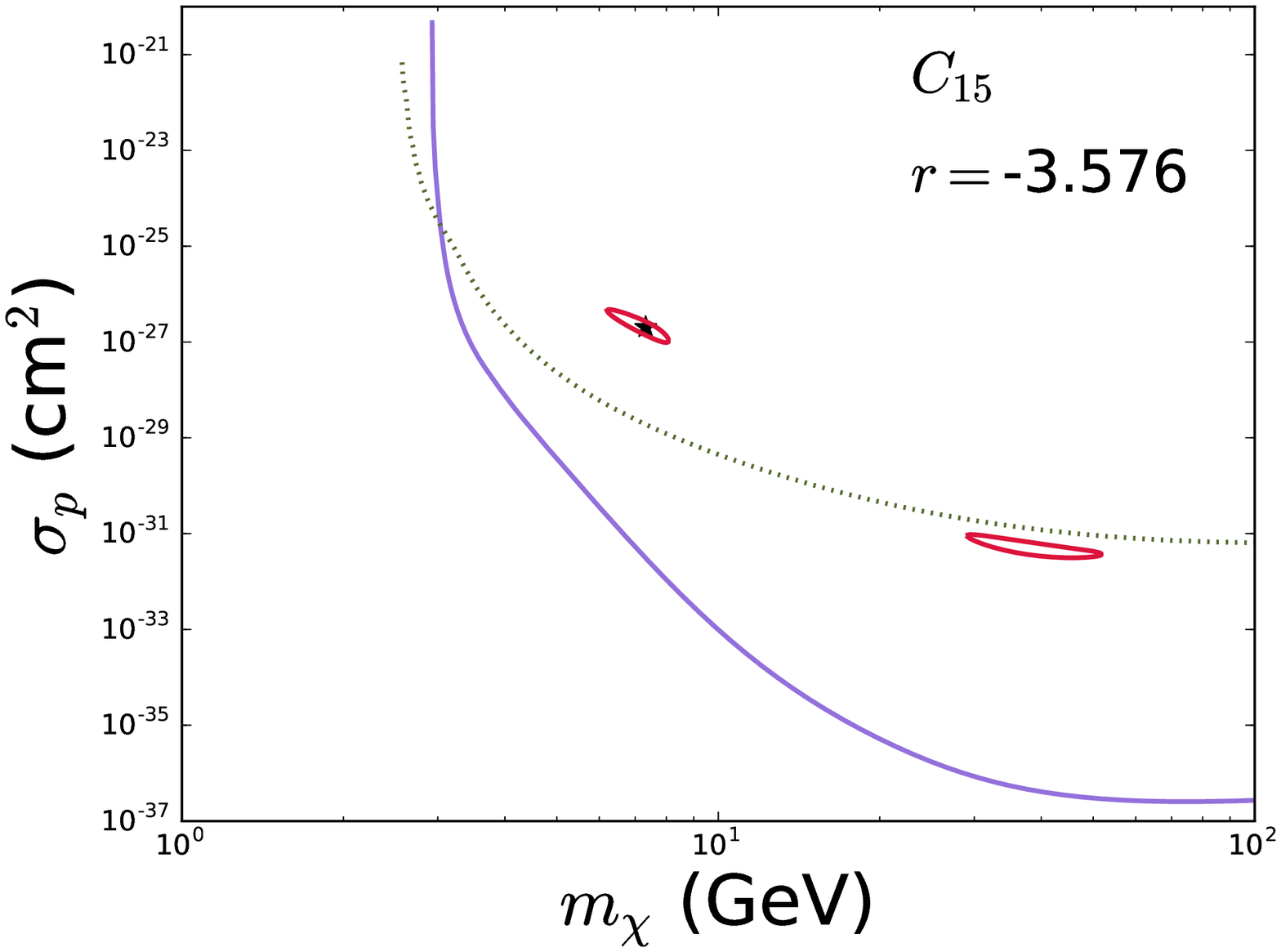}     
\end{center}
\caption{The 5--$\sigma$ best-fit DAMA regions are compared to the
  90\% C.L. upper bounds from XENON1T (solid purple line) and PICO60
  (green dots) in the $m_{\chi}$--$\sigma_p$ plane for each of the
  interaction terms of Eq.(\ref{eq:H}). In each plot the value of $r$
  is fixed to the corresponding absolute minima quoted in
  Table~\ref{tab:best_fit_values} and the star represents the absolute
  best-fit values of $m_{\chi}$ and $\sigma_p$.}
\label{fig:mchi_sigma_exclusions}
\end{figure}
The DAMA collaboration has recently released modulation amplitudes
$S_{m,k}^{exp}\equiv S_{m,[E_k^{\prime},E_{k+1}^{\prime}]}$, with
uncertainties $\sigma_k$, (corresponding to the predictions of
Eq.(\ref{eq:sm})) in the visible energy range 1 keVee$ < E^{\prime}<$
20 keVee in 0.5 keVee energy bins for a total exposure $\simeq$ 2.46
ton year, corresponding to the combination of DAMA/NaI
\cite{dama_1998}, DAMA/LIBRA--phase1 \cite{dama_2008,dama_2010} and
DAMA/LIBRA--phase2 \cite{dama_2018}.  In our analysis we will assume
constant quenching factors $q$=0.3 for sodium, 0.09 for iodine and a
Gaussian energy resolution ${\cal
  G}(E^{\prime},E_{ee})=Gauss(E^{\prime}|E_{ee},\sigma)=1/(\sqrt{2\pi}\sigma)exp(-(E^{\prime}-E_{ee})/2\sigma^2)$
with $\sigma$ = 0.0091 (E$_{ee}$/keVee) + 0.448 $\sqrt{E_{ee}}$/keVee
in keVee. To compare the theoretical predictions to the experimental
data, for each coupling $c_j^p$, $j=1,3,4...15$ we consider 14 energy
bins, of 0.5 keVee width, from 1 keVee to 8 keVee, and one
high--energy control bin from 8 keVee to 16 keVee
($[E_k^{\prime},E_{k+1}^{\prime}]$, $k=1,...,15$ ).  In particular the
combined DAMA phase1--phase2 high--energy spectrum shows a large
positive fluctuation in the interval 16 keVee--20 keVee with
modulation amplitude $S_m=0.0028\pm 0.0006$ cpd/kg/keVee, i.e. a more
than 4 $\sigma$ effect.  DAMA does not provide details about such
excess, but the residual rate in the the 10 keVee--20 keVee,
presumably dominated by the 16 keVee--20 keVee bin, shows random
positive and negative fluctuations during the years~\cite{dama_2018}
so it would be questionable to ascribe it to a possible WIMP
signal. Moreover, diluting such strong positive fluctuation in a
single 8 keVee--20 keVee interval would still yield a 3--$\sigma$
effect that would bias the fitting procedure without making use of its
very peculiar spectral features. For this reason we adopt as a control
high--energy bin the interval 8 keVee$<E^{\prime}<$16 keVee, where the
measured modulation amplitude is $S_m=0.00040\pm 0.00046$
cpd/kg/keVee.

We perform our $\chi^2$ test constructing the quantity:

\begin{equation}
\chi^2(m_{\chi},\sigma_p,r)=\sum_{k=1}^{15} \frac{\left [S_{m,k}-S^{exp}_{m,k}(m_{\chi},\sigma_p,r) \right ]^2}{\sigma_k^2}
  \label{eq:chi2}
  \end{equation}

\noindent and minimize it as a function of $(m_{\chi},\sigma_p,r)$.
In Fig. \ref{fig:chi2_m} we show the result of such minimization at
fixed WIMP mass $m_{\chi}$. From such figure one can see that for each
coupling $c_j^p$ two local minima are obtained. The details of such
minima are provided in Table \ref{tab:best_fit_values}.  The first
thing one can notice from such Table is that all models yield an
acceptable $\chi^2$: in the worst case, i.e. $c_7$,
$(\chi^2)_{min}$=13.94, with $p$--value $\simeq$ 0.30 with 15-3
degrees of freedom. Moreover, for all of them with the exception of
$c_7$ and $c_{15}$ the absolute minimum of the $\chi^2$ is below or
equal to that corresponding the standard SI interaction $c_1$.

In addition, the best fit parameters appear to be less tuned compared
to the SI case. This can be seen in Fig.~\ref{fig:chi2_m_r_planes},
where for each of the effective model couplings we provide the contour
plots of the $\chi^2$ in the $m_{\chi}$--$r$ plane. In such figures
the lines represent contours for $\chi^2-\chi^2_{min}=n^2$: $n=2$ for
the thin (green) solid lines, $n=3$ for the thick (red) solid lines,
$n=4$ for the thin (blue) dotted lines and $n=5$ for the thick (black)
dotted lines. In particular, the regions within 2 and 3 $\sigma$ for
$c_1$ appear strongly tuned to the value $r$=-0.76, corresponding to
a cancellation in the WIMP-iodine cross section, whereas for most of
the other effective interactions the corresponding contour encompasses
a much wider volume of the parameter space.

In agreement to the analysis of Ref.\cite{catena_dama}, also the WIMP
interpretation of the new DAMA data is in conflict to the constraints
from null experiments. To show this in
Fig. \ref{fig:mchi_sigma_exclusions} we compare the best-fit regions
in the $m_{\chi}$--$\sigma_p$ plane to the constraints from XENON1T
and PICO60.  In each plane the value of the $r$ parameter is fixed to
that of the corresponding absolute minimum in Table
\ref{tab:best_fit_values}, and the star represents the absolute
best-fit values of $m_{\chi}$ and $\sigma_p$. Moreover, for clarity we
only show with a solid (red) closed line the best-fit contour for
$\chi^2-\chi^2_{min}=n^2$ with $n$=5, while the solid (purple) and
dotted (green) open curves represent in each plot the 90\% upper bound
from XENON1T and PICO60, respectively. From such figure one can
conclude that all the best-fit solutions are in tension with the null
results of both experiments.

\begin{figure}
\begin{center}
 \includegraphics[width=0.8\columnwidth]{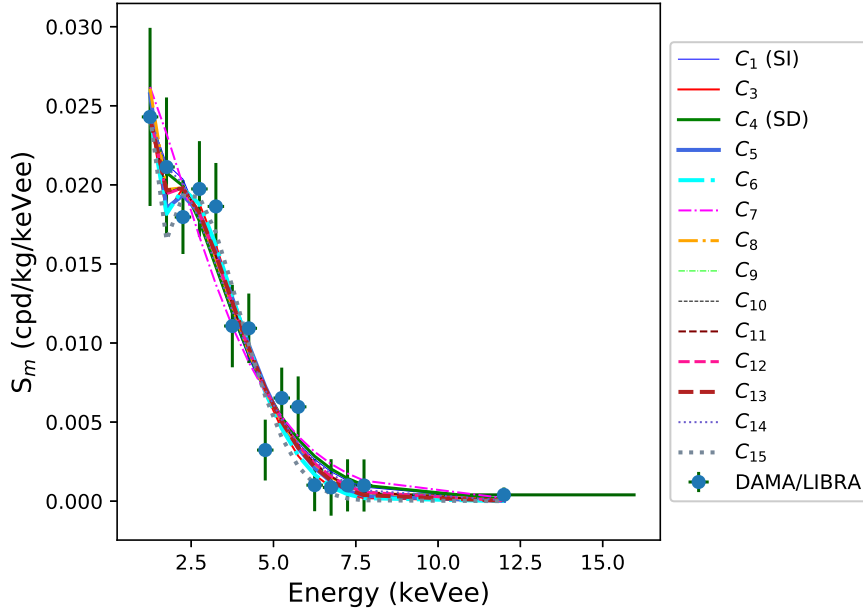}
\end{center}
\caption{DAMA modulation amplitudes as a function of the measured
  ionization energy $E_{ee}$ for the absolute minima of each effective
  model. The points with error bars correspond to the combined data of
  DAMA/NaI \cite{dama_1998}, DAMA/LIBRA--phase1
  \cite{dama_2008,dama_2010} and DAMA/LIBRA--phase2 \cite{dama_2018}.}
\label{fig:sm_tot}
\end{figure}

\begin{figure}[t]
\begin{center}
  \includegraphics[width=0.32\textwidth]{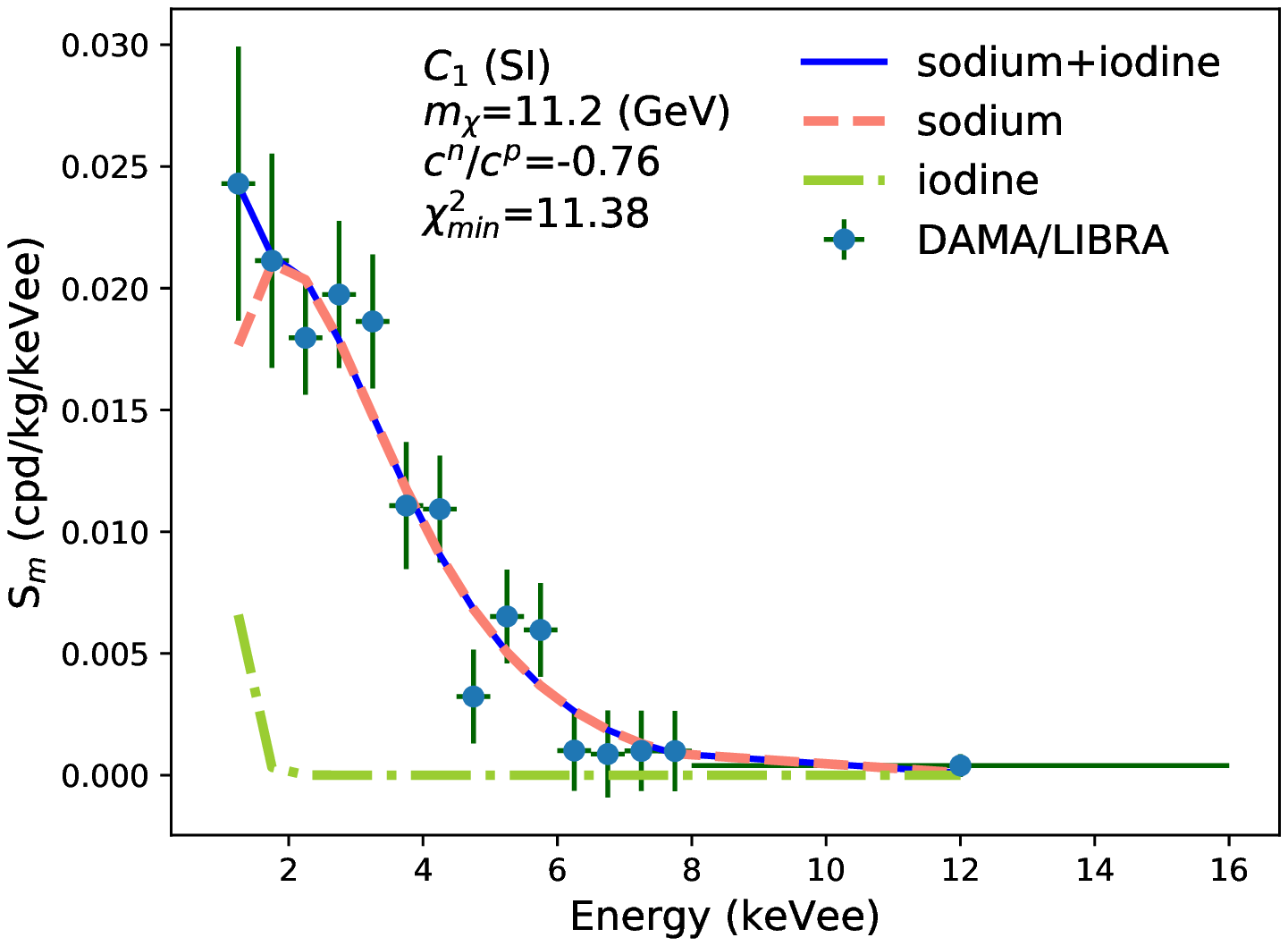}
  \includegraphics[width=0.32\textwidth]{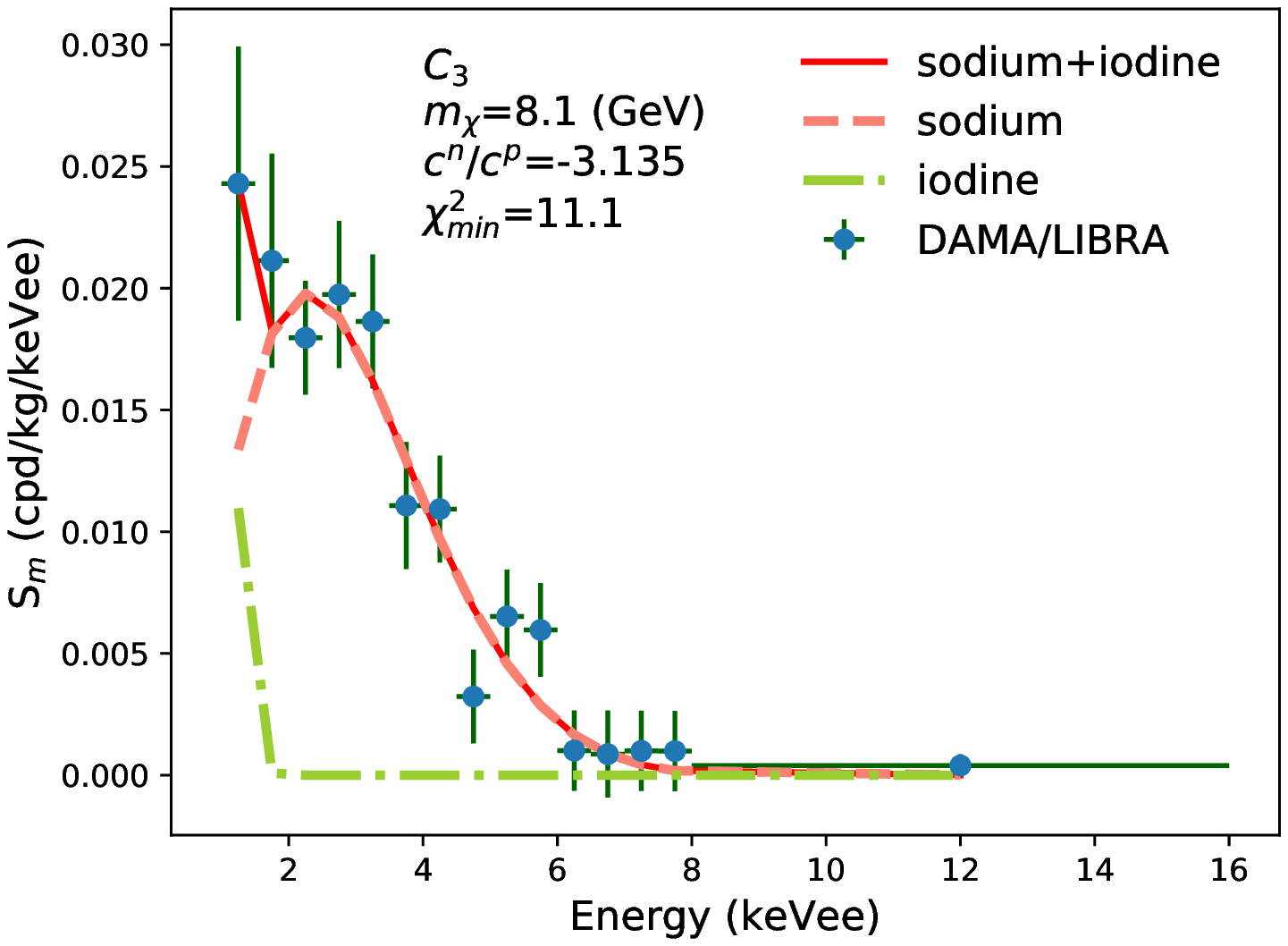}
  \includegraphics[width=0.32\textwidth]{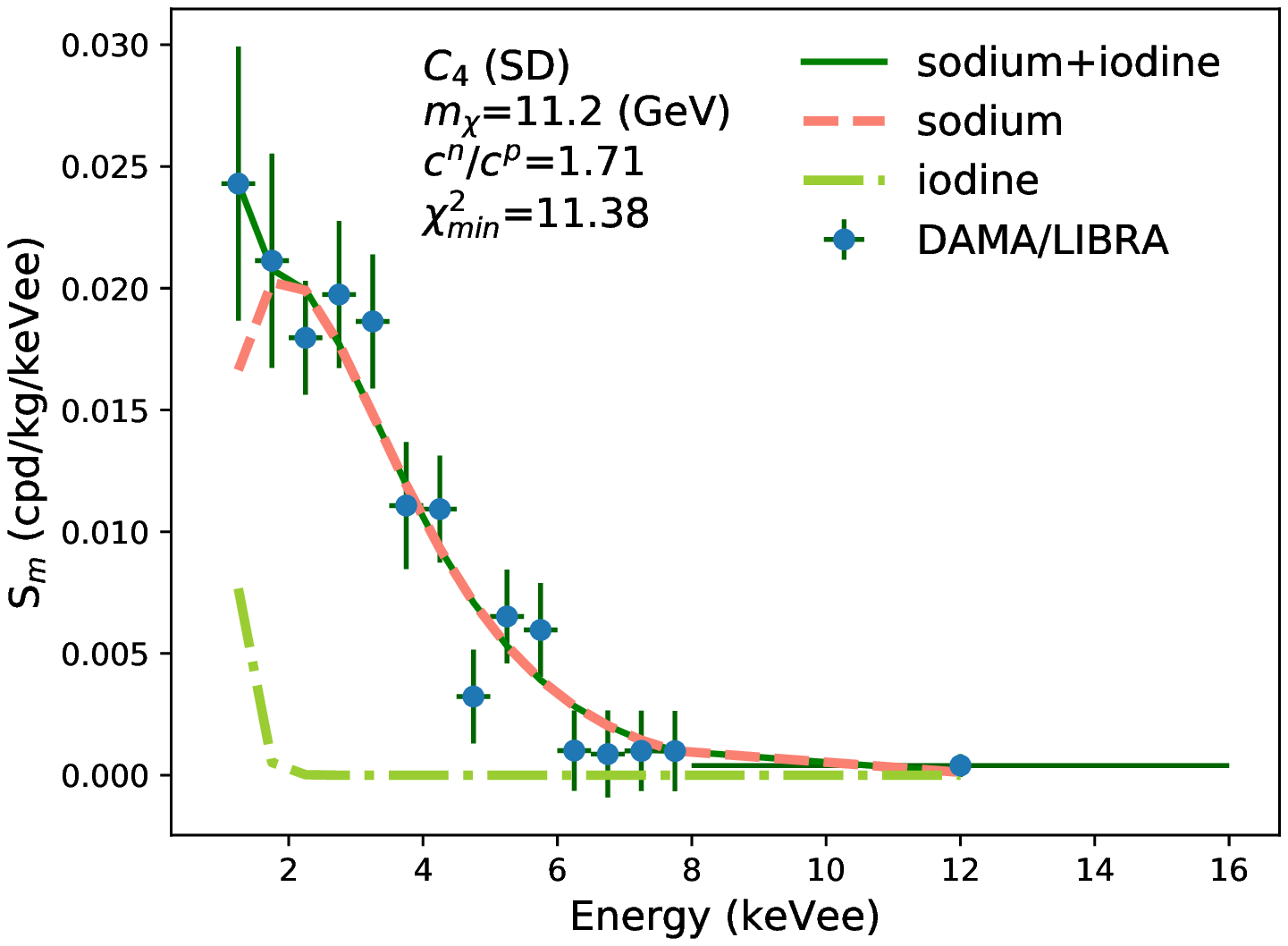}
  \includegraphics[width=0.32\textwidth]{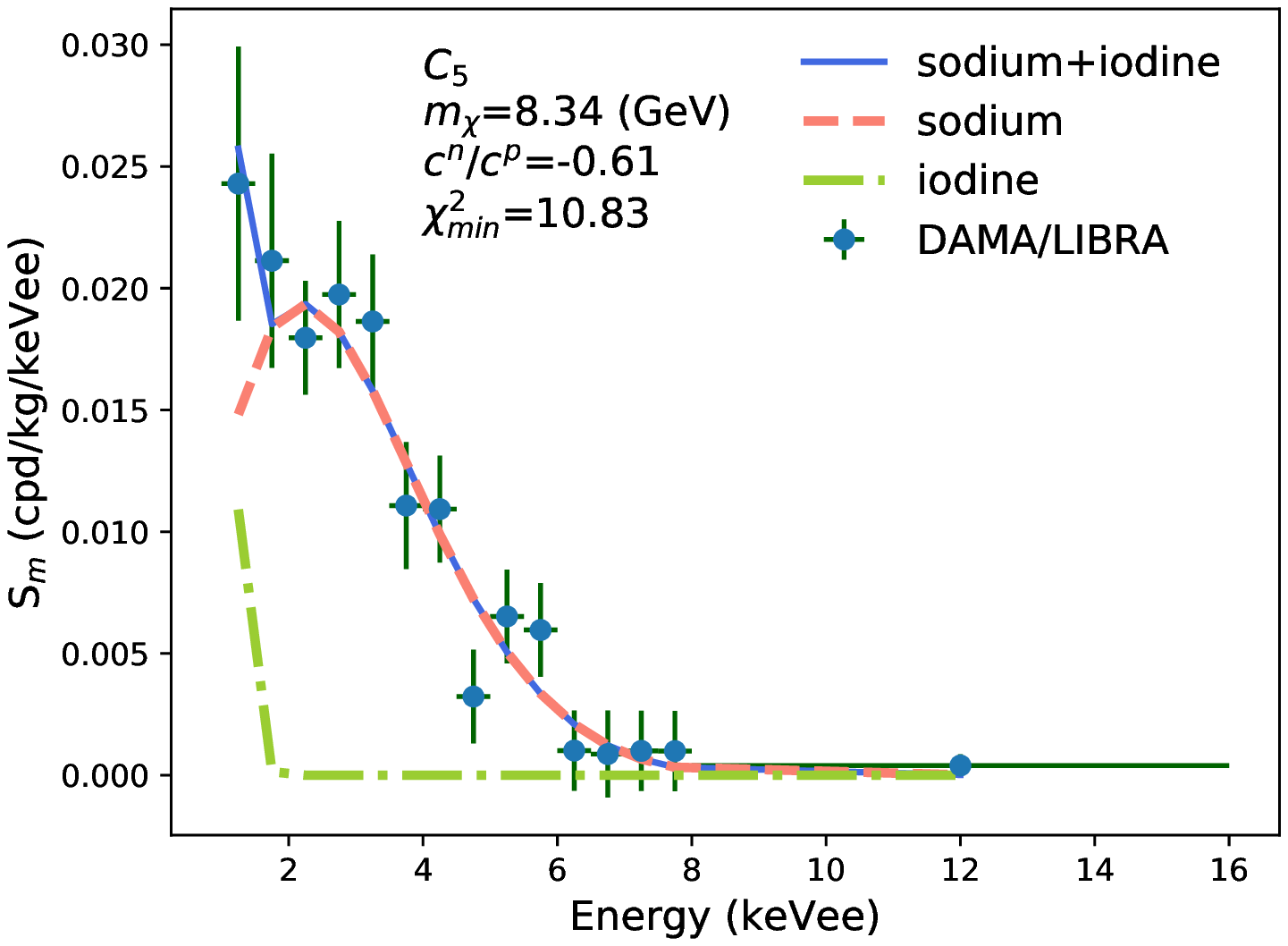}
  \includegraphics[width=0.32\textwidth]{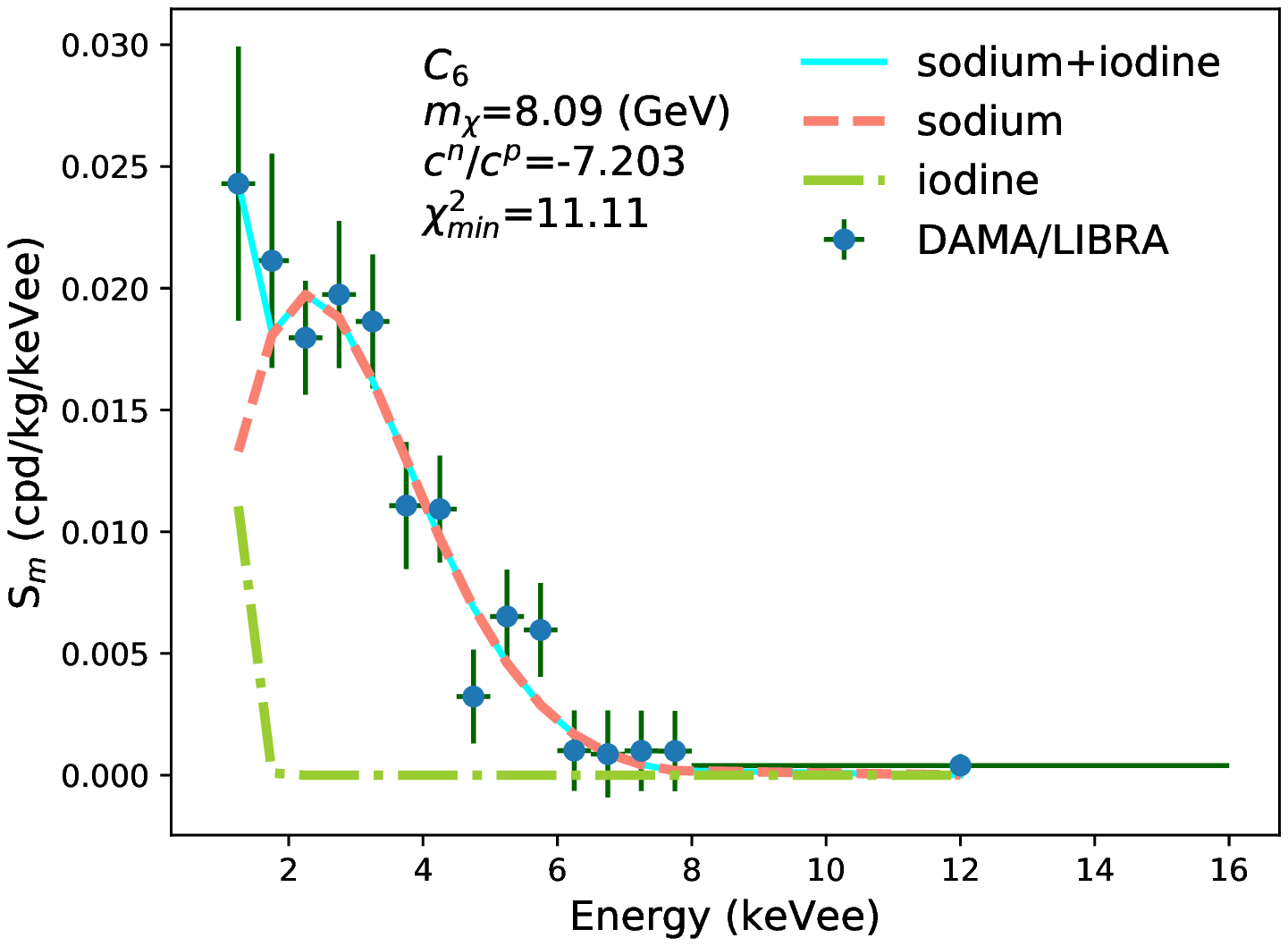}
  \includegraphics[width=0.32\textwidth]{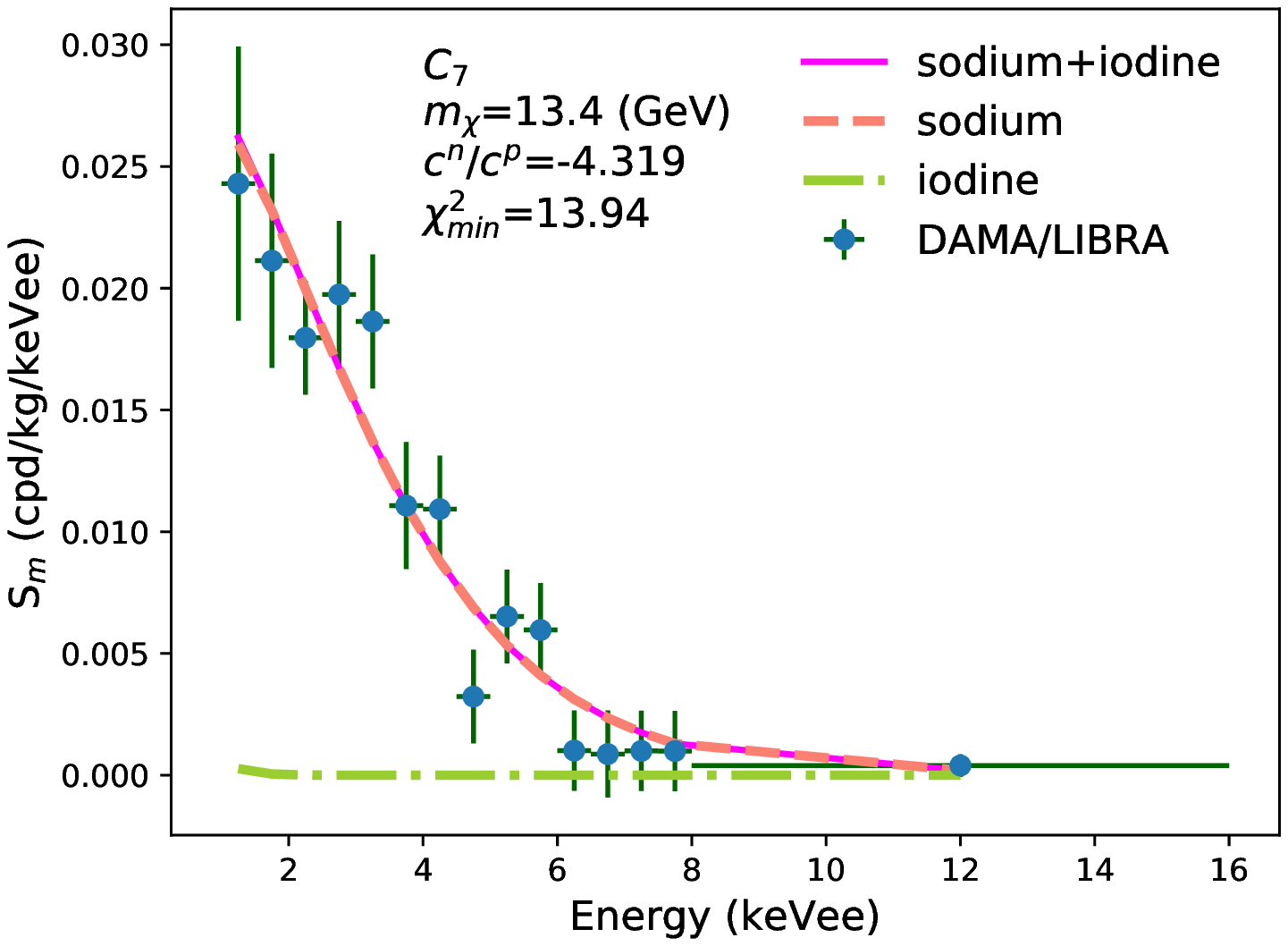}
  \includegraphics[width=0.32\textwidth]{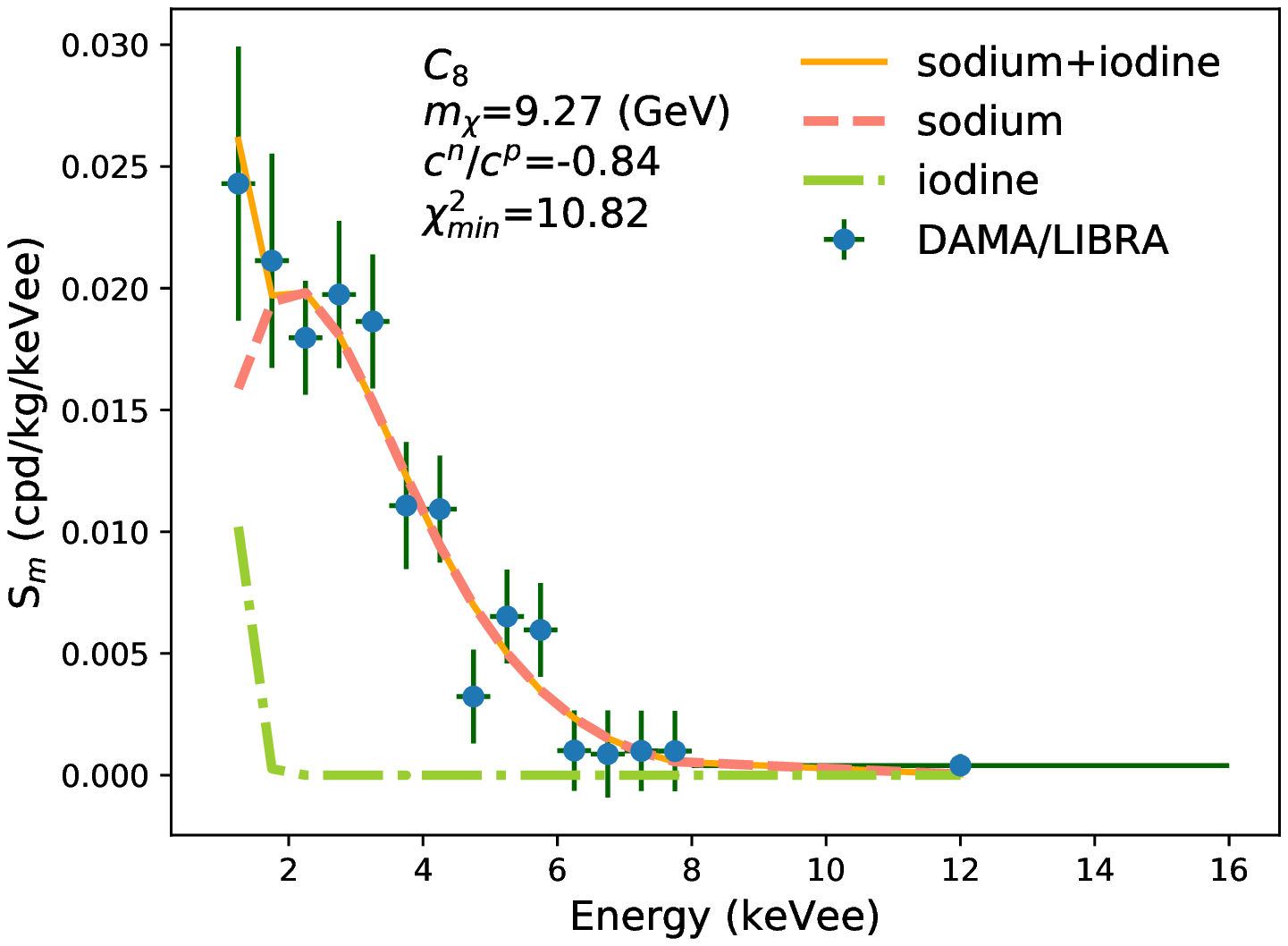}
  \includegraphics[width=0.32\textwidth]{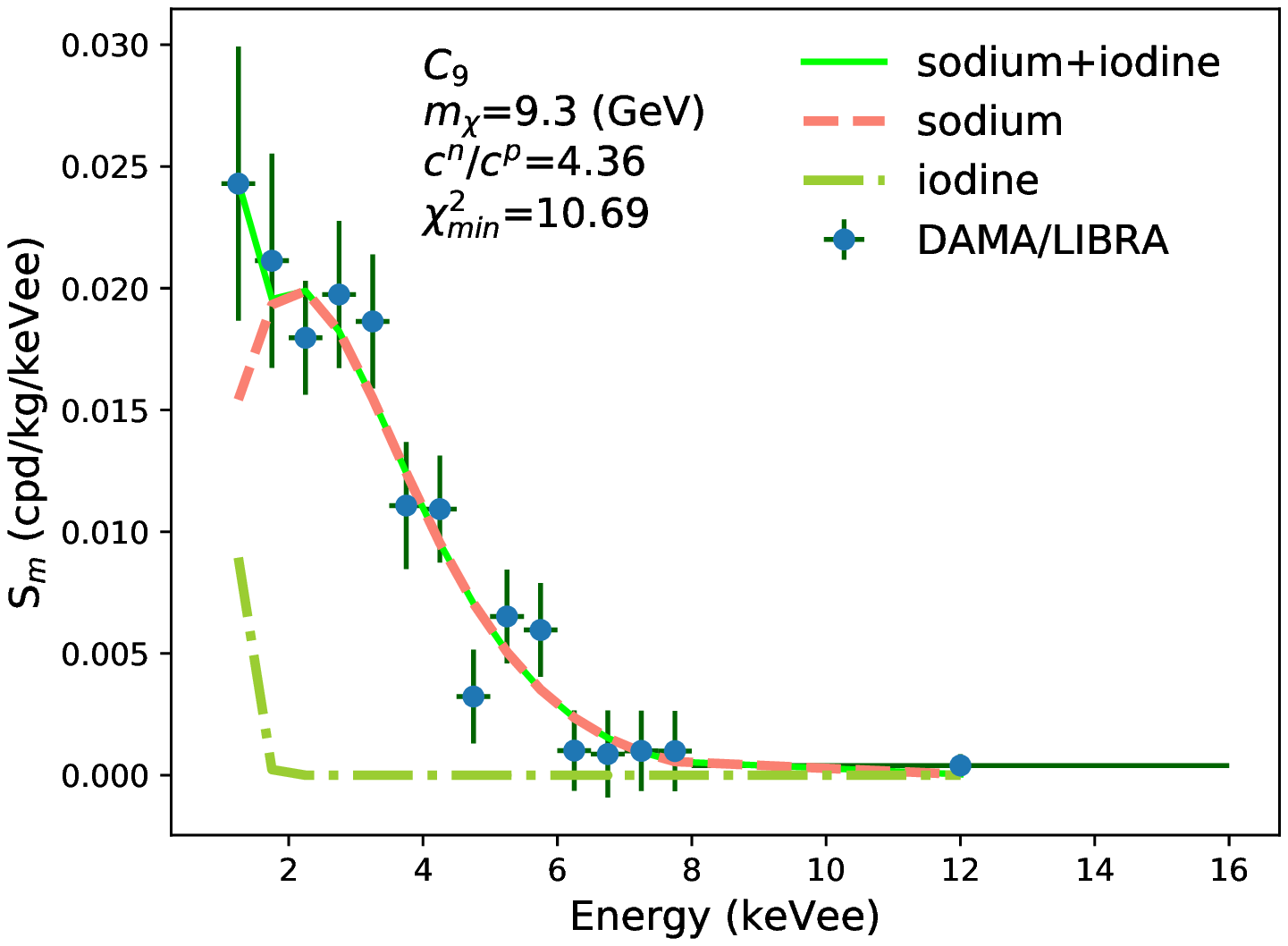}
  \includegraphics[width=0.32\textwidth]{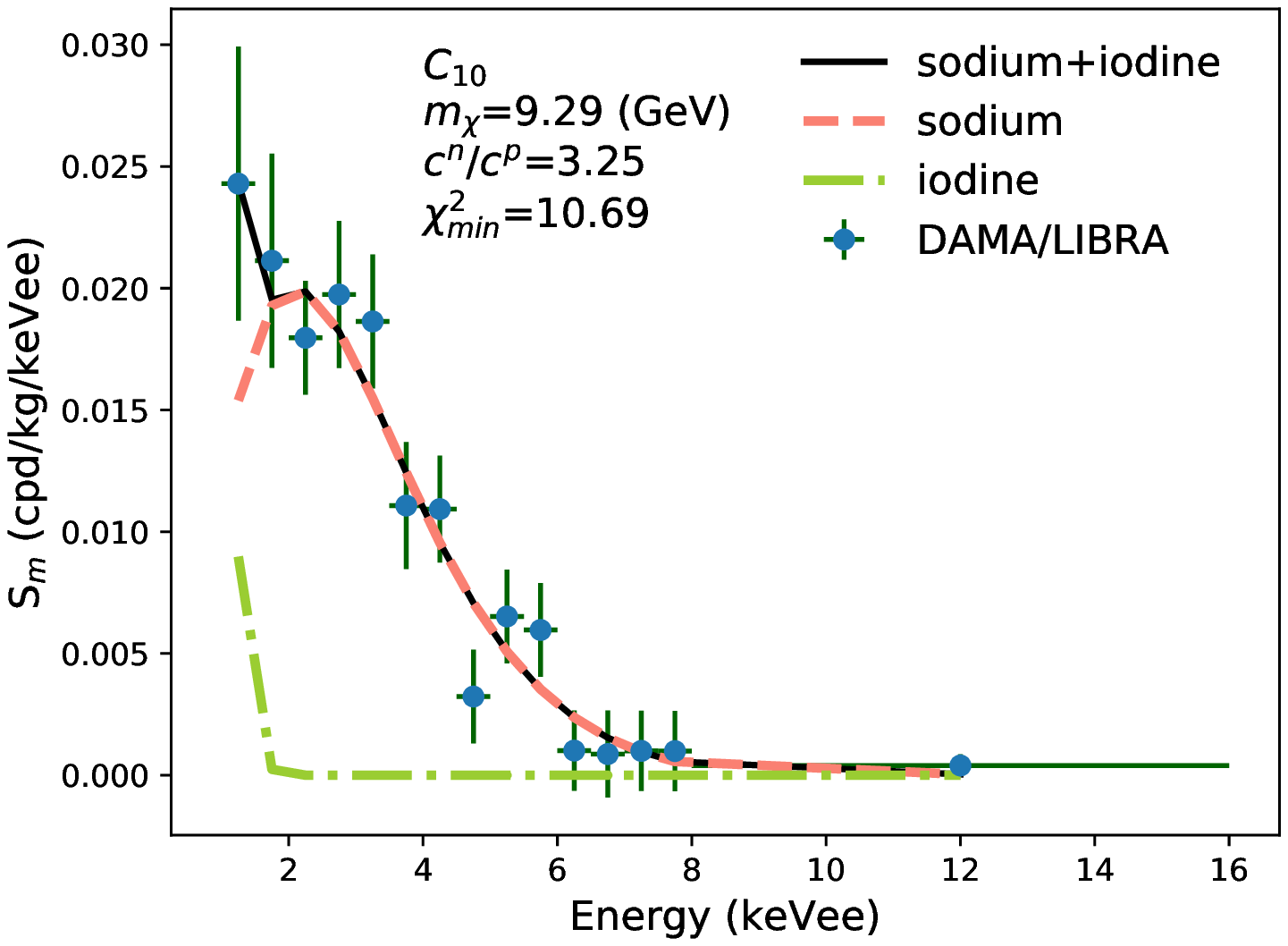}
  \includegraphics[width=0.32\textwidth]{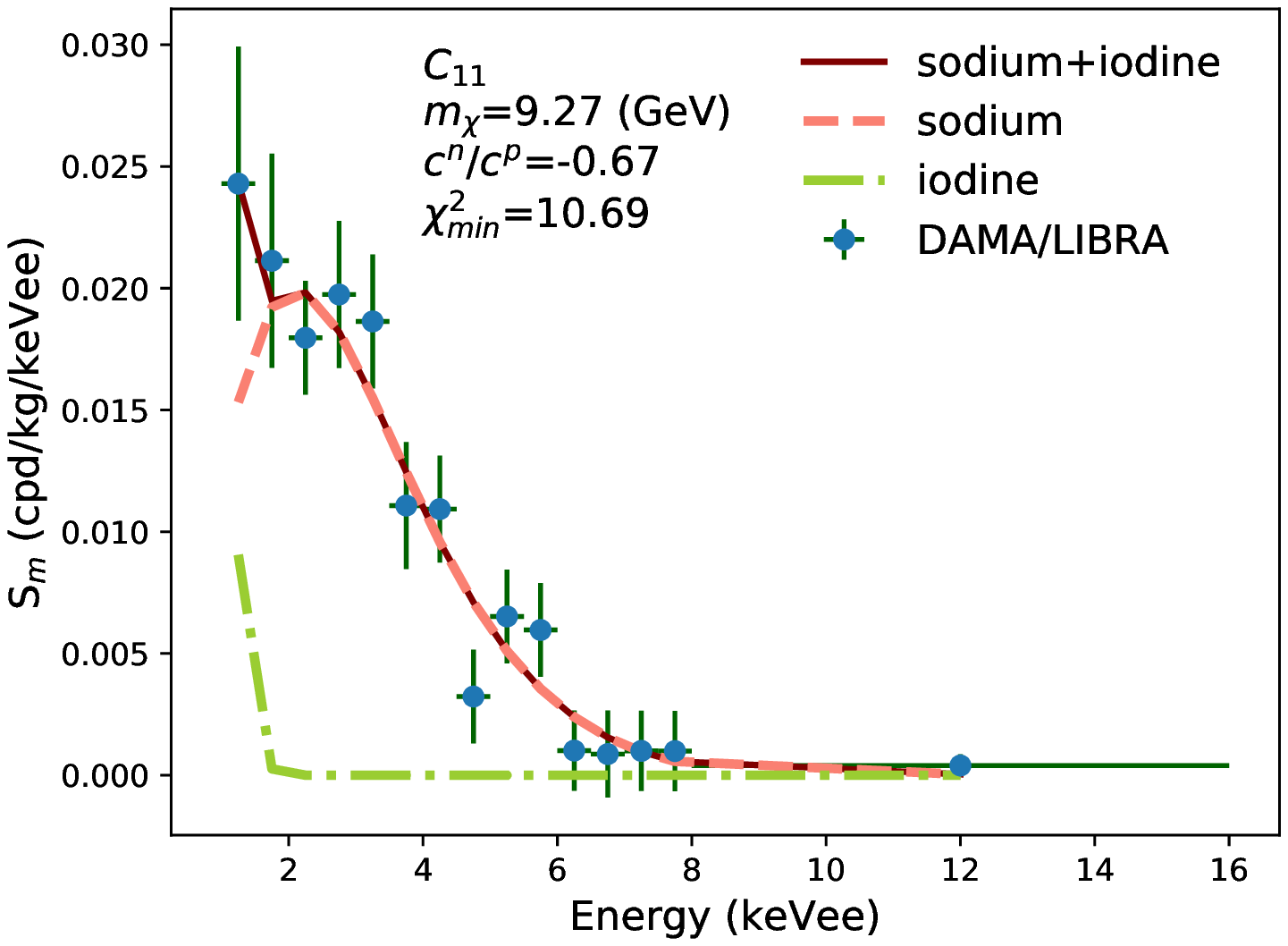}
  \includegraphics[width=0.32\textwidth]{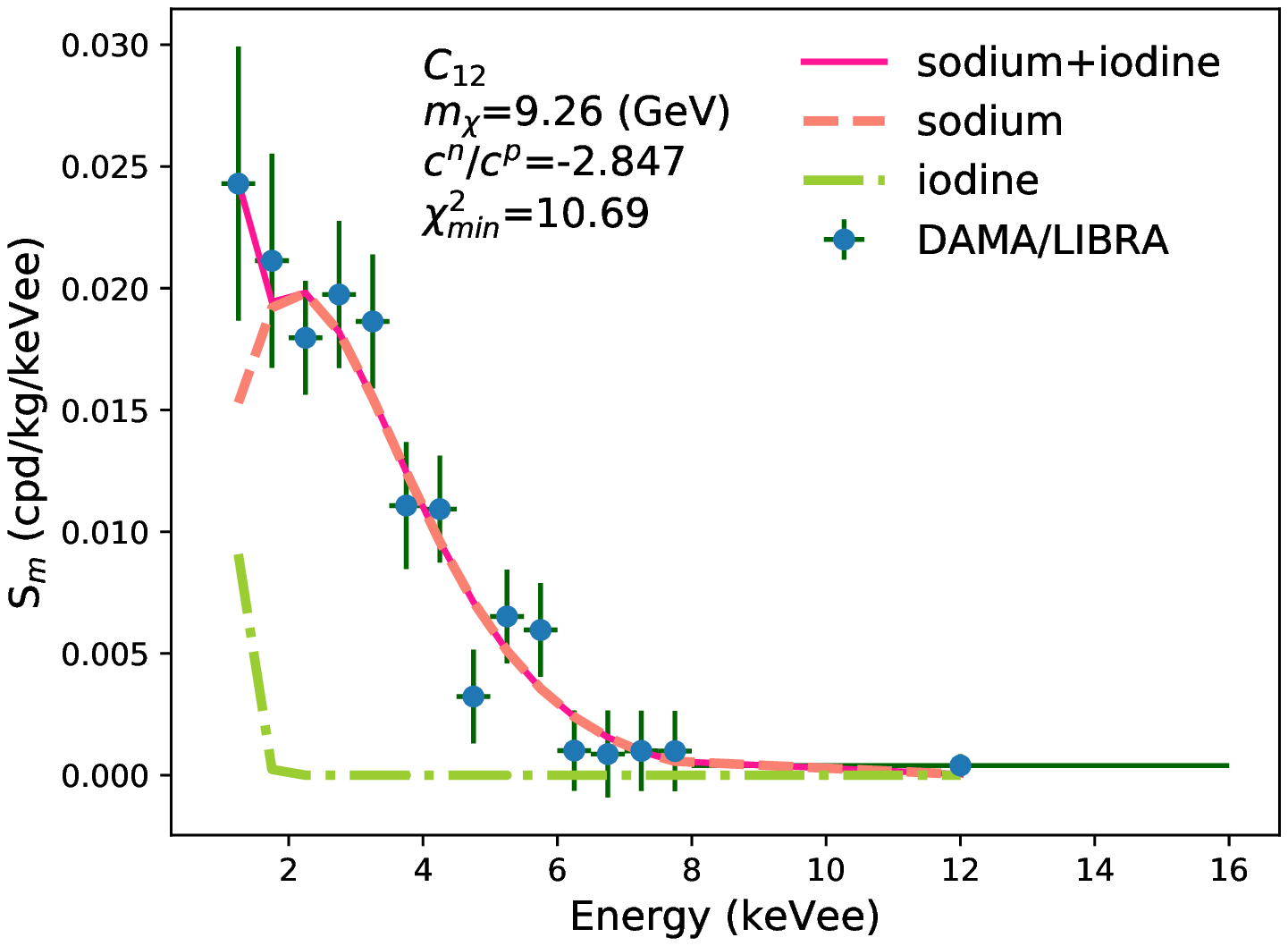}
  \includegraphics[width=0.32\textwidth]{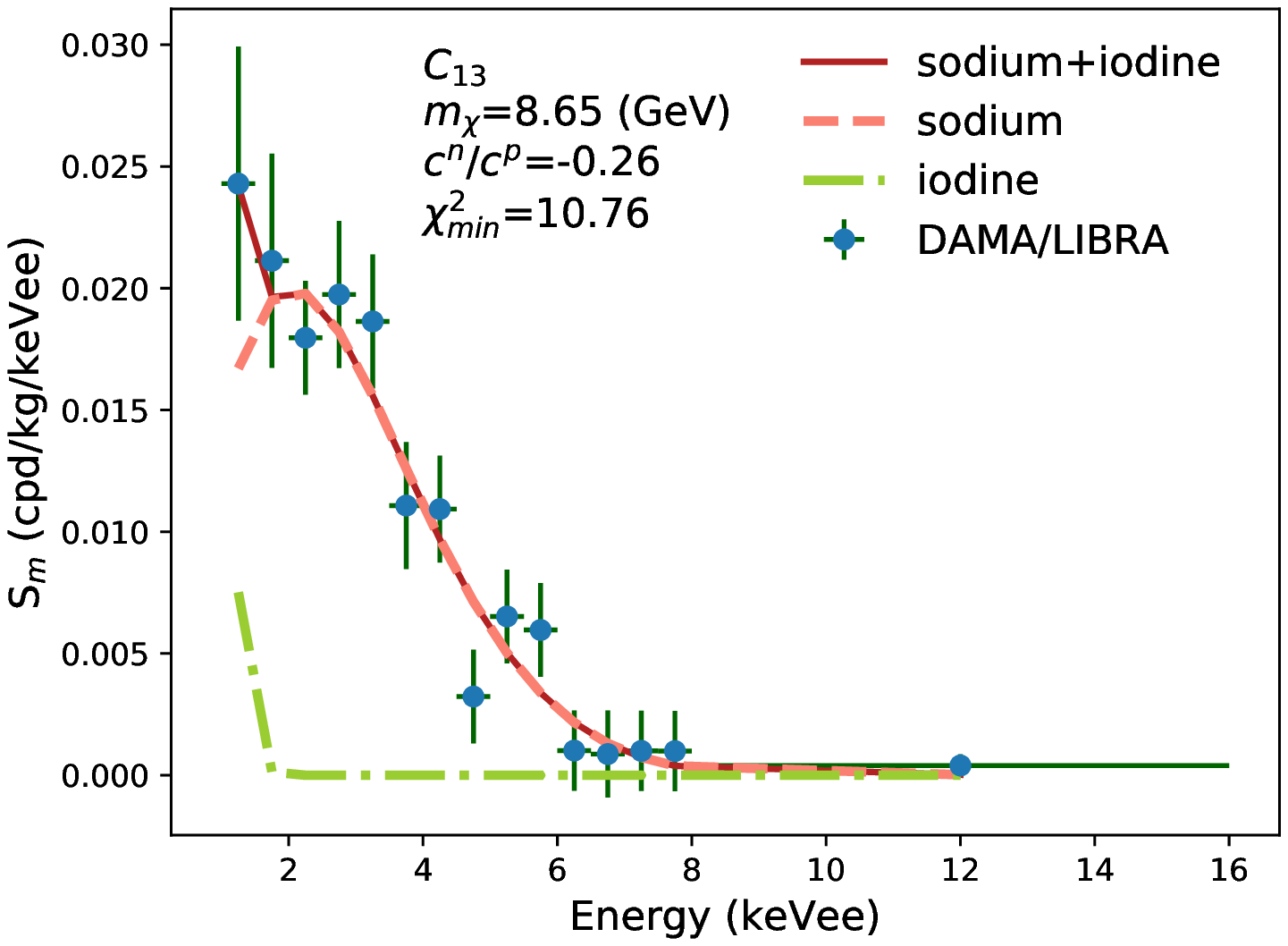}
  \includegraphics[width=0.32\textwidth]{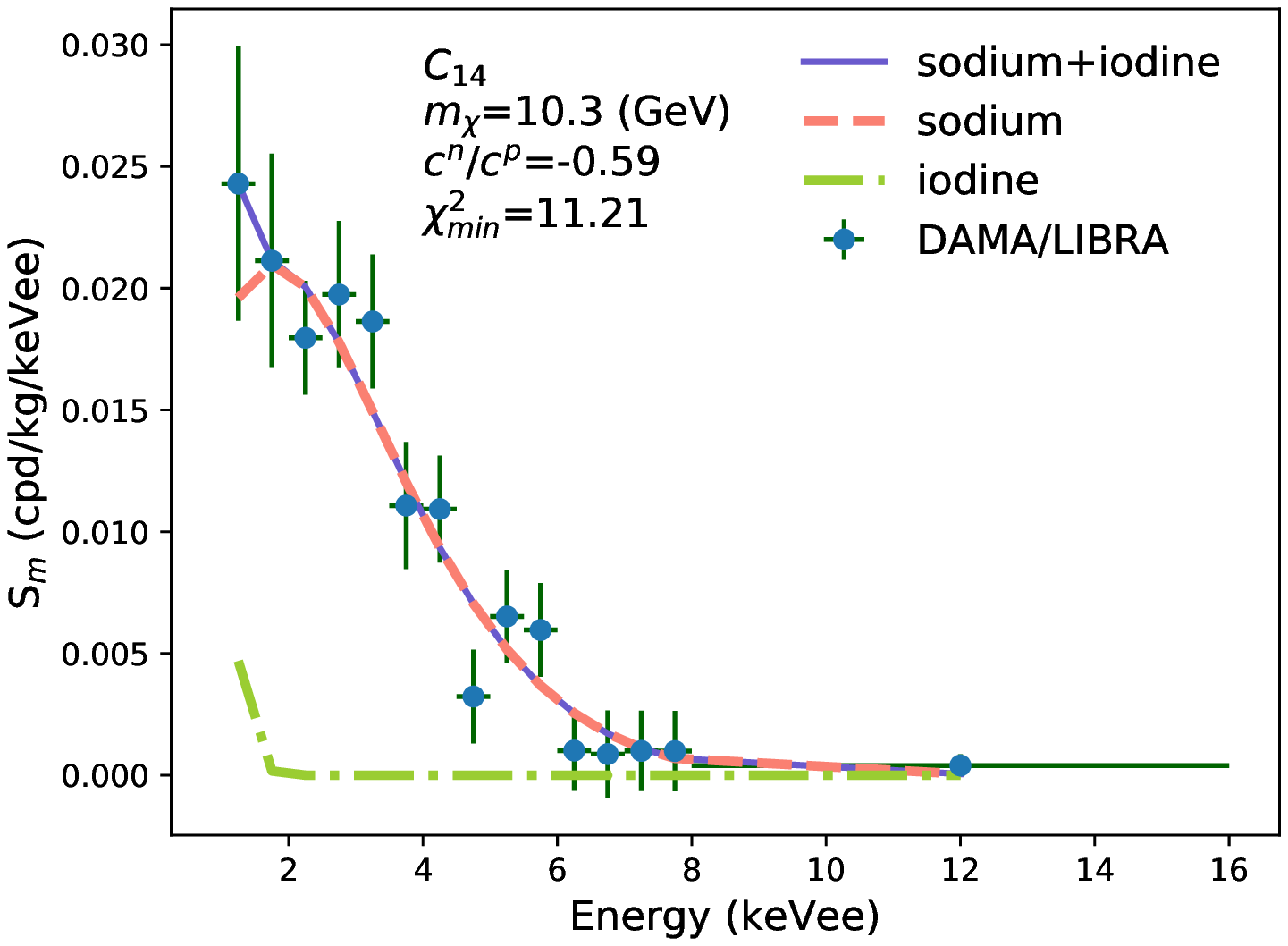}
  \includegraphics[width=0.32\textwidth]{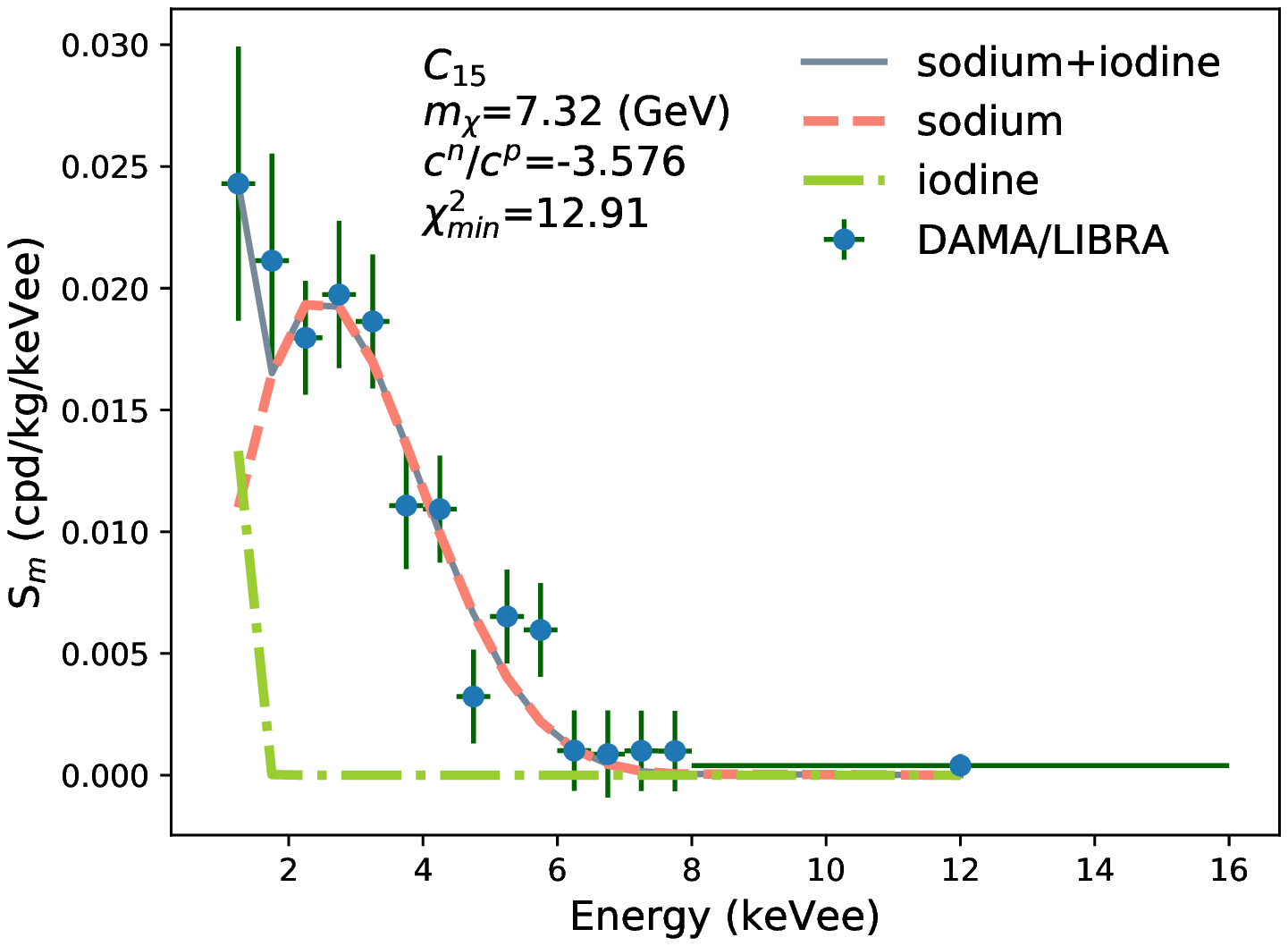}
\end{center}
\caption{DAMA modulation amplitudes as a function of the ionization
  energy $E_{ee}$ for the absolute minimum of each of the interaction
  terms of Eq.(\ref{eq:H}). The DAMA data points with corresponding
  error bars are shown in green color. The salmon (dashed) and
  yellowgreen (dot-dashed) lines show the contributions to the modulation
  amplitude from WIMP scattering off sodium and iodine
  respectively. The combined contributions (sodium+iodine) of each
  model are shown by a solid line.}
\label{fig:sm_contributions}
\end{figure}

To better understand the impact of the new data (and in particular,
the additional two experimental bins below 2 keVee) on the $\chi^2$ in
Fig. \ref{fig:sm_tot} we show the predicted modulation amplitudes for
our best--fit models and compare them to the corresponding
experimental data. Moreover, for each effective coupling the
contributions from WIMP--sodium (dashes) and WIMP--iodine
(dot--dashes) scattering to the total modulation amplitude (solid
line) is provided in Fig. \ref{fig:sm_contributions}.  In particular,
from this latter figure it is evident that, to provide a good fit to
the measured modulation amplitudes, a peculiar pattern for the two
contributions due to WIMP scattering off sodium and iodine is
required, namely a sodium contribution with maximum at approximately 2
keVee and approaching modulation phase inversion at lower energies,
where instead the term due to iodine is steeply increasing. Such
behavior is in agreement to the findings of Ref.\cite{freese_2018} for
the coupling $c_1$ and holds also for all the other interaction terms
of Eq.(\ref{eq:H}). In the case of $c_1$ the iodine contribution is
naturally enhanced compared to that of sodium due to the dependence of
the cross section on the square of the atomic mass number of the
target (see Eq. (\ref{eq:si})). As a consequence, the $r$ parameter
needs to be tuned to suppress the iodine contribution (i.e. close to
the value $r_{Iodine}\simeq -53/(127-53)\simeq$ -0.7, see
Eq.(\ref{eq:si})), since below 2 keVee the measured modulation
amplitudes are increasing only mildly.  In the case of $c_1$ this
inevitably reduces also the sodium contribution, since $-Z/(A-Z)$ is
roughly similar ($\simeq$ -0.9) also for sodium, enhancing the fine
tuning. This is clearly visible in the first panel of
Fig.~\ref{fig:chi2_m_r_planes}.  On the other hand, for all the other
interactions of Eq.(\ref{eq:H}) the value of $r$ corresponding to a
cancellation in the nuclear response function for iodine is normally
unrelated to that for sodium, so that the iodine contribution can be
suppressed without reducing that from sodium in a more natural way.
On top of that, with the exception of the $\Phi^{\prime\prime}$
nuclear response function, all the other ones typically show a milder
enhancement of the iodine signal compared to that for sodium in the
first place. In particular with the exception of $c_7$ and $c_{14}$
and, to a lesser extent, $c_5$ and $c_8$, the contribution of the
scattering amplitude proportional to $v_T^{\perp 2}$ (see
Eq.(\ref{eq:wimp_response_functions}) is completely negligible. On the
other hand for $c_7$ and $c_{14}$ only the term proportional to
$v_T^{\perp 2}$ is present in the cross section, while for $c_5$ and
$c_8$ such term is not negligible (for the choice of parameters
corresponding to the absolute minima of Table
\ref{tab:best_fit_values} it contributes between 10\% and 25\% of the
modulation amplitude in the first bin due to the iodine contribution
through the $M$ nuclear response function).  As a consequence of this,
the interaction terms $c_4$, $c_6$, $c_7$, $c_9$, $c_{10}$ and
$c_{14}$ depend on the spin--dependent nuclear response functions
$\Sigma^{\prime\prime}$ and/or $\Sigma^{\prime}$ which are
proportional, respectively, to the component of the nuclear spin along
the direction of the transferred momentum or perpendicular to it. This
implies only a factor $\simeq$ two hierarchy between the WIMP--iodine
and the WIMP--sodium cross sections.  Moreover, in the case of $c_5$
and $c_8$ the velocity--independent term of the cross section depends
on the $\Delta$ response function, which is proportional to the
nucleon angular momentum content of the nucleus, favoring elements
which have an unpaired nucleon in a non $s$--shell orbital. Both
iodine and sodium have this feature, implying also in this case no
large hierarchy between the cross sections off the two nuclei. Namely,
numerically the isoscalar response function at vanishing momentum
transfer $W^{00}_{T\Delta}(q\rightarrow 0)$ for sodium is a factor
$\simeq$ 0.25 smaller compared to that for iodine. Finally, the
WIMP--nucleus cross section for interaction $c_{13}$ is driven by the
$\tilde{\Phi}^{\prime}$ nuclear response function for which
$W^{00}_{T\tilde{\Phi}^{\prime}}(q\rightarrow 0)$ for sodium turns out
to be a factor $\simeq$ 6.3 larger than that for iodine. The bottom
line is that, compared to the standard SI interaction, for all such
effective models the cross section for scatterings off iodine is
naturally less enhanced or even subdominant, implying a lower fine
tuning of the parameters. On the other hand, interactions $c_3$,
$c_{12}$ and $c_{15}$ are driven by the $\Phi^{\prime\prime}$ nuclear
response function, which is sensitive to the product of the nucleon
spin and its angular momentum.  As a consequence, similarly to the SI
case, such interaction favors heavy elements over light ones, leading
to a large hierarchy between iodine over sodium.  However, as
explained above, in this case the value of $r$ corresponding to a
suppression of the iodine response function is quite different to that
for sodium (for instance, we checked that $r_{iodine}\simeq$ -2.3 and
$r_{sodium}\simeq$ -0.7 for a 2 keVee recoil energy). This implies
that, at variance with SI scattering, the iodine contribution for
$\Phi^{\prime\prime}$ can be suppressed without reducing that for
sodium, and less tuning is needed to obtain a good fit.

We conclude our discussion with a comment about the behavior of the
$\chi^2$ at large $m_{\chi}$. As can be seen from
Fig. \ref{fig:chi2_m} for most models (namely, all those for which the
explicit dependence of the cross section from $v^\perp$ is negligible)
the $\chi^2$ shows a steep rise at large $m_{\chi}$. This can be
understood because in this case the predicted modulation amplitude is
given by the cosine transform of the rate of Eq.(\ref{eq:dr_de}),
which is a function of the $v_{min}$ parameter only, and turns out to
be negative for $v_{min}\lsim$ 200 km/s
\cite{freese_2018}. Moreover, at fixed recoil energy $v_{min}$ is
decreasing with $m_{\chi}$ (and eventually independent on it).  As a
consequence, when $m_{\chi}$ is large one has $v_{min}<$ 200 km/s in
all the energy range of the DAMA signal, implying that the predicted
modulation amplitudes result negative. Since the corresponding
measured modulation amplitudes are all positive this implies a bad fit
to the data and a large $\chi^2$. Moreover, while in DAMA phase1 the
modulation data showed a maximum with energy, this is no longer true
for phase2, where they are monotonically decreasing.  This means that,
to get an acceptable $\chi^2$, the contribution to the expected
modulation rate of the halo function at small values of $v_{min}\lsim$
300 km/s must, if present, be subdominant, in order not to lead to a
spectrum rising with energy (this requirement is even more stringent
in presence of a momentum-dependence of the rate). This means that
effects such as gravitational focusing or energy-dependence of the
phase\cite{gf}, which pertain to the $v_{min}\lsim$ 300 km/s regime of
the WIMP velocity distribution integral, can only affect
configurations with a large $\chi^2$. On the other hand, when the
cross section shows a non--negligible dependence on $v^\perp$,
(i.e. for models $c_5$, $c_8$, $c_7$ and $c_{14}$) the integral of
Eq.(\ref{eq:dr_de}) is dominated by large values of $v>$ 200 km/s
irrespective of $v_{min}$ and positive modulation amplitudes can be
obtained in the energy range of the DAMA signal also at large values
of $m_{\chi}$, implying in such regime a milder increase of the
$\chi^2$.

\section{Conclusions}
\label{sec:conclusions}

The DAMA collaboration has released first results from the upgraded
DAMA/LIBRA-phase2 experiment \cite{dama_2018}.  In the present paper
we have fitted the updated DAMA result for the modulation amplitudes
in terms of a WIMP signal, parameterizing the WIMP--nucleus
interaction in terms of the most general WIMP--nucleus effective
Lagrangian for a WIMP particle of spin 0 or spin 1/2. In particular we
have systematically assumed the dominance of one of the 14 possible
interaction terms of the Hamiltonian of Eq.(\ref{eq:H}) and fitted the
experimental amplitudes to the three parameters $m_{\chi}$ (WIMP
mass), $\sigma_p$ (WIMP--nucleon effective
cross-section) and $c^n/c^p$ (neutron over proton coupling) assuming
for the WIMP velocity distribution a standard Maxwellian.  Our results
show that with only the two exceptions of $c_7$ and $c_{15}$ all the
couplings of the non--relativistic effective Hamiltonian can provide a
better fit compared to the SI case (see Table
\ref{tab:best_fit_values}), and with a reduced fine--tuning of the
parameters (see Fig.~\ref{fig:chi2_m_r_planes}).  This is explained by
the fact that in the new DAMA data the energy threshold has been
lowered from 2 keVee to 1 keVee, and the new energy bins are sensitive
to WIMP--iodine scatterings also for a low WIMP mass.  In the SI case
this requires to highly tune the parameters to suppress the iodine
contribution, in order to avoid an otherwise too steeply increasing
spectrum at low energy of the modulation amplitudes compared to the
data. On the other hand, if the WIMP--nucleus cross section is driven
by other operators the fine tuning required to suppress iodine is
reduced and/or the hierarchy between the WIMP--iodine and the
WIMP--sodium cross section is less pronounced in the first place.
Moreover, we have observed that effective models for which the cross
section depends explicitly on the WIMP incoming velocity show a
different phase of the modulation amplitudes at large values of the
WIMP mass compared to the standard velocity--independent
cross--section, allowing to get a better fit of the DAMA data. As
shown in Fig.~\ref{fig:mchi_sigma_exclusions} all the best fit
solutions are in tension with exclusion plots of both XENON1T and
PICO60.

\acknowledgments This research was supported by the Basic Science
Research Program through the National Research Foundation of
Korea(NRF) funded by the Ministry of Education, grant number
2016R1D1A1A09917964.

\appendix
\section{WIMP response functions}
\label{app:wimp_eft}

We collect here the WIMP particle--physics response functions introduced in Eq.(\ref{eq:squared_amplitude}) and adapted from \cite{haxton1,haxton2}:
\begin{eqnarray}
 R_{M}^{\tau \tau^\prime}\left(v_T^{\perp 2}, {q^2 \over m_N^2}\right) &=& c_1^\tau c_1^{\tau^\prime } + {j_\chi (j_\chi+1) \over 3} \left[ {q^2 \over m_N^2} v_T^{\perp 2} c_5^\tau c_5^{\tau^\prime }+v_T^{\perp 2}c_8^\tau c_8^{\tau^\prime }
+ {q^2 \over m_N^2} c_{11}^\tau c_{11}^{\tau^\prime } \right] \nonumber \\
 R_{\Phi^{\prime \prime}}^{\tau \tau^\prime}\left(v_T^{\perp 2}, {q^2 \over m_N^2}\right) &=& \left [{q^2 \over 4 m_N^2} c_3^\tau c_3^{\tau^\prime } + {j_\chi (j_\chi+1) \over 12} \left( c_{12}^\tau-{q^2 \over m_N^2} c_{15}^\tau\right) \left( c_{12}^{\tau^\prime }-{q^2 \over m_N^2}c_{15}^{\tau^\prime} \right)\right ]\frac{q^2}{m_N^2}  \nonumber \\
 R_{\Phi^{\prime \prime} M}^{\tau \tau^\prime}\left(v_T^{\perp 2}, {q^2 \over m_N^2}\right) &=& \left [ c_3^\tau c_1^{\tau^\prime } + {j_\chi (j_\chi+1) \over 3} \left( c_{12}^\tau -{q^2 \over m_N^2} c_{15}^\tau \right) c_{11}^{\tau^\prime }\right ] \frac{q^2}{m_N^2} \nonumber \\
  R_{\tilde{\Phi}^\prime}^{\tau \tau^\prime}\left(v_T^{\perp 2}, {q^2 \over m_N^2}\right) &=&\left [{j_\chi (j_\chi+1) \over 12} \left ( c_{12}^\tau c_{12}^{\tau^\prime }+{q^2 \over m_N^2}  c_{13}^\tau c_{13}^{\tau^\prime}  \right )\right ]\frac{q^2}{m_N^2} \nonumber \\
   R_{\Sigma^{\prime \prime}}^{\tau \tau^\prime}\left(v_T^{\perp 2}, {q^2 \over m_N^2}\right)  &=&{q^2 \over 4 m_N^2} c_{10}^\tau  c_{10}^{\tau^\prime } +
  {j_\chi (j_\chi+1) \over 12} \left[ c_4^\tau c_4^{\tau^\prime} + \right.  \nonumber \\
 && \left. {q^2 \over m_N^2} ( c_4^\tau c_6^{\tau^\prime }+c_6^\tau c_4^{\tau^\prime })+
 {q^4 \over m_N^4} c_{6}^\tau c_{6}^{\tau^\prime } +v_T^{\perp 2} c_{12}^\tau c_{12}^{\tau^\prime }+{q^2 \over m_N^2} v_T^{\perp 2} c_{13}^\tau c_{13}^{\tau^\prime } \right] \nonumber \\
    R_{\Sigma^\prime}^{\tau \tau^\prime}\left(v_T^{\perp 2}, {q^2 \over m_N^2}\right)  &=&{1 \over 8} \left[ {q^2 \over  m_N^2}  v_T^{\perp 2} c_{3}^\tau  c_{3}^{\tau^\prime } + v_T^{\perp 2}  c_{7}^\tau  c_{7}^{\tau^\prime }  \right]
       + {j_\chi (j_\chi+1) \over 12} \left[ c_4^\tau c_4^{\tau^\prime} +  \right.\nonumber \\
       &&\left. {q^2 \over m_N^2} c_9^\tau c_9^{\tau^\prime }+{v_T^{\perp 2} \over 2} \left(c_{12}^\tau-{q^2 \over m_N^2}c_{15}^\tau \right) \left( c_{12}^{\tau^\prime }-{q^2 \over m_N^2}c_{15}^{\tau \prime} \right) +{q^2 \over 2 m_N^2} v_T^{\perp 2}  c_{14}^\tau c_{14}^{\tau^\prime } \right] \nonumber \\
     R_{\Delta}^{\tau \tau^\prime}\left(v_T^{\perp 2}, {q^2 \over m_N^2}\right)&=& {j_\chi (j_\chi+1) \over 3} \left( {q^2 \over m_N^2} c_{5}^\tau c_{5}^{\tau^\prime }+ c_{8}^\tau c_{8}^{\tau^\prime } \right)\frac{q^2}{m_N^2} \nonumber \\
 R_{\Delta \Sigma^\prime}^{\tau \tau^\prime}\left(v_T^{\perp 2}, {q^2 \over m_N^2}\right)&=& {j_\chi (j_\chi+1) \over 3} \left (c_{5}^\tau c_{4}^{\tau^\prime }-c_8^\tau c_9^{\tau^\prime} \right) \frac{q^2}{m_N^2}.
\label{eq:wimp_response_functions}
\end{eqnarray}

\section{Constraints}
  \label{app:exp}
In the present analysis we include the constraints from
XENON1T~\cite{xenon_1t} and PICO60~\cite{pico60}.

\subsection{XENON1T}

For XENON1T we have assumed zero WIMP candidate events in the range 3
PE$\le S_1 \le$ 30 PE in the lower half of the signal band, as shown
in figure 2 of Ref.\cite{xenon_1t} for the primary scintillation
signal S1 (directly in Photo Electrons, PE) for an exposure of 34.2
days and a fiducial volume of 1042 kg of xenon. We have used the
efficiency taken from Fig. 1 of~\cite{xenon_1t}, a light collection
efficiency $g_1$=0.144, while for the light yield $L_y$ we have used
the NEST model of Ref. \cite{nest} with an electric field E=120 V/cm
and the parameters of Table 1 with the exception of the Lindhard
parameter k=0.15, to reproduce the combined energy curves of Fig. 2b of~\cite{xenon_1t}.

For XENON1T we have modeled the energy resolution
combining a Poisson fluctuation of the observed primary signal $S_1$
compared to $<S_1>$ and a Gaussian response of the photomultiplier
with $\sigma_{PMT}=0.5$, so that:

\begin{equation}
{\cal G}_{Xe}(E_R,S)=\sum_{n=1}^{\infty}
Gauss(S|n,\sqrt{n}\sigma_{PMT})Poiss(n,<S(E_R)>),
\label{eq:g_xe}  
\end{equation}

\noindent with $Poiss(n,\lambda)=\lambda^n/n!exp(-\lambda)$.

\subsection{PICO60}

PICO60~\cite{pico60} uses $C_3F_8$ as the target. Only the threshold
$E_{th}$=3.3 keV was analyzed, with a total exposure of 1167.0 kg day
and 0 event detected. We use for fluorine and carbon the nucleation
probabilities of Fig. 4 of~\cite{pico2l}.


\begin{thebibliography}{99}



\bibitem{planck}
{\bf Planck} Collaboration, P.~A.~R. Ade et~al., {\it {Planck 2013 results.
  XVI. Cosmological parameters}},  {\em Astron. Astrophys.} {\bf 571} (2014)
  A16, [\href{http://arxiv.org/abs/1303.5076}{{\tt arXiv:1303.5076}}].

\bibitem{dama_2008}
{\bf DAMA} Collaboration, R.~Bernabei et~al., {\it {First results from
  DAMA/LIBRA and the combined results with DAMA/NaI}},  {\em Eur. Phys. J.}
  {\bf C56} (2008) 333--355, [\href{http://arxiv.org/abs/0804.2741}{{\tt
  arXiv:0804.2741}}].

\bibitem{dama_2010}
{\bf DAMA, LIBRA} Collaboration, R.~Bernabei et~al., {\it {New results from
  DAMA/LIBRA}},  {\em Eur. Phys. J.} {\bf C67} (2010) 39--49,
  [\href{http://arxiv.org/abs/1002.1028}{{\tt arXiv:1002.1028}}].

\bibitem{dama_2013}
R.~Bernabei et~al., {\it {Final model independent result of
  DAMA/LIBRA-phase1}},  {\em Eur. Phys. J.} {\bf C73} (2013) 2648,
  [\href{http://arxiv.org/abs/1308.5109}{{\tt arXiv:1308.5109}}].

\bibitem{xenon_1t}
{\bf XENON} Collaboration, E.~Aprile et~al., {\it {First Dark Matter Search
  Results from the XENON1T Experiment}},  {\em Phys. Rev. Lett.} {\bf 119}
  (2017), no.~18 181301, [\href{http://arxiv.org/abs/1705.06655}{{\tt
  arXiv:1705.06655}}].

\bibitem{lux}
{\bf LUX} Collaboration, D.~S. Akerib et~al., {\it {First results from the LUX
  dark matter experiment at the Sanford Underground Research Facility}},  {\em
  Phys. Rev. Lett.} {\bf 112} (2014) 091303,
  [\href{http://arxiv.org/abs/1310.8214}{{\tt arXiv:1310.8214}}].

\bibitem{xenon100}
{\bf XENON100} Collaboration, E.~Aprile et~al., {\it {Dark Matter Results from
  225 Live Days of XENON100 Data}},  {\em Phys. Rev. Lett.} {\bf 109} (2012)
  181301, [\href{http://arxiv.org/abs/1207.5988}{{\tt arXiv:1207.5988}}].

\bibitem{xenon10}
{\bf XENON10} Collaboration, J.~Angle et~al., {\it {A search for light dark
  matter in XENON10 data}},  {\em Phys. Rev. Lett.} {\bf 107} (2011) 051301,
  [\href{http://arxiv.org/abs/1104.3088}{{\tt arXiv:1104.3088}}]. [Erratum:
  Phys. Rev. Lett.110,249901(2013)].

\bibitem{kims}
S.~C. Kim et~al., {\it {New Limits on Interactions between Weakly Interacting
  Massive Particles and Nucleons Obtained with CsI(Tl) Crystal Detectors}},
  {\em Phys. Rev. Lett.} {\bf 108} (2012) 181301,
  [\href{http://arxiv.org/abs/1204.2646}{{\tt arXiv:1204.2646}}].

\bibitem{kims_modulation}
Y.~Kim, {\it {Recent progress in KIMS experiment}},  {\em talk given at
  13$^{th}$ International Conference on Topics in Astroparticle and Underground
  Physics, September 8--13 2013, Asilomar, California USA (TAUP2013)}.

\bibitem{cdms_ge}
{\bf CDMS-II} Collaboration, Z.~Ahmed et~al., {\it {Results from a Low-Energy
  Analysis of the CDMS II Germanium Data}},  {\em Phys. Rev. Lett.} {\bf 106}
  (2011) 131302, [\href{http://arxiv.org/abs/1011.2482}{{\tt
  arXiv:1011.2482}}].

\bibitem{cdms_lite}
{\bf SuperCDMS} Collaboration, R.~Agnese et~al., {\it {Search for Low-Mass
  Weakly Interacting Massive Particles Using Voltage-Assisted Calorimetric
  Ionization Detection in the SuperCDMS Experiment}},  {\em Phys. Rev. Lett.}
  {\bf 112} (2014), no.~4 041302, [\href{http://arxiv.org/abs/1309.3259}{{\tt
  arXiv:1309.3259}}].

\bibitem{super_cdms}
{\bf SuperCDMS} Collaboration, R.~Agnese et~al., {\it {Search for Low-Mass
  Weakly Interacting Massive Particles with SuperCDMS}},  {\em Phys. Rev.
  Lett.} {\bf 112} (2014), no.~24 241302,
  [\href{http://arxiv.org/abs/1402.7137}{{\tt arXiv:1402.7137}}].

\bibitem{cdms_2015}
{\bf SuperCDMS} Collaboration, R.~Agnese et~al., {\it {Improved WIMP-search
  reach of the CDMS II germanium data}},  {\em Phys. Rev.} {\bf D92} (2015),
  no.~7 072003, [\href{http://arxiv.org/abs/1504.05871}{{\tt
  arXiv:1504.05871}}].

\bibitem{simple}
M.~Felizardo et~al., {\it {Final Analysis and Results of the Phase II SIMPLE
  Dark Matter Search}},  {\em Phys. Rev. Lett.} {\bf 108} (2012) 201302,
  [\href{http://arxiv.org/abs/1106.3014}{{\tt arXiv:1106.3014}}].

\bibitem{coupp}
{\bf COUPP} Collaboration, E.~Behnke et~al., {\it {First Dark Matter Search
  Results from a 4-kg CF$_3$I Bubble Chamber Operated in a Deep Underground
  Site}},  {\em Phys. Rev.} {\bf D86} (2012), no.~5 052001,
  [\href{http://arxiv.org/abs/1204.3094}{{\tt arXiv:1204.3094}}]. [Erratum:
  Phys. Rev.D90,no.7,079902(2014)].

\bibitem{picasso}
{\bf PICASSO} Collaboration, S.~Archambault et~al., {\it {Constraints on
  Low-Mass WIMP Interactions on $^{19}F$ from PICASSO}},  {\em Phys. Lett.}
  {\bf B711} (2012) 153--161, [\href{http://arxiv.org/abs/1202.1240}{{\tt
  arXiv:1202.1240}}].

\bibitem{pico2l}
{\bf PICO} Collaboration, C.~Amole et~al., {\it {Dark Matter Search Results
  from the PICO-2L C$_3$F$_8$ Bubble Chamber}},  {\em Phys. Rev. Lett.} {\bf
  114} (2015), no.~23 231302, [\href{http://arxiv.org/abs/1503.00008}{{\tt
  arXiv:1503.00008}}].

\bibitem{pico60}
{\bf PICO} Collaboration, C.~Amole et~al., {\it {Dark Matter Search Results
  from the PICO-60 C$_3$F$_8$ Bubble Chamber}},  {\em Phys. Rev. Lett.} {\bf
  118} (2017), no.~25 251301, [\href{http://arxiv.org/abs/1702.07666}{{\tt
  arXiv:1702.07666}}].

\bibitem{dama_2018}
R.~Bernabei et~al., {\it {First model independent results from
  DAMA/LIBRA-phase2}},  \href{http://arxiv.org/abs/1805.10486}{{\tt
  arXiv:1805.10486}}.

\bibitem{kelso}
C.~Kelso, C.~Savage, M.~Valluri, K.~Freese, G.~S. Stinson, and J.~Bailin, {\it
  {The impact of baryons on the direct detection of dark matter}},  {\em JCAP}
  {\bf 1608} (2016) 071, [\href{http://arxiv.org/abs/1601.04725}{{\tt
  arXiv:1601.04725}}].

\bibitem{green}
A.~M. Green, {\it {Astrophysical uncertainties on direct detection
  experiments}},  {\em Mod. Phys. Lett.} {\bf A27} (2012) 1230004,
  [\href{http://arxiv.org/abs/1112.0524}{{\tt arXiv:1112.0524}}].

\bibitem{bottino_low_mass1}
A.~Bottino, F.~Donato, N.~Fornengo, and S.~Scopel, {\it {Lower bound on the
  neutralino mass from new data on CMB and implications for relic
  neutralinos}},  {\em Phys. Rev.} {\bf D68} (2003) 043506,
  [\href{http://arxiv.org/abs/hep-ph/0304080}{{\tt hep-ph/0304080}}].

\bibitem{bottino_low_mass2}
A.~Bottino, F.~Donato, N.~Fornengo, and S.~Scopel, {\it {Light neutralinos and
  WIMP direct searches}},  {\em Phys. Rev.} {\bf D69} (2004) 037302,
  [\href{http://arxiv.org/abs/hep-ph/0307303}{{\tt hep-ph/0307303}}].

\bibitem{freese_2018}
S.~Baum, K.~Freese, and C.~Kelso, {\it {Dark Matter implications of
  DAMA/LIBRA-phase2 results}},  \href{http://arxiv.org/abs/1804.01231}{{\tt
  arXiv:1804.01231}}.

\bibitem{Kahlhoefer:2018}
F.~Kahlhoefer, F.~Reindl, K.~Schäffner, K.~Schmidt-Hoberg, and S.~Wild, {\it
  {Model-independent comparison of annual modulation and total rate with direct
  detection experiments}},  \href{http://arxiv.org/abs/1802.10175}{{\tt
  arXiv:1802.10175}}.

\bibitem{dobrescu_eft}
B.~A. Dobrescu and I.~Mocioiu, {\it {Spin-dependent macroscopic forces from new
  particle exchange}},  {\em JHEP} {\bf 11} (2006) 005,
  [\href{http://arxiv.org/abs/hep-ph/0605342}{{\tt hep-ph/0605342}}].

\bibitem{reece_eft}
J.~Fan, M.~Reece, and L.-T. Wang, {\it {Non-relativistic effective theory of
  dark matter direct detection}},  {\em JCAP} {\bf 1011} (2010) 042,
  [\href{http://arxiv.org/abs/1008.1591}{{\tt arXiv:1008.1591}}].

\bibitem{haxton1}
A.~L. Fitzpatrick, W.~Haxton, E.~Katz, N.~Lubbers, and Y.~Xu, {\it {The
  Effective Field Theory of Dark Matter Direct Detection}},  {\em JCAP} {\bf
  1302} (2013) 004, [\href{http://arxiv.org/abs/1203.3542}{{\tt
  arXiv:1203.3542}}].

\bibitem{haxton2}
N.~Anand, A.~L. Fitzpatrick, and W.~C. Haxton, {\it {Weakly interacting massive
  particle-nucleus elastic scattering response}},  {\em Phys. Rev.} {\bf C89}
  (2014), no.~6 065501, [\href{http://arxiv.org/abs/1308.6288}{{\tt
  arXiv:1308.6288}}].

\bibitem{catena_dama}
R.~Catena, A.~Ibarra, and S.~Wild, {\it {DAMA confronts null searches in the
  effective theory of dark matter-nucleon interactions}},  {\em JCAP} {\bf
  1605} (2016), no.~05 039, [\href{http://arxiv.org/abs/1602.04074}{{\tt
  arXiv:1602.04074}}].

\bibitem{catena}
R.~Catena and B.~Schwabe, {\it {Form factors for dark matter capture by the Sun
  in effective theories}},  {\em JCAP} {\bf 1504} (2015), no.~04 042,
  [\href{http://arxiv.org/abs/1501.03729}{{\tt arXiv:1501.03729}}].

\bibitem{v0_koposov}
S.~E. Koposov, H.-W. Rix, and D.~W. Hogg, {\it {Constraining the Milky Way
  potential with a 6-D phase-space map of the GD-1 stellar stream}},  {\em
  Astrophys. J.} {\bf 712} (2010) 260--273,
  [\href{http://arxiv.org/abs/0907.1085}{{\tt arXiv:0907.1085}}].

\bibitem{vesc_2014}
T.~Piffl et~al., {\it {The RAVE survey: the Galactic escape speed and the mass
  of the Milky Way}},  {\em Astron. Astrophys.} {\bf 562} (2014) A91,
  [\href{http://arxiv.org/abs/1309.4293}{{\tt arXiv:1309.4293}}].

\bibitem{gf}
S.~K. Lee, M.~Lisanti, A.~H.~G. Peter, and B.~R. Safdi, {\it {Effect of
  Gravitational Focusing on Annual Modulation in Dark-Matter Direct-Detection
  Experiments}},  {\em Phys. Rev. Lett.} {\bf 112} (2014), no.~1 011301,
  [\href{http://arxiv.org/abs/1308.1953}{{\tt arXiv:1308.1953}}].

\bibitem{freese_review}
K.~Freese, M.~Lisanti, and C.~Savage, {\it {Colloquium: Annual modulation of
  dark matter}},  {\em Rev. Mod. Phys.} {\bf 85} (2013) 1561--1581,
  [\href{http://arxiv.org/abs/1209.3339}{{\tt arXiv:1209.3339}}].

\bibitem{dama_1998}
R.~Bernabei et~al., {\it {Searching for WIMPs by the annual modulation
  signature}},  {\em Phys. Lett.} {\bf B424} (1998) 195--201.

\bibitem{nest}
B.~Lenardo, K.~Kazkaz, A.~Manalaysay, J.~Mock, M.~Szydagis, and M.~Tripathi,
  {\it {A Global Analysis of Light and Charge Yields in Liquid Xenon}},  {\em
  IEEE Trans. Nucl. Sci.} {\bf 62} (2015), no.~6 3387--3396,
  [\href{http://arxiv.org/abs/1412.4417}{{\tt arXiv:1412.4417}}].

  
\end{thebibliography}
\end{document}